\documentclass[usenatbib]{mnras}
\usepackage{graphicx}
\usepackage{float}
\usepackage{longtable}
\usepackage{threeparttable}
\usepackage{amsmath}
\usepackage{color}
\usepackage{booktabs}
\usepackage{subcaption}
\usepackage{afterpage}

\newcommand{\Nc}{N$_{\mathrm{C}}$}

\title[Probing PAH size and charge]{Probing the size and charge of Polycyclic Aromatic Hydrocarbons}
\author[A. Maragkoudakis et al.]
{A. Maragkoudakis$^{1}$\thanks{E-mail: amaragko@uwo.ca}, E. Peeters$^{1,2,3}$ and A. Ricca$^{3,4}$\\
	$^{1}$Department of Physics and Astronomy, University of Western Ontario, London, ON, N6A 3K7, Canada\\
	$^{2}$Centre for Planetary Science and Exploration, University of Western Ontario, London, Ontario N6A 5B7, Canada \\
	$^{3}$Carl Sagan Center, SETI Institute, 189 Bernardo Ave., Mountain View, CA 94043, USA\\
	$^{4}$Space Science \& Astrobiology Division, Astrophysics Branch, Mail Stop 245-6, NASA Ames Research Center, Moffett Field, CA 94035, USA}

\begin{document}
	
\date{}
	
\maketitle

\begin{abstract}

We present a new method to accurately describe the ionization fraction and the size distribution of polycyclic aromatic hydrocarbons (PAHs) within astrophysical sources. To this purpose, we have computed the mid-infrared emission spectra of 308 PAH molecules of varying sizes, symmetries, and compactness, generated in a range of radiation fields. We show that the intensity ratio of the solo CH out-of-plane bending mode in PAH cations and anions (referred to as the “11.0” \micron{} band, falling in the 11.0--11.3 \micron{} region for cations and anions) to their 3.3 \micron{} emission, scales with PAH size, similarly to the scaling of the 11.2/3.3 ratio with the number of carbon atoms (\Nc{}) for neutral molecules. Among the different PAH emission bands, it is the 3.3 \micron{} band intensity which has the strongest correlation with \Nc{}, and drives the reported PAH intensity ratio correlations with \Nc{} for both neutral and ionized PAHs. The 6.2/7.7 intensity ratio, previously adopted to track PAH size, shows no evident scaling with \Nc{} in our large sample. We define a new diagnostic grid space to probe PAH charge and size, using the (11.2+11.0)/7.7 and (11.2+11.0)/3.3 PAH intensity ratios respectively. We demonstrate the application of the (11.2+11.0)/7.7 -- (11.2+11.0)/3.3 diagnostic grid for galaxies M82 and NGC~253, for the planetary nebula NGC~7027, and the reflection nebulae NGC~2023 and NGC~7023. Finally, we provide quantitative relations for PAH size determination depending on the ionization fraction of the PAHs and the radiation field they are exposed to.

\end{abstract}


\section{Introduction}

The family of polycyclic aromatic hydrocarbon (PAH) molecules is considered the prime candidate for the prominent emission features in the 3--20 \micron{} spectral range in a plethora of astrophysical sources, including reflection and planetary nebulae, carbon-rich evolved stars, the interstellar medium (ISM), star-forming complexes (H\,$\textsc{ii}$ regions) or entire galaxies (e.g. \citealt{Hony2001}; \citealt{Verstraete2001}; \citealt{Peeters2002}; \citealt{Smith07b}; \citealt{Gordon2008}). The strong emission bands at 3.3, 6.2, 7.7, 8.6, and 11.2 \micron{}, accompanied by sets of weaker features, are the product of radiative cooling of PAHs through their vibrational modes after the absorption of FUV ($ < 13.6 $ eV) photons (\citealt{Leger1984}; \citealt{Allamandola1985}). For a comprehensive review of the interstellar PAH properties, see \cite{Tielens2008}.

PAHs have been extensively utilized in the literature to probe the local physical conditions of their irradiating sources as well as the characteristics of the surrounding environments. Due to their large cross sections, PAH species have a direct impact on the ionization balance of Photo-Dissociation Regions (PDRs) through their interactions with electrons and cations  \citep{LeppDalgarno1988, BakesTielens1994}, and are also considered to be an effective route for molecular hydrogen formation (\citealt{Bauschlicher1998}; \citealt{Mannella2012}; \citealt{Thrower2012}; \citealt{Foley2018}). PAH emission has been calibrated from spectroscopic and photometric bands and used for the estimation of star-formation rates on both galaxy-wide and spatially resolved galactic scales (e.g. \citealt{Peeters2004}; \citealt{Farrah2007}; \citealt{Calzetti2007}; \citealt{Shipley2016}; \citealt{Maragkoudakis2018b}), particularly in the cases of metal-rich and dust-rich environments. In addition, the existence of a galaxy-wide correlation between PAH and CO emission \citep[e.g.][]{Regan2006, Cortzen2019} makes PAHs alternative tracers of the molecular gas in star-forming galaxies.

Variations between the relative strengths, peak positions, and spectral profiles of PAH emission bands have been well documented among different sources, or within a given source (e.g. \citealt{BregmanTemi2005}; \citealt{Smith07b}; \citealt{Galliano2008}; \citealt{Peeters2017}). These variations reflect the range in structures, size, and charge state of the multiple PAH sub-populations, which are set by the physical conditions of the environment. The proper characterization of these variations, through the precise determination of the quantities directly linked to them (i.e. size, charge state, structure), is crucial for determining and constraining the nature of the emitters as well as the physical properties of the irradiating source that governs PAH emission.

In this paper, we utilized the NASA Ames PAH IR Spectroscopic Database\footnote{\href{https://www.astrochemistry.org/pahdb/}{https://www.astrochemistry.org/pahdb/}.} (PAHdb; \citealt{Bauschlicher2010};  \citealt{Boersma2014a}; \citealt{Bauschlicher2018}) to properly characterize the charge and size distribution of astrophysical PAHs by introducing a new diagnostic charge -- size grid space and presenting quantitative relations for PAH size determination for neutral and cationic species and mixtures of them. This paper is structured as follows: Section \ref{sec:Sample} describes our sample selection criteria. The emission models and spectra calculations are outlined in Section \ref{sec:Spectra}. Our PAH size relations, charge -- size grid, and results are presented in Section \ref{sec:Results}. We showcase the astrophysical implications of our method in Section \ref{sec:astroapp}, and discuss method limitations in Section \ref{sec:Limitations}. Finally, a summary of our results and conclusions are given in Section \ref{sec:Summary}.

\section{Sample} \label{sec:Sample}

In this paper, we utilized version 3.00 of the NASA Ames PAHdb consisting of 3139 theoretical absorption spectra, computed using density functional theory (DFT). We constructed our sample of neutral and singly charged cationic molecule pairs based on the following criteria: 

\begin{enumerate}
	\item Contain more than 20 carbon atoms.
	\item Have no nitrogen, oxygen, magnesium or iron atoms.
	\item Have solo C-H bonds\footnote{This criterion excludes the most compact PAHs such as coronene (C$_{24}$H$_{12}$) and hexabenzocoronene (C$_{42}$H$_{18}$).}.
\end{enumerate}

Since we aim to examine the charge and size properties of PAHs within an astrophysical context, we selected PAHs that consist of more than 20 carbon atoms. PAHs with \Nc{} (number of carbon atoms) $ > 20 $ can typically survive destruction via photodissociation when exposed to the strong UV radiation fields of the sources in which they reside (e.g. \citealt{Allain1996}). The second criterion is imposed in order to establish a consistent sample of PAH populations. The 11.2 and 11.0 $ \mu $m emission features in PAH spectra are due to C--H out-of-plane (CH$_{oop}$) modes of solo hydrogens in neutral and cationic species respectively, and therefore applying the third criterion ensures the inclusion of molecules producing these key diagnostic bands for our analysis. 

Given the above criteria, 81 pairs of neutral and singly charged cationic molecules were selected from the v3.00 PAHdb library. To further extend our sample, we included the neutral and cationic pairs of the straight edge molecules\footnote{Now available in version 3.20 of the PAHdb.} presented in \cite{Ricca2018}, as well as their corresponding anions. In total, our final sample consisted of 308 molecules, 133 neutral--cationic pairs and 42 anions, with a size distribution between 22 and 216 carbons. 

\section{Model and spectra} \label{sec:Spectra}

To generate PAH emission spectra from the DFT-computed absorption spectra, we considered PAHs subjected to radiation fields with average photon energies of 6, 8, 10, and 12 eV, as well as the interstellar radiation field (ISRF) model from \citet*{Mathis1983}, where the entire radiation field spectrum is considered instead of the average absorbed photon energy for each PAH (see Appendix \ref{sec:equations}). In all cases, the entire emission cascade is taken into account as the PAH relaxes from the highest excitation level to the vibrational ground state (see \citealt{Boersma2011}). Throughout the paper, we present results from emission spectra of PAHs generated with the ISRF model, and showcase results from emission spectra generated with single-photon absorption of 6, 8, 10, and 12 eV photons in Fig. \ref{fig:multigrid}, with more results given in Appendix \ref{sec:effectonspec}, Fig. \ref{fig:speccomp}, Appendix \ref{sec:scal-rel}, and Tables \ref{tab:fitparams6ev} -- \ref{tab:binparams12ev}. We discuss the effects of different radiation fields in Section \ref{sec:emmodels}.

Because astronomical PAH spectra emanate from highly vibrationally excited molecules, anharmonicity effects are imprinted in their spectra. To account for such effects, a small ($\sim$ 15 cm$ ^{-1}$) redshift to the peak positions of the emission bands relative to the corresponding band positions in absorption is typically applied throughout the literature. However, a recent study of 20 small PAHs (\Nc{} $= 10 - 18$) by \cite{Mackie2018} show that the magnitude and direction of ``pseudo-shifts" due to anharmonic effects are both band and molecule dependent and are typically less than 15 cm$ ^{-1}$. Our integration ranges include the bands with and without a 15 cm$ ^{-1}$ shift. Consequently there is no implicit requirement for a 15 cm$ ^{-1}$ redshift application to account for anharmonicity effects. As such, no redshift was applied to our calculated spectra. We note however, that the conclusions presented in this paper do not depend on a redshift application or not (see Appendix \ref{sec:zspec}, Fig. \ref{fig:Z_comp}).

We convolved the calculated PAH spectral emission bands using Lorentzian line profiles with a full width at half maximum of 15 cm$ ^{-1} $. While a truly isolated harmonic oscillator has a Lorentzian emission profile, astronomical PAH band profiles are distinct from either Lorentzian or Gaussian approximations. However, extensive comparisons have shown that the adopted emission profile does not affect observed PAH trends such as correlations between relative intensities of different features (\citealt{Smith07b}; \citealt{Galliano2008}), and minor to negligible differences are observed in the resulting spectra when using different emission profiles \citep{Shannon2019}.

\subsection{Emission line measurements} \label{sec:EmissionLines}

\begin{figure*}
	\begin{center}
		\includegraphics[keepaspectratio=true,scale=0.45]{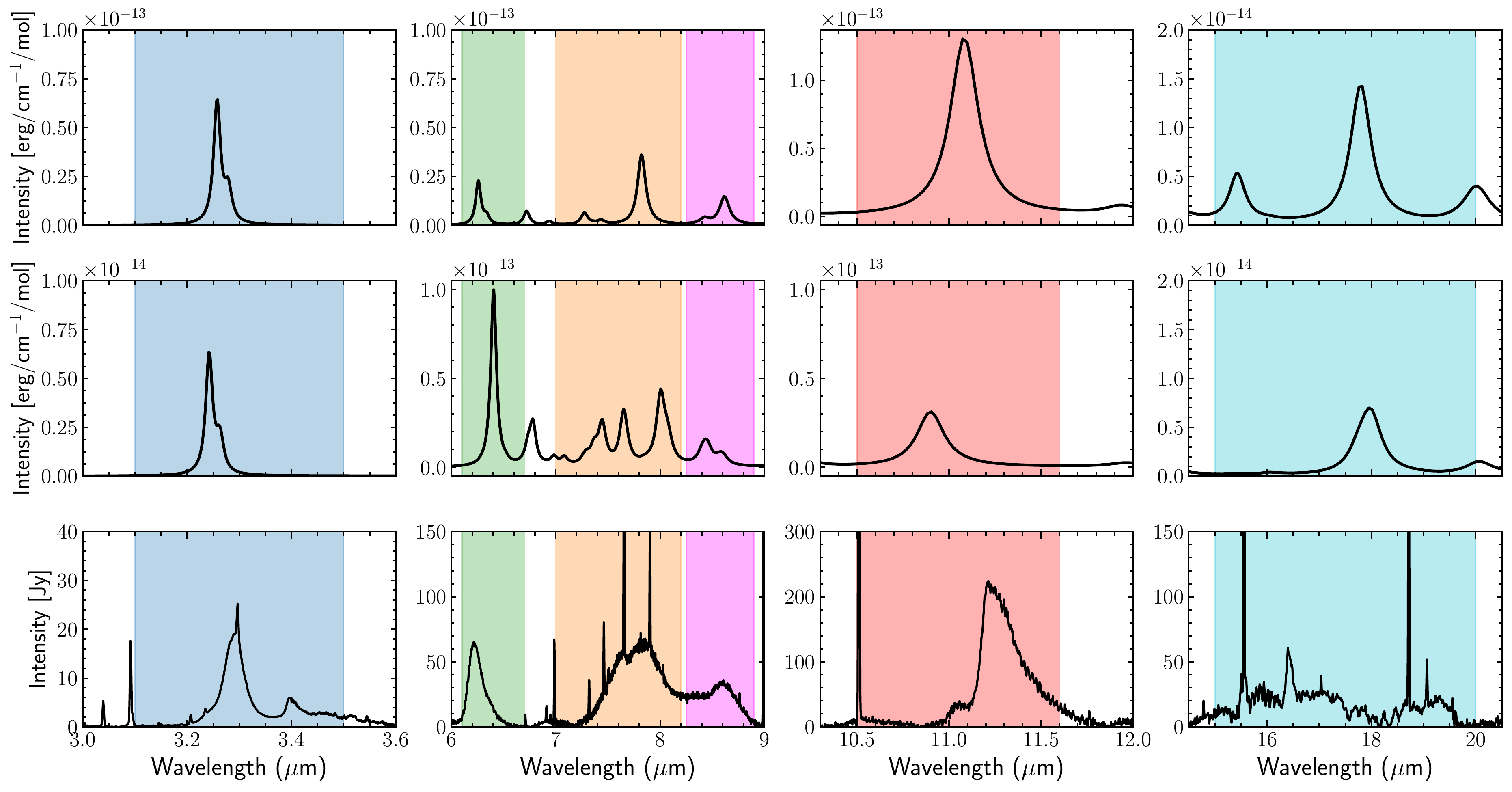}
		\caption[]{Example PAHdb emission spectra of C$_{54}$H$_{18}$ (top row) and C$_{54}$H$_{18}^{+}$ (middle row) as calculated with the ISRF radiation field, and the ISO-SWS spectrum of NGC~7027 shown for comparison (bottom row). The panels present in order from left to right: the 3.3 \micron{} emission region (1st panel), the 6.2, 7.7, and 8.6 \micron{} emission region (2nd panel), the 11.0 and 11.2 \micron{} emission region (3rd panel), and the $\Sigma_{(15-20)}$ \micron{} emission region (4th panel). The respective integration ranges (given in Table \ref{tab:bandrange}) are indicated with the colored rectangles. The PAHdb flux density units correspond to erg/cm$^{-1}$ per mole. The observed flux density (Jy) is proportional to the product of the PAHdb flux density, the PAH photon absorption rate (s$ ^{-1} $ per PAH molecule; \citealt{Pech2002}), and source-dependent factors such as distance, source geometry, and PAH number density.}
		\label{fig:exampspec}
	\end{center}
\end{figure*}

We measured the 3.3, 6.2, 7.7, 8.6, 11.0, and 11.2 $ \mu $m PAH emission band fluxes as the sum of fluxes between predefined wavebands given in Table \ref{tab:bandrange} and shown in Fig. \ref{fig:exampspec}. No separate treatment for plateau emission was considered. We note that the plateau emission in astronomical observations exhibit a spatial morphology distinct from the bands located on top of the plateau emission and may originate in larger-sized species \citep{Peeters2012, Peeters2017}. We adopted the same waveband ranges for neutral and cationic PAHs. The prominent features between 11.0 and 11.5 $ \mu $m are assigned to CH$_{oop}$ bending modes of solo hydrogens. Neutral PAHs contribute principally to the 11.2 $ \mu $m band and cations to the 11.0 $ \mu $m band, although the peak position in each case varies depending on the presence of even-carbon or odd-carbon PAHs (see \citealt{Ricca2018}) and on the PAH edge structure (\citealt{Bauschlicher2008,Bauschlicher2009}). Adjacent solo hydrogens will peak at shorter wavelength than solo hydrogens next to duo, trio and quartet hydrogens \citep{Hony2001}. In order to account for peak position shifts and proper measurement of the full 11.0 and 11.2 $ \mu $m intensity profiles, we applied a wider integration range for these features (see Fig. \ref{fig:exampspec}), defined based on the minimum and maximum profile shifts in our sample spectra. Throughout the paper, we maintain the terms ``11.2" and ``11.0", typically used in the literature, to describe the respective emission features in the spectra of neutrals and cationic PAHs, although their peak position may be shifted and their integration ranges are identical as explained above.

In various figures throughout the paper we give indicative observed astronomical ranges of PAH intensity ratios, based on ISO-SWS observations obtained from \cite{Peeters2002} and \cite{vanDiedenhoven2004}. This observational sample consists of diverse objects, including reflection nebulae (RNe), H$\, \textsc{ii}$ regions, young stellar objects, post-asymptotic giant branch stars, planetary nebulae, and galaxies, and therefore not all of the objects are necessarily exposed to an ISRF.

\begin{table}
    \begin{center}
	    \caption{Wavebands of individual or combined PAH features for neutral and cationic PAHdb molecules.}
	    \label{tab:bandrange}
		\begin{threeparttable}
			\begin{tabular}{lccccc}
				\hline 
				PAH feature & & & &  & Waveband (\micron) \\ 
				\hline
				3.3	& & & & &$ 3.1 - 3.5 $ \\ 
				6.2	& & & & &$ 6.1 - 6.8 $ \\ 
				7.7$^{a}$	& & & & &$ 7.0 - 8.2 $ \\ 
				8.6	& & & & &$ 8.2 - 8.9 $ \\ 
				11.2 (11.0)$^{b}$ & & &  & &$ 10.5 - 11.6 $ \\ 
				$ \Sigma_{(6-9)}^{c} $ & & & & &$ 6.0 - 9.0 $ \\
				$ \Sigma_{(7-9)}^{c} $ & & & & &$ 7.0 - 9.0 $ \\
				$ \Sigma_{(15-20)}^{c} $ & & & & &$ 15.0 - 20.0 $ \\
				\hline
			\end{tabular} 
			\begin{tablenotes}
			    \item[a] The complex of interlaced PAH features generally referred to as the 7.7 \micron{} complex.
			    \item[b] The C-H out-of-plane bending modes of neutrals (11.2) and charged (11.0) PAHs.
				\item[c] The sum of fluxes between the indicated wavelength range.
			\end{tablenotes}
		\end{threeparttable}
		\end{center}
\end{table}

\section{Results and Discussion} \label{sec:Results}

\subsection{Tracing PAH size} \label{sec:pahsize}

The spectral emission characteristics of PAHs are directly linked to their size and charge state, as well as the radiation field to which they are exposed to (see Appendix \ref{sec:effectonspec}, Fig. \ref{fig:speccomp}). PAH charge has a strong effect on the intrinsic band strengths with ionized PAHs exhibiting strong emission between 6 and 9 $ \mu $m (\citealt{Allamandola1999}; Appendix \ref{sec:effectonspec} and Fig. \ref{fig:speccomp} of this paper). Likewise, PAH relative band ratios are influenced by PAH size, with larger PAH contributing most of their emission to longer wavelengths bands while smaller PAHs contributing dominantly to the emission at shorter wavelengths \citep*{ATB1989, Schutte1993}. To a first degree, the emission spectrum of a PAH can be approximated by multiplying the intrinsic intensity in each mode with a blackbody at an average emission temperature of the molecule\footnote{We do not adopt this approximation in our analysis.}. Due to the higher number of vibrational modes in larger PAHs compared to smaller molecules, they attain a lower excitation temperature.

\subsubsection{The 11.2/3.3 ratio} \label{sec:pahsizeneut}

The intensity ratio of 11.2/3.3 is considered the most robust tracer of size for neutral PAH molecules (\citealt{ATB1989}; \citealt{Jourdain1990}; \citealt{Schutte1993}; \citealt{Mori2012}; \citealt{Ricca2012}; \citealt{Croiset2016}). Specifically, \cite{Ricca2012} showed that the 11.2/3.3 ratio scales with the number of carbon atoms for the compact PAH families of coronene and ovalene (comprised of 7 and 4 molecules respectively). In addition, using the PAHdb to calculate the intrinsic emission spectrum for two excitation levels at 6 and 9 eV, \cite{Ricca2012} displayed that the 11.2/3.3 ratio increases with higher absorbed photon energy. Similarly, \cite{Croiset2016} presented the scaling of the 11.2/3.3 ratio with \Nc{} using 27 molecules (with \Nc{} $ < 150 $) from the PAHdb using an average absorbed photon energy of 6.5 eV. The 27 molecules in \citealt{Croiset2016} included the coronene and ovalene families from \cite{Ricca2012}, along with the families of anthracene, tetracene, and pyrene, hence focusing mostly on very compact molecules including only a few non-compact ones.

\begin{figure*}
	\begin{center}
		\hspace*{-0.5cm}\includegraphics[keepaspectratio=true,scale=0.50]{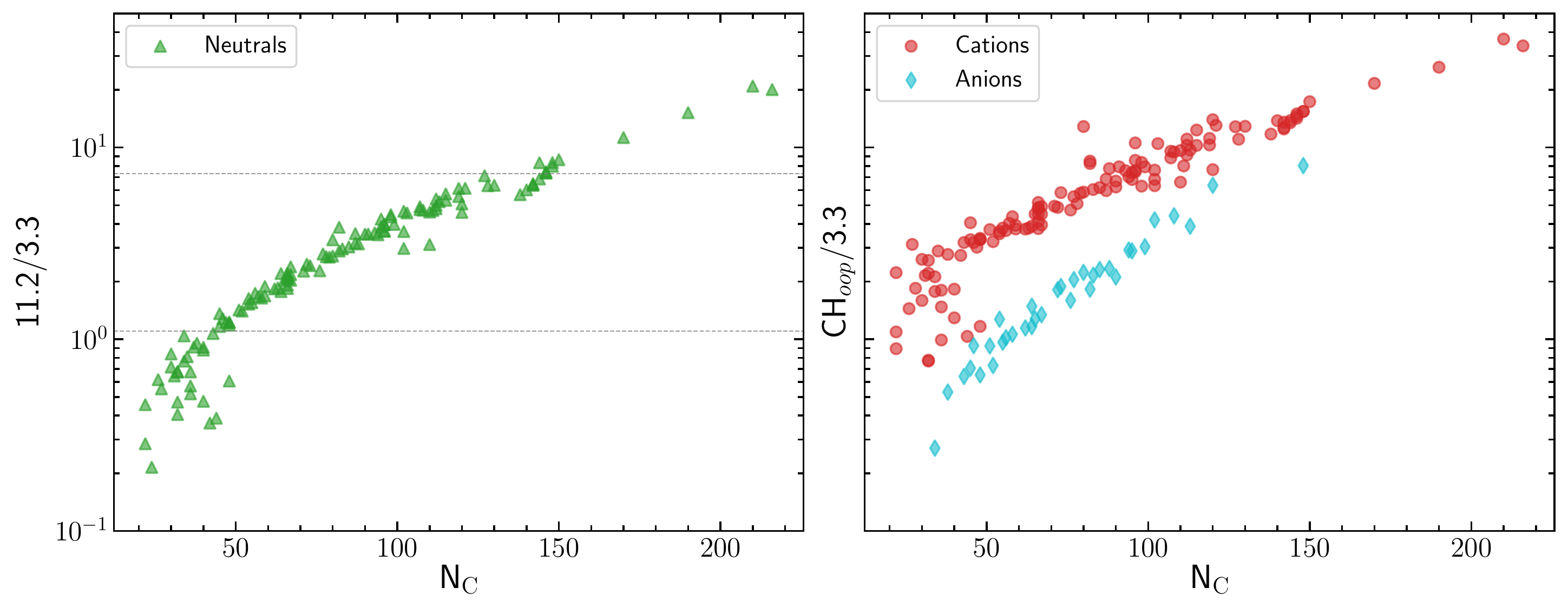}
		\caption{The intensity ratios of 11.2/3.3 for neutral PAHs (left-panel plot) and CH$_{oop}/3.3$ for cations and anions (right-panel plot), as a function of N$ _{\mathrm{C}} $ for the ISRF case. Dashed lines indicate observed astronomical ranges based on ISO-SWS data (see Section \ref{sec:EmissionLines}).}
		\label{fig:goodratiosAll}
	\end{center}
\end{figure*}

Here, we expand on the previous works by employing a much larger sample of neutral PAHdb molecules, with no constraints on \Nc, symmetry, or compactness, thus resulting in an all-inclusive and detailed characterization of the 11.2/3.3 -- \Nc{} correlation. Fig. \ref{fig:goodratiosAll} (left panel) presents the emission ratio of 11.2/3.3 as a function of \Nc{} for the neutral molecules in our sample, with the average solo CH$ _{oop} $ feature peaking at $ 11.09 \pm 0.13 $ \micron. The 11.2/3.3 ratio spans across $\sim$ 2.5 orders of magnitude, thoroughly sampling molecules with number of carbon atoms between 20 and 150, and expanding up to 216 carbons, offering a comprehensive description of the 11.2/3.3 -- \Nc{} correlation. Only the 5 smallest molecules in our sample, with \Nc{} $ < 25 $, appear to deviate from the 11.2/3.3 -- \Nc{} correlation, but overall the 11.2/3.3 intensity ratio proves to be the most efficient tracer of neutral PAH size. Comparison with observational 11.2/3.3 ratios (Fig. \ref{fig:goodratiosAll}, left panel) suggests that astrophysical PAH sizes range between 50 $<$ \Nc{} $<$ 140, consistent with \cite{Allain1996} and \cite{Ricca2012}.

\subsubsection{The 11.0/3.3 ratio} \label{sec:pahsizecat}

The theoretical behavior of the 11.0/3.3 intensity ratio for PAH cations is important when modeling PAH emission, but in practice this is impossible to observe in space. This is because the 3.3 \micron{} emission feature, while emanating from both neutral and cationic populations, it is $ \sim 10$ times stronger in neutrals than cations (Fig. \ref{fig:exampspec}), and as a result the observed 3.3 \micron{} flux will be dominated by the neutral PAHs unless the PAH ionization fraction is above $ \sim 80\% $. As a consequence, the size distribution of cationic PAHs have not been investigated thoroughly in the literature. With the average solo CH$ _{oop} $ feature of our sample's cations peaking at $ 11.00 \pm 0.12 $ \micron, our study reveals the scaling of the 11.0/3.3 intensity ratio of cation PAH molecules with the number of their carbon atoms (Fig. \ref{fig:goodratiosAll}; right panel). The correlation is similar to the 11.2/3.3 -- \Nc{} correlation, but shifted upwards due to the higher contrast between the 11.0 and the 3.3 $ \mu $m intensities for cations, compared to that of the 11.2 and 3.3 $ \mu $m intensities for neutral PAHs. Similarly to the 11.2/3.3 -- \Nc{} relation, smaller molecules (25--50 \Nc{}) show a scatter in the 11.0/3.3 -- \Nc{} space. Because the observed astrophysical measurements of the 11.0/3.3 ratio consist of a mixture of both neutrals and ionic PAHs, which are inseparable in the case of the 3.3 \micron{} feature as explained above, we make no comparison of the 11.0/3.3 ranges of cations with observations.

We further investigated intensity ratios sensitive to size for anion PAH molecules. Although the fraction of anions is expected to be exceedingly small within PDRs given the Far-Ultraviolet (FUV) radiation field from the illuminating star, PAH anions should be more abundant in the diffuse ISM \citep{BakesTielens1994}. We used 42 straight-edge anion molecules from \cite{Ricca2018} and recovered a well-defined correlation between their intensity ratio of the solo CH$ _{oop} $ bending mode emission (with the average peak being at $ 11.34 \pm 0.15 $) to their 3.3 \micron{} emission, with \Nc.

\subsubsection{The 6.2/7.7 ratio} \label{sec:62_77}

The wealth of spectroscopic observations from \textit{Spitzer}/IRS covering the 5 -- 38 \micron{} range did not provide access to the 3.3 \micron{} emission band, which was regarded as sensitive to PAH size \citep{ATB1989, Schutte1993}. Therefore, alternate size indicators had to be employed, and the brightest ionic bands at 6.2 and 7.7 \micron{}, which were less affected by the 9.7 \micron{} silicate absorption, appeared as the best candidates. As such, a popular approach in the literature for determining average PAH sizes, especially for extragalactic astrophysical sources \citep[e.g.][]{ODowd2009, Diamond-Stanic2010, Sandstrom2012, Stierwalt2014, Maragkoudakis2018b}, is based on the usage of the \cite{Draine2001} (hereafter DL01) models, and specifically the employment of the 7.7/11.2 and 6.2/7.7 ratios (DL01; their Fig. 16) to track PAH charge and size respectively. The DL01 model utilizes the estimation of cross-sections per C-atom for neutral and ionized PAHs based on experimental data \citep{Allamandola1999, HudginsAllamandola1999} and astronomical observations \citep{Boulanger1998} and numerical equations to calculate the number of C- and H-atoms for a PAH molecule of a given size. These are then exposed to radiation fields of varying strengths, followed by the calculation of their emission spectrum.

\begin{figure}
		\includegraphics[keepaspectratio=true,scale=0.56]{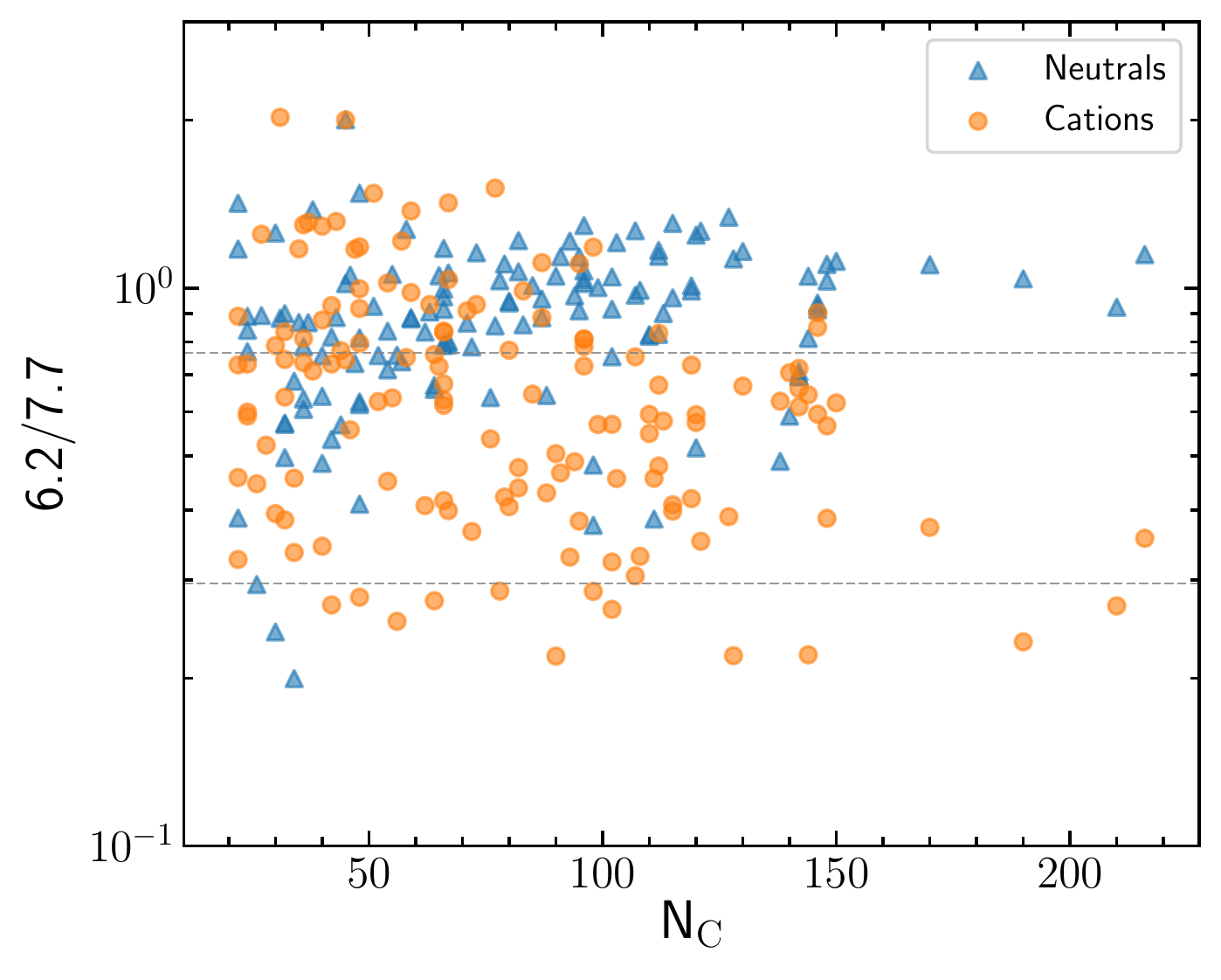}
		\caption{The intensity ratio of 6.2/7.7 (commonly used in DL01 plot to track PAH size) as a function of \Nc{} for the neutral and cationic PAHdb molecules for the ISRF case. Dashed lines indicate observed astronomical ranges based on ISO-SWS data (see Section \ref{sec:EmissionLines}).}
		\label{fig:62_77_Nc}
\end{figure}

The effectiveness of the 6.2/7.7 intensity ratio to serve as a PAH size tracer can be subjected to criticism, considering that the 6.2 and 7.7 \micron{} emission features lie close in wavelength, and hence originate from similar-sized species, as discussed in Section \ref{sec:pahsize}. Furthermore, since the 6.2 and 7.7 \micron{} emission mostly emanates from ionized PAH populations the 6.2/7.7 ratio appears to be a controversial size tracer for the case of neutral PAHs. \cite{Maragkoudakis2018b} presented initial evidence that the 6.2/7.7 ratio does not effectively track PAH size using a small sample of PAHdb molecules. Here, we demonstrate the ambiguity of the 6.2/7.7 ratio as a tracer of PAH size in Fig. \ref{fig:62_77_Nc}, where the 6.2/7.7 intensity ratio is plotted as a function of \Nc{} for both cases of neutral and cationic PAHdb molecules. The presence of significant scatter raises strong concerns about the effectiveness of the 6.2/7.7 ratio to trace PAH size. In addition, observed astronomical ranges are consistent with cationic PAHs of all sizes, rather than neutral PAHs (Fig. \ref{fig:62_77_Nc}). The DL01 PAH models adopt an absorption cross section depending on the number of carbon atoms, the H/C ratio based on fixed prescriptions, and the charge state of the PAH, without accounting for the effects of PAH symmetry or structure. PAH structure determines the amount of H atoms relative to C atoms and the number of peripheral H atoms which are isolated, doubly adjacent, or triply adjacent, therefore influencing the relative strengths of the various C--H and C--C modes \citep[e.g.][]{Peeters2017}. However, a comprehensive understanding on the effects of symmetry and structure on the calculated PAH spectra is yet to be achieved, and further research is needed to fully characterize their systematics. 

The observed 6.2/7.7 ratio variations can be attributed to a number of factors. Specifically, as discussed by \cite{Ricca2012}, PAH structure contributes to the variance of the 7.7/6.2 ratio. Furthermore, several authors reported that the 7.7 \micron{} component is comprised of two subcomponents at 7.6 and 7.8 \micron{} \citep{Bregman1989, Cohen1989, Molster1996, Roelfsema1996, Peeters1999}. Furthermore, \cite{BregmanTemi2005} and \cite{Peeters2017} presented evidence that the 7.7 $ \mu $m complex has at least 2 components with very distinct spatial distribution and different band assignments. Consequently, the complexity of the 7.7 \micron{} emission feature will have a direct impact in the variance of the 6.2/7.7 emission ratio.

\subsubsection{Other PAH ratios} \label{sec:other-ratios}

\begin{figure*}
	\begin{center}
		\includegraphics[keepaspectratio=true,scale=0.45]{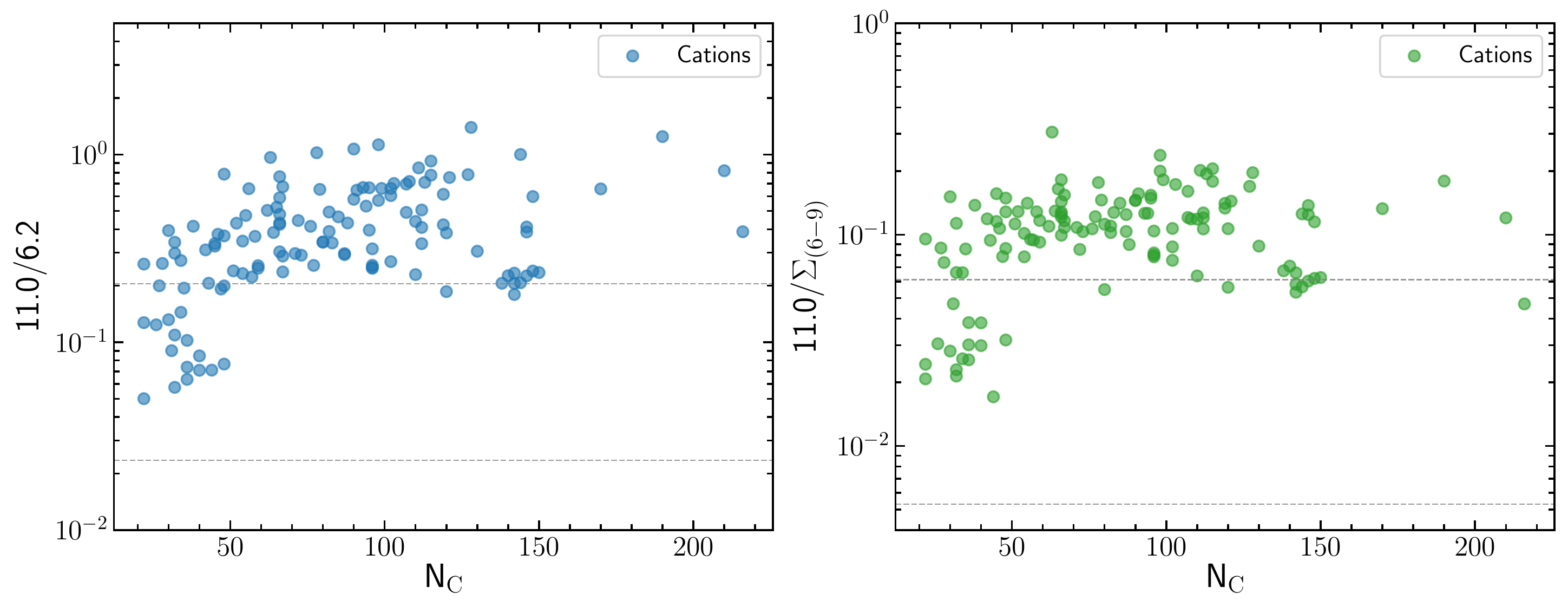}
		\caption{Cation intensity ratios as a function of \Nc\, for the ISRF case. Left panel: the ratio of 11.0/6.2; right: the ratio of 11.0 over the sum of fluxes between 6--9 $ \mu $m ($ 11.0/\Sigma_{(6-9)} $). Dashed lines indicate observed astronomical ranges based on ISO-SWS data (see Section \ref{sec:EmissionLines}).}
		\label{fig:badratiosAll}
	\end{center}
\end{figure*}

We examined the ratio of the 11.0 $\micron$ emission to all prominent features in the spectra of cationic and anionic PAHs, including the sum of fluxes at $\Sigma_{(6-9)}$, $\Sigma_{(7-9)}$ (excluding the 6.2 $\micron$ emission feature), and $\Sigma_{(15-20)}$. The only ratio having a tight correlation with Nc is the 11.0/3.3 ratio for cations.  Similarly, the 11.0/3.3 ratio has a tight correlation with Nc for anions. Fig. \ref{fig:badratiosAll} presents as example the ratios of 11.0/6.2 and 11.0/$ \Sigma_{(6-9)} $ for cations as a function of \Nc. Although an overall increase of 11.0/6.2 with increasing \Nc{} is observed, a non-negligible fraction of different sized molecules have similar intensity ratios, and furthermore the intensity ratio levels at larger sizes. Likewise, for the majority of molecules the 11.0/$ \Sigma_{(6-9)} $ ratio appears to be insensitive to the increase of \Nc{}. The observed astronomical intensity ratio ranges are much lower than the calculated ratios (Fig. \ref{fig:badratiosAll}). However, this is consistent with \cite{Ricca2012} stating that the weakness of the 11.0 \micron{} band in astronomical spectra -compared to model calculations- results from severe dehydrogenation of PAH cations. Given that our sample excludes fully dehydrogenated PAHs, higher 11.0 \micron{} fluxes and ratios are expected in such a case. This should only affect the 11.0 \micron{} feature as the 6.2 and 6 -- 9 \micron{} emission features are dominated by C-C stretching modes. Lastly, examination of intensity ratio combinations between the 6.2, 7.7, and 8.6 \micron{} features with other emission bands revealed no scaling with \Nc. As a result, the 11.0/3.3 ratio is found to be the best proxy for size in the cases of cationic and anionic PAH species.

\subsection{The 3.3 $ \mu $m -- \Nc{} relation} \label{sec:33-Nc}

\begin{figure*}
	\begin{center}
		\includegraphics[keepaspectratio=true,scale=0.38]{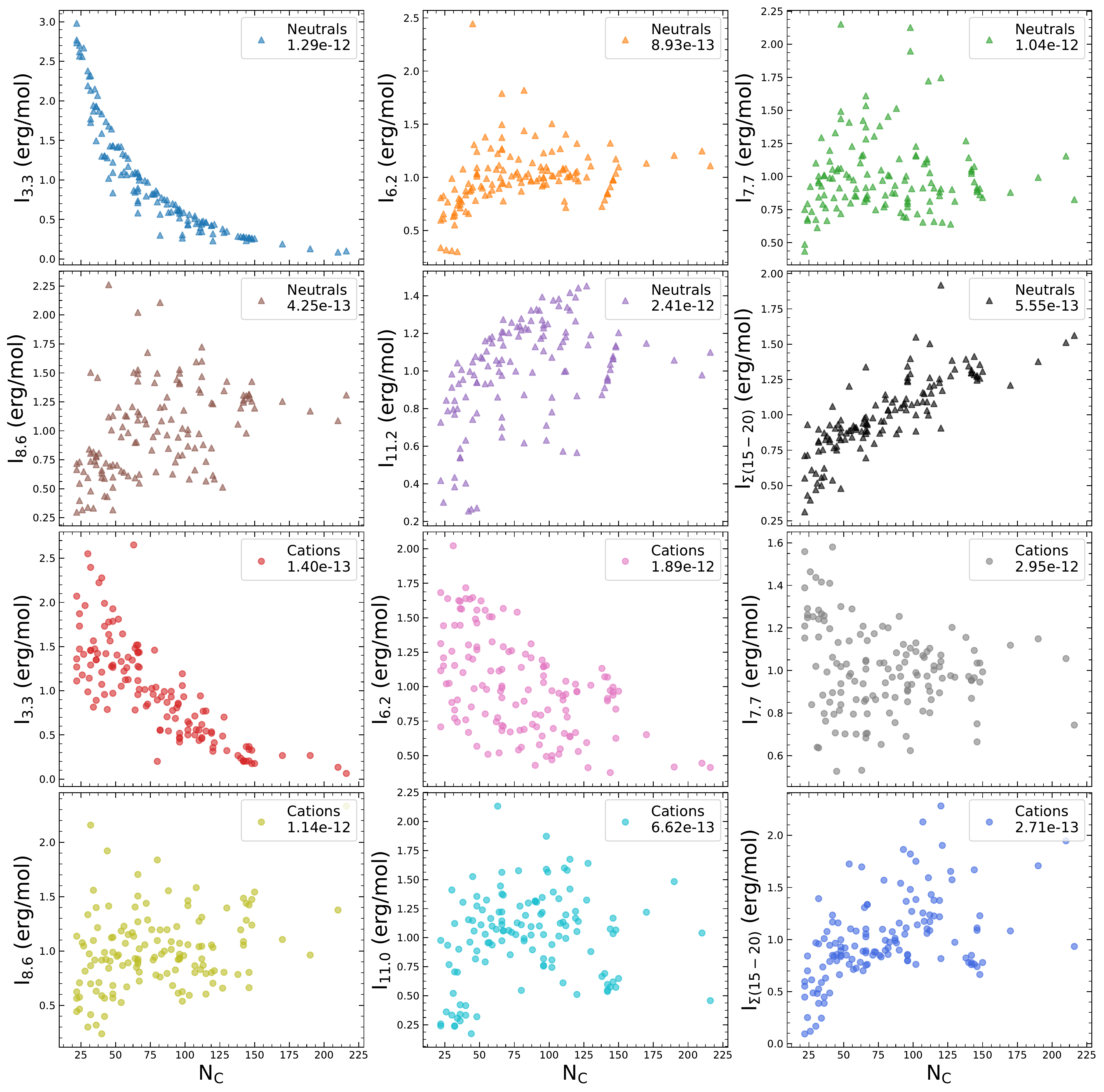}
		\caption{PAH intensities calculated based on the ISRF model, normalized to their mean values (indicated in each panel), as a function of \Nc. Top two rows present the intensities of the 3.3, 6.2, 7.7, 8.6, 11.2, and $\Sigma_{(15-20)}$ micron features as a function of \Nc{} for neutral PAHs, and the bottom two rows present the corresponding intensities as a function of \Nc{} for cationic PAHs.}
		\label{fig:fluxcomp}
	\end{center}
\end{figure*}

In an effort to uncover the main driver behind the 11.2/3.3 and 11.0/3.3 correlations with \Nc, we further examined the relation between the individual intensities of the emission features present in these ratios, as a function of the number of carbon atoms, for both neutral and cationic molecules. In addition, we examined the scaling of all the remaining prominent spectral emission bands (at 6.2, 7.7, and 8.6 \micron), as well as the sum of fluxes between $ 15-20 $ \micron{} ($\Sigma_{(15-20)}$), as a function of \Nc{}. Fig. \ref{fig:fluxcomp} presents the results. Evidently, it is the 3.3 $ \mu $m emission that stands out with the strongest dependence on \Nc{} (decreasing intensity with increasing \Nc) and the least scatter compared to the other emission features, for both neutral and cationic PAHs. 

Weaker correlations with \Nc{} can be also identified in certain cases. In neutral molecules, the 11.2 \micron{} emission increases with \Nc{}, albeit with noticeable scatter. The observed dependencies of the 11.2 and 3.3 \micron{} emission with \Nc{} are in accordance with the results of \cite{Schutte1993} describing the contribution of a PAH feature to the total emission as a function of \Nc{} (their Fig. 9). Furthermore, the intensity of $\Sigma_{(15-20)}$ shows a linear increase with increasing \Nc{} for neutral molecules, but dampens for cations although a trend is still observed. Finally, the 6.2 \micron{} emission in cations shows a weaker dependence with \Nc{} but with significant scatter compared to the 3.3 \micron{} emission.

\begin{figure*}
	\begin{center}
		\hspace*{-0.5cm}\includegraphics[keepaspectratio=true,scale=0.42]{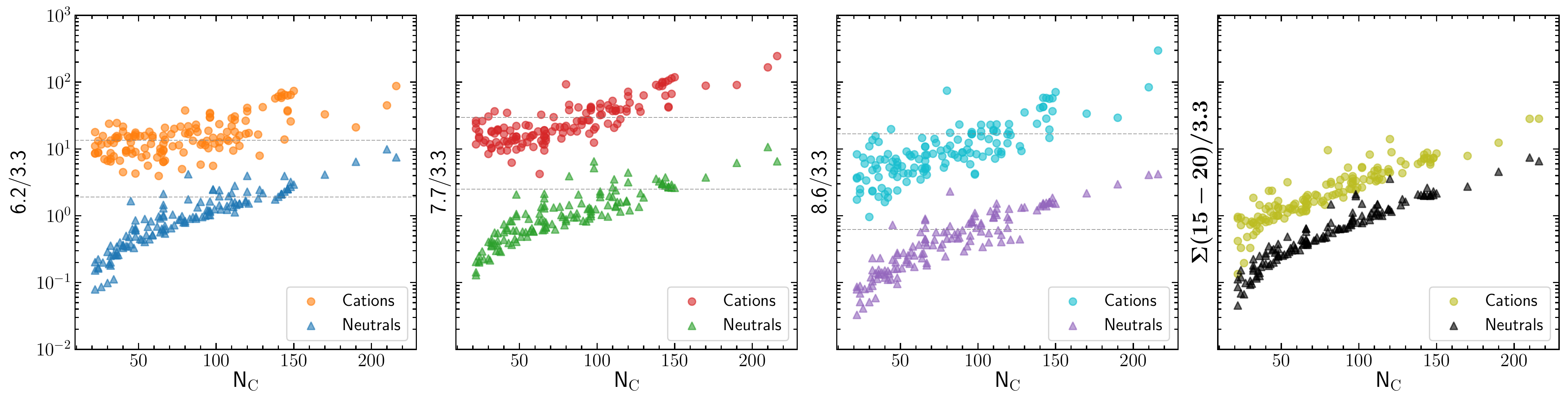}
		\caption{Alternate PAH intensity ratios scaling with \Nc{} for the ISRF case. From left to right, the ratios of 6.2/3.3, 7.7/3.3, 8.6/3.3, and $\Sigma_{(15-20)}$/3.3 as a function of \Nc{} for neutral and cationic PAHs indicated with circles and triangles respectively. Dashed lines, where present, indicate observed astronomical ranges based on ISO-SWS data (see Section \ref{sec:EmissionLines}).}
		\label{fig:newratiosAll}
	\end{center}
\end{figure*}

Given the remarkable sensitivity of the 3.3 $ \mu $m emission to \Nc, we further examined the scaling of the relative intensities of the prominent 6.2, 7.7 and 8.6 $ \mu $m emission components and the sum of fluxes between $ 15-20 $ \micron\, with the 3.3 $ \mu $m intensity. Fig. \ref{fig:newratiosAll} presents the scaling of the 6.2/3.3, 7.7/3.3, 8.6/3.3, and $\Sigma_{(15-20)}$/3.3 intensity ratios as a function of \Nc{} for neutral and cationic PAHs. All four ratios examined display a correlation with \Nc{} across 2 orders of magnitude and can serve as alternate size tracers, although their scaling with \Nc{} is less pronounced and more scatter is induced than in the case of the 11.2/3.3 and 11.0/3.3 ratios. Overall, the ratios of neutral molecules present a firmer scaling with \Nc{} and less scatter compared to cations, given the most notable dependence of the 3.3 and 11.2 \micron{} emissions with \Nc. Astronomical observed ranges fall in between the cationic and neutral PAH ranges, which is expected considering that the 6.2, 7.7, and 8.6 \micron{} features are attributed predominantly to ions while the 3.3 \micron{} feature is attributed to neutral PAHs. We also note that \cite{Boersma2010} presented an estimate for PAH sizes based on the ratio of 15--20 \micron{} C--C--C/6--9 \micron{} C--C emission (their Figure 19 and Section 4.4), and \cite{Tappe2012} extrapolated their PAH size estimation to calculate PAH sizes in the supernova remnant N132D in the Large Magellanic Cloud. However, despite the almost linear scaling of the $\Sigma_{(15-20)}$ emission with \Nc{}, the $\Sigma_{(15-20)}$/$\Sigma_{(6-9)}$ ratio presents a weaker dependence with \Nc{} compared to the well-defined scaling of the $\Sigma_{(15-20)}$/3.3 ratio. We ascribe the difference with the results from \citealt{Boersma2010} to the different calculation of the band intensities. Specifically, these authors apply a fixed ratio for the intrinsic cross-section ratio of 0.013 for all sizes while in this paper, no such assumptions was made.

\subsection{The 11.2(11.0)/7.7 -- 11.2(11.0)/3.3 plane} \label{sec:Newplane}

The well-established correlation between the 11.2/3.3 and 11.0/3.3 intensity ratios with \Nc{}, combined with the efficiency of the 11.2/7.7 and 11.0/7.7 ratios to separate neutral and charged PAH populations, constitutes the 11.2(11.0)/7.7 -- 11.2(11.0)/3.3\footnote{The 11.2(11.0)/7.7 term indicates the usage of the 11.2/7.7 and 11.0/7.7 ratios for the case of neutral and cationic PAHs respectively, and accordingly the 11.2(11.0)/3.3 term indicates the usage of the 11.2/3.3 ratio for neutrals and the 11.0/3.3 ratio for cations. We remind the reader that the same integration ranges are used for neutral and cationic species, but we maintain the terms ``11.2" and ``11.0" in accordance with the literature (see Section \ref{sec:EmissionLines}).} plane ideal for the examination of the degree of ionization and the size distribution of astrophysical sources. The joint power of the 11.2/7.7 and 11.0/7.7 ratios to track PAH charge is illustrated when considering the relative strength of the features in each ratio. Specifically, the strength of the coupled C--C stretching and C--H  in-plane bending modes, which produce the 7.7 $ \mu $m complex emission, increases considerably upon ionization (e.g. \citealt{Allamandola1999}), having comparable strength to the C--H out-of-plane bending modes, which produce the 11.0 $ \mu $m feature. In contrast, the strength of the 11.2 emission is dominating over the 7.7 emission in neutral PAHs (see Appendix \ref{sec:effectonspec} and Fig. \ref{fig:speccomp}). As a result, the strength of the 11.2/7.7 ratio in neutral PAHs will always be higher compared to the 11.0/7.7 ratio in cations, leading to an effective separation of the two populations. 

\cite{Maragkoudakis2018b} have initially presented the 11.2(11.0)/7.7 -- 11.2(11.0)/3.3 plane and questioned whether the 6.2/7.7 ratio is effective in tracking PAH size, using a small sample of 34 symmetric pure PAH molecules (including the families of coronene and ovalene) with $ 24 < $ \Nc{} $ < 170 $ drawn from the v3.00 PAHdb libraries. In this paper we vastly expand the work of \cite{Maragkoudakis2018b} by: i) using a much larger sample of 133 neutral and cation PAH pairs (266 PAH molecules), including straight edge molecules from the next release of the PAHdb (v3.10); ii) using molecules with a large diversity of symmetries; and iii) covering a broad range of molecule sizes: $ 22 < ${} \Nc{} $ < 216 $. 

\begin{figure*}
	\begin{center}
		\includegraphics[keepaspectratio=true,scale=0.50]{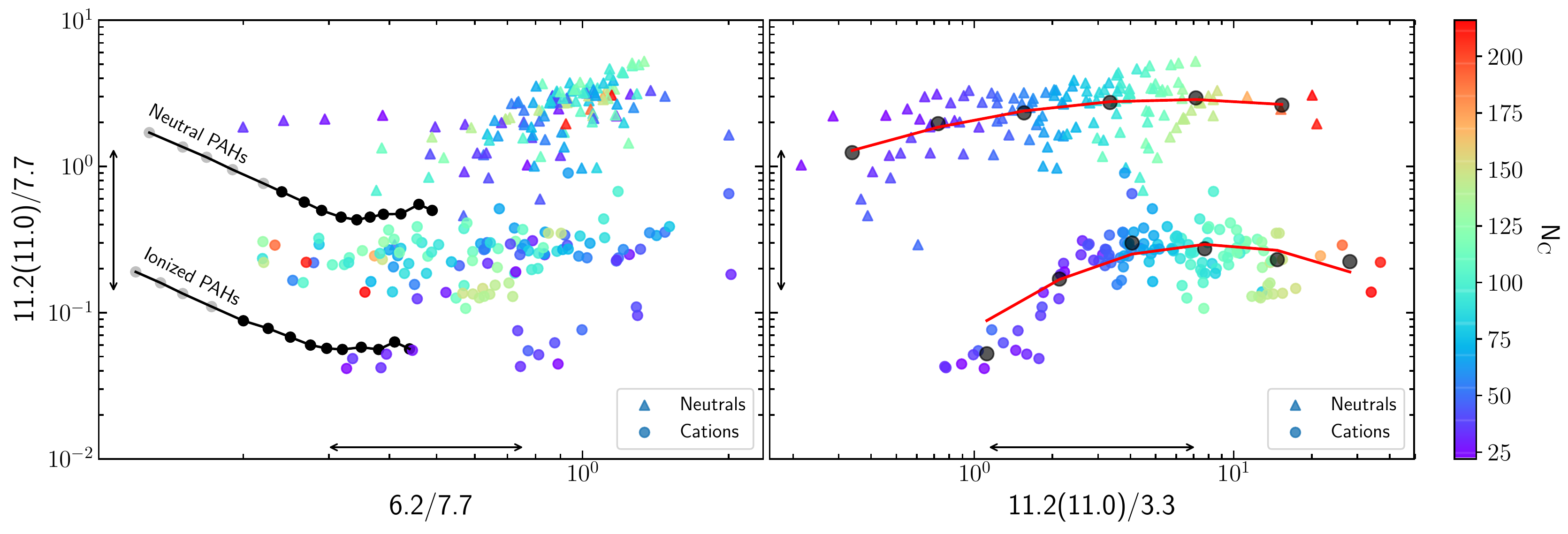}
		\caption{Comparison between the diagnostic diagrams of \protect\cite{Draine2001} (left panel) and the new PAH charge -- size diagram defined in \protect\cite{Maragkoudakis2018b} and this work (right panel). Both panels are populated with the PAHdb sample molecules calculated with the ISRF, with triangles and circles representing neutral and cationic PAHs respectively, color-coded based on the number of their carbon atoms (\Nc). Black circles in the right-panel plot are the median 11.2(11.0)/7.7 values in equally spaced 11.2(11.0)/3.3 bins. The red curves are one dimensional log parabola model fits to the binned values (equation \ref{eq:logparab}). For comparison, the points on the DL01 tracks (left panel) that correspond to molecule sizes between 18 $ < ${} \Nc{} $ < 240$ based on the DL01 models are colored in black, while the gray points correspond to molecules with \Nc $ > 240 $ according to the DL01 models. Black arrows indicate observed astronomical ranges of the 11.2/7.7, 6.2/7.7, and 11.2/3.3 ratios based on ISO-SWS data.}
		\label{fig:newdiagram}
	\end{center}
\end{figure*}

A comparison between the DL01 11.2/7.7 -- 6.2/7.7 diagram and the new 11.2(11.0)/7.7 -- 11.2(11.0)/3.3 diagnostic plane is presented in Fig. \ref{fig:newdiagram}. In the DL01 11.2(11.0)/7.7 -- 6.2/7.7 diagram (Fig. \ref{fig:newdiagram}; left panel), the PAHdb molecules are displaced from the DL01 tracks in both terms of charge and size. Specifically, there is an overall inter-mixture of different sized molecules throughout the 6.2/7.7 range for both populations of neutral and cationic PAHs. Note that the size distribution in the DL01 tracks increases from larger to smaller 6.2/7.7 values, i.e. from right to left. The bulk of neutral molecules are located at high 6.2/7.7 values, outside of the DL01 tracks, while a handful of small-sized PAHdb molecules can be seen in 6.2/7.7 values that correspond to \Nc $ > 240 $ in the DL01 models. Approximately half of the sample cations are located outside the DL01 tracks to higher 6.2/7.7 values, while the rest are distributed throughout the range of the DL01 tracks with no specific order in terms of \Nc, with a subset found at 6.2/7.7 values that correspond to \Nc $ > 240 $ in the DL01 models. Additionally, only a small fraction of the DL01 tracks falls within observed astronomical ranges of the 6.2/7.7 ratio (Fig. \ref{fig:newdiagram}; left panel), while the majority of PAHdb cations coincide with the observed ranges. Finally, the PAHdb cations have the largest PAHs at lower 6.2/7.7 values (towards the left-hand side of the plot) compared to the largest neutral PAHs found at higher 6.2/7.7 values (towards the right-hand side of the plot), in contrast to the DL01 models where the PAH size increases at lower 6.2/7.7 values for both neutral and cationic PAHs.

Furthermore, although the same radiation field (ISRF, \citealt{Mathis1983}) was used to generate the emission spectra of our sample molecules and the DL01 tracks, there is an overall offset for all molecules to higher 11.2(11.0)/7.7 values compared to the locations of the pure cationic and pure neutral molecule tracks in the DL01 models, as seen in Fig. \ref{fig:newdiagram}. Specifically, the majority of cations are occupying the space between the ionized and neutral tracks, while neutral molecules are located above the track of the purely neutral molecules. We note that part of the observed offset between the DL01 models and our results may be due to the usage of enhancement factors applied by DL01, based on comparison with laboratory data of a few symmetric and mostly small (\Nc{} $ < 32 $) PAHs, in order to better fit observed PAH spectra.

In the new diagnostic 11.2(11.0)/7.7 -- 11.2(11.0)/3.3 plane (Fig. \ref{fig:newdiagram}; right panel) a distinct and smooth transition from smaller to larger molecules with increasing 11.2(11.0)/3.3 values is displayed, for both neutral and cationic PAHs. Outliers that reside between the pure neutral and cationc tracks are mostly due to two classes of PAHs: PAHs containing defects (five- and seven-membered rings; e.g. PAHdb UIDs 591, 595, 603, 722, 727) and partially dehydrogenated PAHs (e.g. PAHdb unique identifiers (UIDs) 693, 696, 700, 714). Additional outliers include very elongated PAHs and non-planar PAHs which will be less stable than more compact and planar PAHs. In order to define diagnostic tracks for the size distribution of both populations, average values of 11.2(11.0)/7.7 are calculated in equally-sized 11.2(11.0)/3.3 bins, and the binned data (black points in the right panel of Fig. \ref{fig:newdiagram}) are fitted with a one dimensional log parabola model defined as:
\begin{equation} \label{eq:logparab}
f(x) = A\left(\frac{x}{x_{0}}\right)^{-\alpha-\beta\,log\,\left(\frac{x}{x_{0}}\right)}
\end{equation}
where A is the amplitude, $ \alpha $ is the power-law index, $ \beta $ is the curvature of the parabola, and $ x_{0} $ the reference point for the normalization. All fit parameters, \Nc{} values, and intensity ratios for the binned points are given in Tables \ref{tab:fitparams} and \ref{tab:binparams}.

As seen in Fig. \ref{fig:newdiagram}, astronomical observed ranges of the 11.2/7.7 ratio fall in between the cationic and neutral PAH tracks in both the DL01 and PAHdb defined tracks, highlighting the efficiency of the 11.2/7.7 ratio to track the PAH ionization fraction. Similarly, the astronomical 11.2/3.3 ranges include the majority of both neutral and cationic PAHdb molecules with \Nc{} up to $ \sim 125$.

\subsection{The PAH charge -- size grid} \label{sec:grid}

The 11.2(11.0)/7.7 -- 11.2(11.0)/3.3 diagnostic plane, combined with the corresponding size distribution tracks presented in Section \ref{sec:Newplane}, can characterize the charge state and the average size of purely neutral or cationic PAH species. Astrophysical PAH families though contain a mixture of neutral and cation molecules. As such, the 11.2(11.0)/7.7 -- 11.2(11.0)/3.3 plane can only be indicative of the charge state and size of PAHs within astrophysical sources, depending on the relative location of the source within the pure neutral and cation tracks.

\begin{figure}
	\begin{center}
		\hspace*{-0.8cm}
		\includegraphics[keepaspectratio=true,scale=0.52]{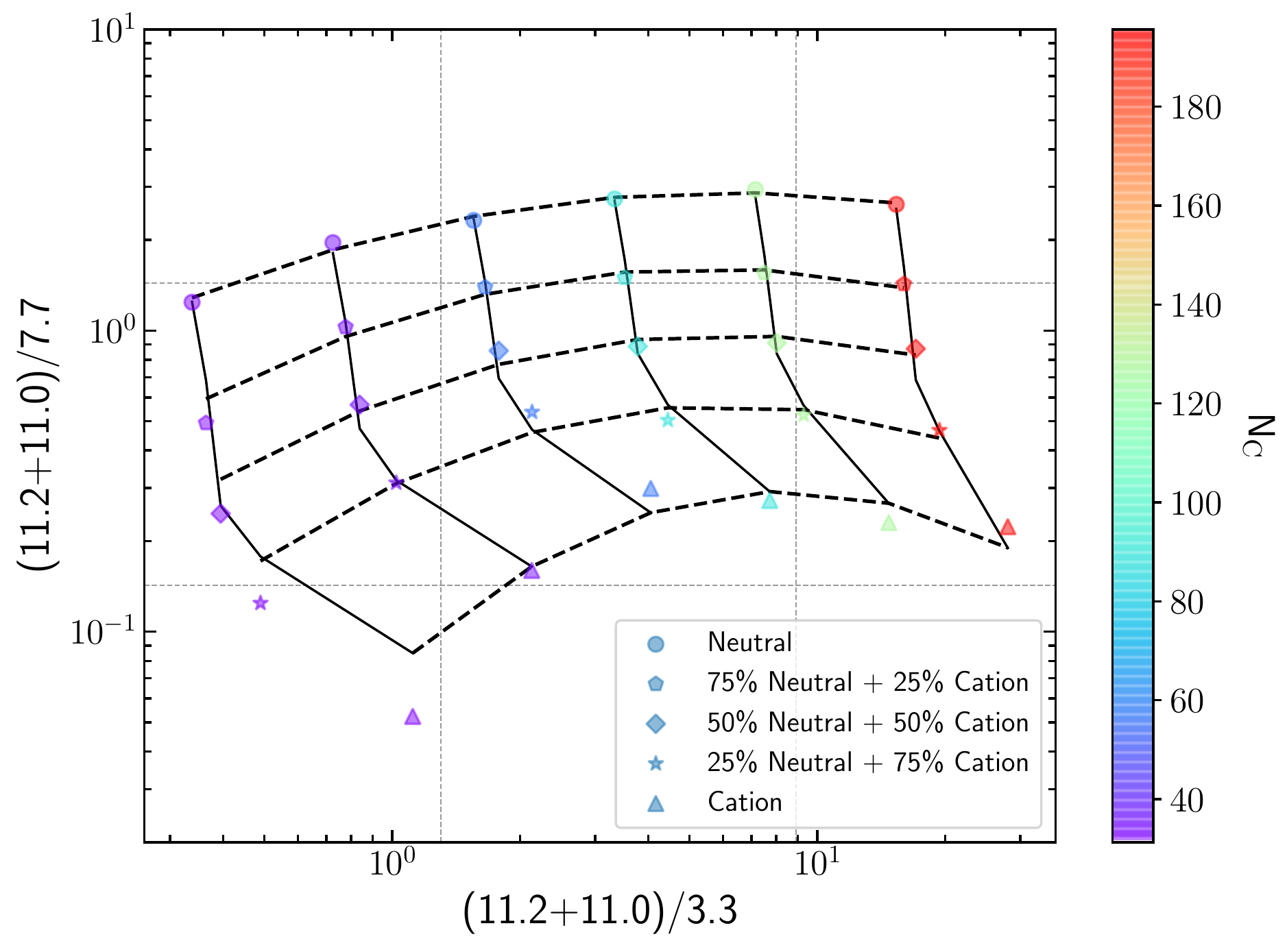}
		\caption{The charge -- size grid from PAH spectra generated with the ISRF model. The points are the median (11.2+11.0)/7.7 values in equally-sized (11.2+11.0)/3.3 bins for spectra of a given ionization fraction as described in Section \ref{sec:grid}. Points of different shapes correspond to different ionization fractions. Dashed lines are one dimensional log parabola model fits to the binned values (equation \ref{eq:logparab}), and solid lines are third-degree polynomial fits (equation \ref{eq:polynom}) to the binned values between groups of points with similar \Nc{} among different ionization fractions. Points are color-coded based on the average \Nc{} of the points along each solid-line track. Grey dashed lines indicate observed astronomical ranges based on ISO-SWS data (see Section \ref{sec:EmissionLines}).}
		\label{fig:grid}
	\end{center}
\end{figure}

In order to provide a more comprehensive description of the charge state and size distribution of astrophysical PAHs, we expanded the 11.2(11.0)/7.7 -- 11.2(11.0)/3.3 diagnostic diagram, constructing a grid in the (11.2+11.0)/7.7 -- (11.2+11.0)/3.3 space (Fig. \ref{fig:grid}). Aside from the purely neutral and cationic molecule spectra, we synthesized spectra that correspond to populations with a mixture of neutral and cationic PAH components, thus different ionization fractions. The synthesized spectra were created by combining the individual spectra of neutral--cation molecule pairs within predefined weighting schemes. Specifically, we have created spectra with neutral--cation relative contributions of: i) 75\% -- 25\% (N75C25), ii) 50\% -- 50\% (N50C50), and iii) 25\% -- 75\% (N25C75) respectively. Following the methodology of Section \ref{sec:Newplane}, we calculated average (11.2+11.0)/7.7 values in equally-sized (11.2+11.0)/3.3 bins for the purely neutral and cationic PAHs, as well as for each group of synthesized spectra. The respective binned data were fitted with a one dimensional log parabola model as described by equation \ref{eq:logparab}, defining a set of five tracks across the (11.2+11.0)/3.3 axis (dashed lines in Fig. \ref{fig:grid}). The tracks connecting binned data of similar \Nc{} between groups of different ionization fractions, are calculated by fitting the respective points using a third-degree polynomial of the form:
\begin{equation} \label{eq:polynom}
P = \sum_{i=0}^{3} Ci*x^{i}
\end{equation}
The points are color-coded based on the average \Nc{} of the similar bins among the different ionization fraction tracks. The fit parameters and average \Nc{} values are given in Tables \ref{tab:fitparams} and \ref{tab:binparams}. Comparison with (11.2+11.0)/3.3 ranges from ISO-SWS observations suggest that the average astrophysical PAH sizes range approximately between $ 40 < $ \Nc{} $ < 140 $, with the smallest and largest PAH species not being present. This PAH charge -- size grid is a powerful tool to probe the average PAH charge and size distribution within astrophysical sources as well as compare these properties among different sources (see Sections \ref{sec:emmodels} and \ref{sec:astroapp}).

In a different approach, \cite{Mori2012} presented two model grids for the 3.3/11.2 and 7.7/11.2 intensities, constructed for a mixture of neutral and ionized PAHs of different ionized fractions, assuming the \cite{Draine:07} PAH cross sections, a power-law size distribution of PAHs with $20 <$ \Nc{} $< 4000$ and $40 <$ \Nc{} $< 4000$, and exposure to a blackbody radiation field of different temperatures. For a given minimum PAH size (\Nc{} of 20 or 40), these grids explore the influence of the PAH ionization fraction and the radiation field on the considered PAH intensity ratios. Based on the two grids, these authors attributed the observed intensity variation in their sample to destruction of small PAHs and a varying degree of ionization. They furthermore concluded that the 3.3 \micron, PAH band provides clues on the size distribution (i.e. the minimum PAH size) and/or the excitation conditions of PAHs. In contrast, here we emphasize the influence of PAH size and the PAH ionization fraction on these ratios given a radiation field.

\begin{table}
		\caption{The ISRF charge--size grid log parabola and polynomial fit parameters (presented in Fig. \ref{fig:grid}) as described by equations \ref{eq:logparab} and \ref{eq:polynom}.}
		\label{tab:fitparams}
		\hspace*{-0.8cm}
		\begin{tabular}{@{}lcccc}
			\hline
			\multicolumn{5}{c}{ISRF Log parabola parameters (eq. \ref{eq:logparab})} \\
			Charge state & A & x$ _{0} $ & $ \alpha $ & $ \beta $ \\
			\hline
			Neutral & 0.81 & 0.16 & -0.70 & 0.10 \\
			75\% Neutral + 25\% Cation & 1.28 & 1.50 & -0.35 & 0.13 \\
			50\% Neutral + 50\% Cation & 0.66 & 1.24 & -0.47 & 0.15 \\
			25\% Neutral + 75\% Cation & 0.21 & 0.61 & -0.86 & 0.19 \\
			Cation & 0.03 & 0.52 & -1.68 & 0.30 \\
			\hline
			\multicolumn{5}{c}{ISRF $3^{\mathrm{rd}}$ degree polynomial parameters (eq. \ref{eq:polynom})} \\
			$\overline{\mathrm{N_{C}}}$ & c0 & c1 & c2 & c3 \\
			\hline
			31 & 30.65 & -164.93 & 277.30 & -137.97 \\
			35 & 48.87 & -126.55 & 103.78 & -25.87 \\
			56 & 72.40 & -90.53 & 36.28 & -4.52 \\
			89 & 101.35 & -61.44 & 12.06 & -0.75 \\
			128 & 136.22 & -40.20 & 3.87 & -0.12 \\
			196 & 178.09 & -25.67 & 1.22 & -0.02 \\
			\hline
		\end{tabular} 
\end{table}

\begin{table*}
    \begin{center}
    \begin{minipage}{125mm}
                \caption{ISRF charge -- size grid average points parameters.}
                \label{tab:binparams}
                \begin{tabular}{ccccccccc}
                \toprule
                \multicolumn{3}{c}{Neutrals} & \multicolumn{3}{c}{N75C25} & \multicolumn{3}{c}{N50C50} \\
                \cmidrule(lr){1-3} \cmidrule(lr){4-6} \cmidrule(lr){7-9}
                \Nc{} & $\frac{11.2}{3.3}$ & $\frac{11.2}{7.7}$ & \Nc{} & $\frac{(11.2+11.0)}{3.3}$ & $\frac{(11.2+11.0)}{7.7}$  & \Nc{} & $\frac{(11.2+11.0)}{3.3}$ & $\frac{(11.2+11.0)}{7.7}$ \\
                \cmidrule(lr){1-3} \cmidrule(lr){4-6} \cmidrule(lr){7-9}    
                31 & 0.34 & 1.24 & 31 & 0.36 & 0.49 & 30 & 0.39 & 0.25 \\
                35 & 0.73 & 1.96 & 35 & 0.78 & 1.03 & 35 & 0.84 & 0.57 \\
                56 & 1.56 & 2.32 & 56 & 1.65 & 1.39 & 56 & 1.78 & 0.86 \\
                86 & 3.34 & 2.74 & 87 & 3.52 & 1.50 & 87 & 3.78 & 0.89 \\
                126 & 7.15 & 2.94 & 127 & 7.50 & 1.55 & 129 & 8.03 & 0.91 \\
                196 & 15.34 & 2.63 & 196 & 15.98 & 1.43 & 192 & 17.05 & 0.87 \\
                \cmidrule(lr){1-3} \cmidrule(lr){4-6} \cmidrule(lr){7-9}
                \multicolumn{3}{c}{N25C75} & \multicolumn{3}{c}{Cations} & \multicolumn{3}{c}{} \\
                \cmidrule(lr){1-3} \cmidrule(lr){4-6}
                \Nc{} & $\frac{(11.2+11.0)}{3.3}$ & $\frac{(11.2+11.0)}{7.7}$ & \Nc{} & $\frac{11.0}{3.3}$ & $\frac{11.0}{7.7}$ \\
                \cmidrule(lr){1-3} \cmidrule(lr){4-6}
                31 & 0.49 & 0.12 & 34 & 1.12 & 0.05 & & & \\
                34 & 1.02 & 0.31 & 34 & 2.13 & 0.16 & & & \\
                56 & 2.13 & 0.54 & 58 & 4.06 & 0.30 & & & \\
                89 & 4.45 & 0.50 & 96 & 7.73 & 0.27 & & & \\
                127 & 9.29 & 0.52 & 129 & 14.73 & 0.23 & & & \\
                196 & 19.38 & 0.47 & 196 & 28.06 & 0.22 & & & \\
                \bottomrule
                \end{tabular}
    \end{minipage}
    \end{center}
\end{table*}

\subsection{PAH size scaling relations} \label{sec:calibrations}

The location of a source in the PAH charge -- size grid indicates its ionization fraction and provides an estimate of the average PAH size based on the \Nc{} of the binned values at the proximity of the source (see Table \ref{tab:binparams}). For a precise determination of the PAH size though, explicit sets of relations are required describing the scaling between 11.2/3.3, 11.0/3.3, and (11.2+11.0)/3.3 ratios with \Nc{}, for neutral, cationic, and mixed PAH populations respectively. The ionization balance, which describes the relative contribution of neutral and cationic PAHs, depends on the photoionization rate and the electron recombination rate of PAHs. These, in turn, depend on the strength of the local UV field ($G_{\mathrm{0}}$\footnote{Given in units of the Habing field ($1.2 \times 10^{-4}$ erg cm$^{-2}$ s$^{-1}$ sr$^{-1}$).}), the electron density $n_{e}$, and the gas temperature $T$ (e.g. \citealt{BakesTielens1994}). Consequently, the emitted PAH spectrum depends on the local physical conditions, and variations in the measured emission bands and intensity ratios are expected upon variations in the physical conditions. Therefore, for a proper \Nc{} determination specific scaling relations are required depending on the ionization fraction of the PAHs within a source.

\begin{figure*}
	\begin{center}
		\includegraphics[keepaspectratio=true,scale=0.40]{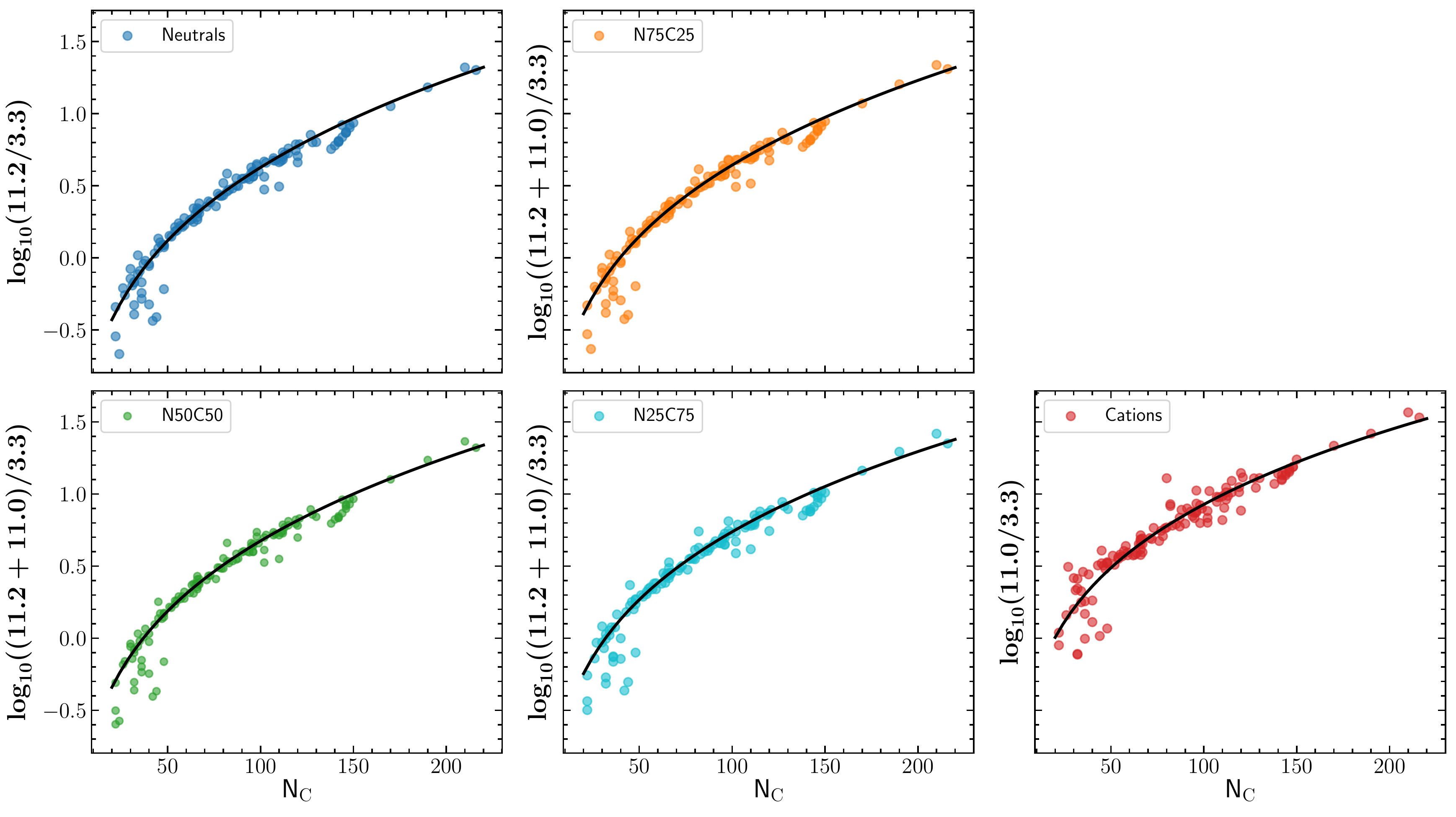}
		\caption{The intensity ratios of (11.2+11.0)/3.3 as a function of N$ _{\mathrm{C}} $ for the different ionization fractions (described in Section \ref{sec:grid}) for PAH spectra generated with an ISRF model. For pure neutral and cationic molecules the intensity ratios correspond to the 11.2/3.3 and 11.0/3.3 ratios respectively. The black curve is a power-law fit described by equation \ref{eq:powerlaw}, with the fit parameters given in Table \ref{tab:calibr}.}
		\label{fig:calib}
	\end{center}
\end{figure*}

Figure \ref{fig:calib} presents the scaling of the $\log(11.2/3.3)$, $\log((11.2+11.0)/3.3)$, and $\log(11.0/3.3)$ with \Nc{} for the different ionization fractions, parameterized with a power-law fit of the form:
\begin{equation} \label{eq:powerlaw}
\log_{10}\mathrm{I_{R}} = c_{o} + \mathrm{N_{C}}^{a}
\end{equation}
where I$_{\mathrm{R}}$ is the corresponding intensity ratio. Depending on the ionization fraction of the source as determined from the charge -- size grid, the corresponding relation will provide the precise average \Nc{} of a source. All size relation parameters are given in Table \ref{tab:calibr}.

\begin{table}
    \begin{center}
    \begin{minipage}{63mm}
			\caption{Power-law fit parameters for the different PAH charge states, as described by equation \ref{eq:powerlaw} for the ISRF.}
			\label{tab:calibr}
			\begin{tabular}{lcc}
				\hline 
				Charge state & c$_{o}$ & $\alpha$ \\ 
				\hline
			Neutral	& -2.550 & 0.251 \\
			75\% Neutral + 25\% Cation & -2.490 & 0.248 \\
			50\% Neutral + 50\% Cation & -2.430 & 0.246 \\
			25\% Neutral + 75\% Cation & -2.310 & 0.242 \\
			Cation & -2.010 & 0.234 \\
				\hline
			\end{tabular} 
	\end{minipage}
	\end{center}
\end{table}

\subsection{Different emission models} \label{sec:emmodels}

The emitted PAH spectrum of a source depends on the local physical conditions (e.g. $G_{\mathrm{0}}$, $T$, $n_{e}$; see Section \ref{sec:calibrations},), with the emission band profiles undergoing shifts in the peak position, variations in width, or revealing substructure. On the other hand, peak position shifts and intensity variations in the intrinsic PAH spectra are observed with increasing PAH size (e.g. \citealt{Bauschlicher2008},  \citealt{Bauschlicher2009}, \citealt{Ricca2012} and references therein). To allow a robust interpretation of the ionization fraction and average PAH size of a source, an appropriate charge -- size grid and set of size scaling relations are required, corresponding to the radiation field the PAHs are exposed to. To that end, we produced the emission spectra for the sample molecules as exposed to radiation fields with average photon energies of 6, 8, 10 and 12 eV, and calculated their respective charge -- size grids and the corresponding 11.2/3.3, (11.2+11.0)/3.3, 11.0/3.3 -- \Nc{} size scaling relations.

\begin{figure*}
	\begin{subfigure}{0.475\textwidth}
		\centering
		\includegraphics[keepaspectratio=true,scale=0.45]{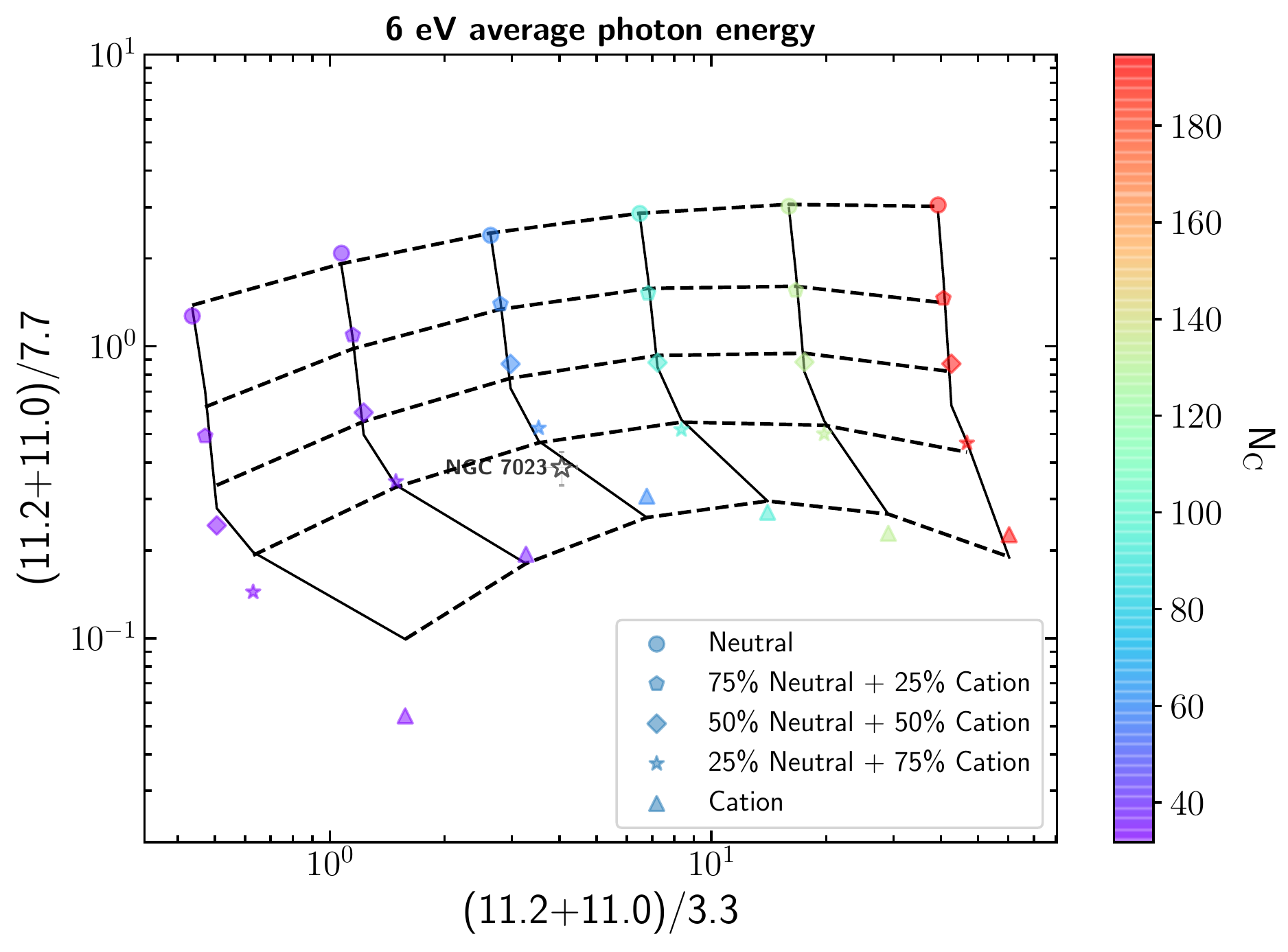}
	\end{subfigure}%
	\hfill
	\begin{subfigure}{0.475\textwidth}
		\centering
		\includegraphics[keepaspectratio=true,scale=0.45]{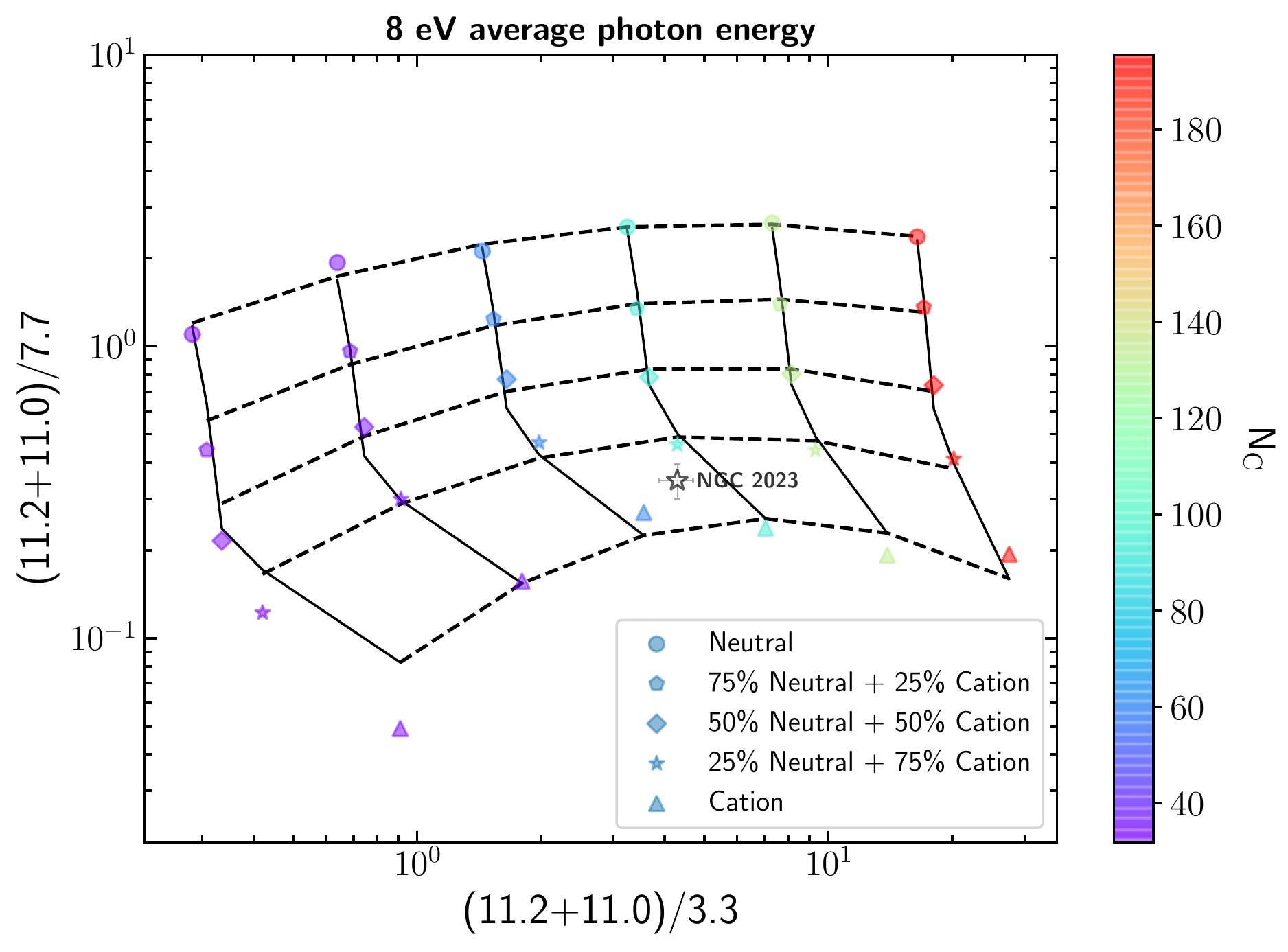}\\
	\end{subfigure}
	\vskip\baselineskip
	\begin{subfigure}{0.475\textwidth}
		\centering
		\includegraphics[keepaspectratio=true,scale=0.45]{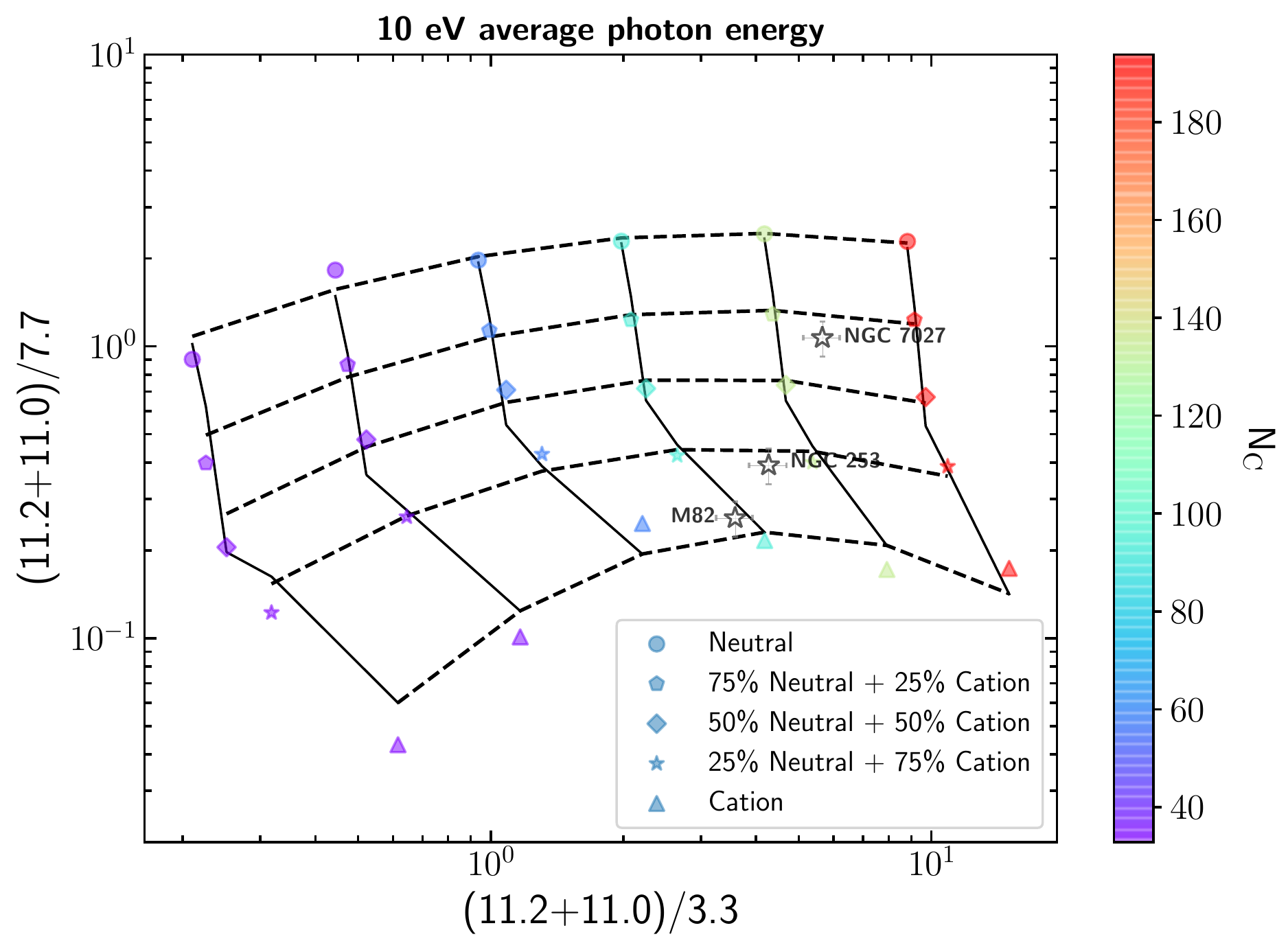}
	\end{subfigure}%
	\hfill
	\begin{subfigure}{0.475\textwidth}
		\centering
		\includegraphics[keepaspectratio=true,scale=0.45]{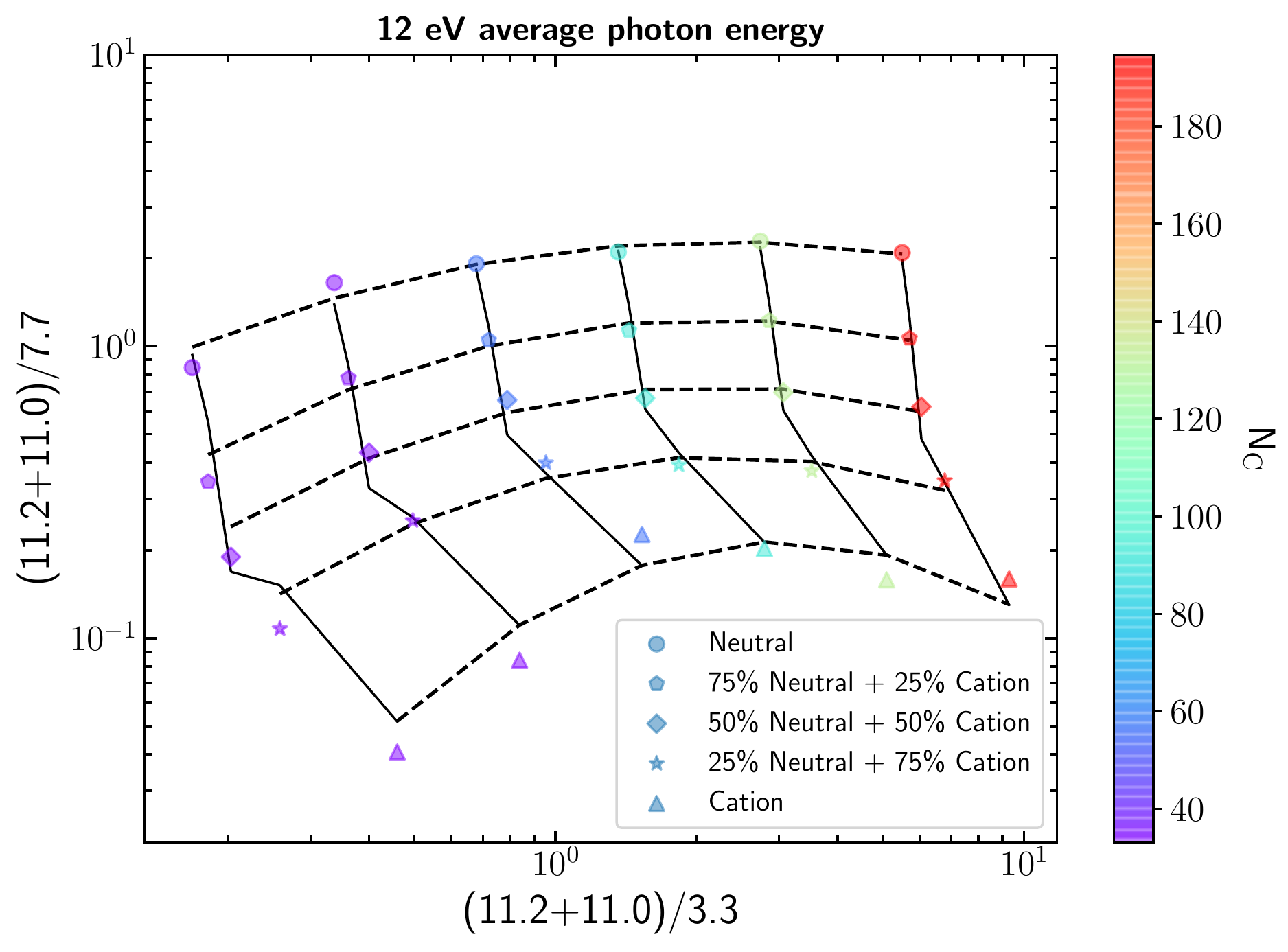}\\
	\end{subfigure}
	\caption{The charge -- size grid for radiation fields with average photon energies of 6 eV (top left panel), 8 eV (top right panel), 10 eV (bottom left panel), and 12 eV (bottom right panel). The points are the median (11.2+11.0)/7.7 values in equally-sized (11.2+11.0)/3.3 bins for spectra of a given ionization fraction as described in Section \ref{sec:grid}. Points of different shapes correspond to different ionization fractions. Dashed lines are one dimensional log parabola model fits to the binned values (equation \ref{eq:logparab}), and solid lines are third-degree polynomial fits (equation \ref{eq:polynom}) to the binned values between groups of points with similar \Nc{} among different ionization fractions. Points are color-coded based on the average \Nc{} of the points along each solid-line track. Astrophysical sources (described in Section \ref{sec:astroapp}) are plotted on the corresponding charge -- size grids based on their estimated average UV photon energies (see Sections \ref{sec:RNe} - \ref{sec:galaxies}).}
	\label{fig:multigrid}
\end{figure*}

Figure \ref{fig:multigrid} present the charge -- size grids for the different emission models, each populated with astrophysical sources (Section \ref{sec:astroapp}) of similar radiation fields, and Tables \ref{tab:fitparams6ev}-\ref{tab:fitparams12ev} describe their respective fit parameters. The 11.2/3.3, (11.2+11.0)/3.3, 11.0/3.3 -- \Nc{} size scaling relations are given in Tables \ref{tab:binparams6ev} - \ref{tab:binparams12ev}. As the absorbing photon energy increases, the grid is shifted to lower (11.2+11.0)/7.7 and (11.2+11.0)/3.3 values, with the (11.2+11.0)/3.3 ratio presenting the largest shifts. This is expected because the intensity gain with increasing absorption photon energy is higher for shorter wavelength emission features (3.3 \micron{} and 7.7 \micron{} complex) than for the higher wavelength features (11.0 or 11.2 \micron{}), as demonstrated in Fig. \ref{fig:speccomp}.

\section{Astrophysical application} \label{sec:astroapp}
	
We demonstrate the application of the (11.2+11.0)/7.7 -- (11.2+11.0)/3.3 (i.e., PAH charge -- size) grid on astrophysical sources of diverse environments, including galaxies, reflection and planetary nebulae, with the purpose to characterize the charge state and average size of their PAH populations. We used high-resolution spectroscopic observations from the Short Wavelength Spectrometer (SWS) on-board the Infrared Space Observatory (ISO), where the key 3.3 $ \mu $m emission band is sampled (unlike with \textit{Spitzer}-IRS). The sources used, along with their corresponding measurements, were taken from \cite{vanDiedenhoven2004} and consist of the two archetypical starburst galaxies M82 and NGC~253, the planetary nebula NGC~7027, and the RNe NGC~2023 and NGC~7023. No multiplicative scaling was required to match the ISO-SWS spectral segments covering the 3.3, 6.2, 7.7, 11.0, and 11.2 \micron{} features, while any additive offset between spectral segments does not affect the feature fluxes after continuum removal. For each source, we used the appropriate charge -- size grid corresponding to the average radiation field of the source (Fig. \ref{fig:multigrid}). Next, we identified the track of a given ionization fraction at the proximity of the source, which effectively describes the source's average ionization fraction. Based on the chosen track, the appropriate relation for the average PAH size estimation was then determined.

\subsection{NGC~2023 and NGC~7023} \label{sec:RNe}

NGC~2023 and NGC~7023 are two of the most well-studied RNe (e.g. \citealt{Fleming2010}; \citealt{Peeters2012}; \citealt{Boersma2014b}; \citealt{Stock2016}; Knight et al. 2019, submitted, hereafter K19) located at distances of 403$\pm$4 pc \citep{Kounkel2018} and 361$\pm$6 pc \citep{Gaia2016, Gaia2018} respectively. Since the average photon energy of the radiation field is 6.5 eV for NGC~7023 (\citealt{Croiset2016}; K19) and 7.3 eV for NGC~2023 (K19), we utilize the 6 and 8 eV charge -- size grids for each object respectively. In both grids the sources are located in between the N25C75 and pure cationic tracks (Fig. \ref{fig:multigrid}), indicating the contribution of mostly ionized PAH populations in their output spectrum. Choosing the N25C75 track as the most representative ionization fraction in both cases, we used equation \ref{eq:powerlaw} and the N25C75 track power-law parameters (Tables \ref{tab:calibr6ev} and \ref{tab:calibr8ev}) to estimate the average PAH size in NGC~7023 and NGC~2023. 

The average \Nc{} in NGC~7023 is estimated to be $ 65 \pm 15 $ and $ 94 \pm 19$ for NGC~2023, both consistent within their uncertainties with previous studies. Specifically, the PAH sizes reported by \cite{Croiset2016} and K19 for NGC~7023 range between 50 -- 80 \Nc{} and 50 -- 70 \Nc{} respectively across a large FOV centered on the NW PDR. Similarly, K19 reported a PAH size range between 75 and 130 \Nc{} for NGC~2023 across a large FOV centered on the dense shell south-southwest (-11", -78") of the exciting star HD~37903 corresponding to the H$_2$ emission peak. The aperture of the ISO-SWS data used in this paper lies well within the FOV of these studies (see K19, their Fig. 1 and 2). 
Variations in the calculated \Nc{} among different studies are expected, considering the different methods and apertures used in each case. For instance, the 3.3 \micron{} measurements in \citealt{Croiset2016} and K19 are based on photometric observations obtained from the Stratospheric Observatory for Infrared Astronomy (SOFIA, \citealt{Young2012}), as opposed to the high resolution ISO-SWS spectral measurements in our analysis. Furthermore, the ISO-SWS (14"x20") is much larger than the spatial resolution of the data used by \citealt{Croiset2016} and K19.

\subsection{NGC~7027} \label{sec:PN}

NGC~7027 is a young and carbon-rich planetary nebula excited by a hot white dwarf with T$ _{eff} \sim 2 \times 10^{5}$ K \citep{Latter2000}. The overall nature and morphology of NGC~7027 is rather complicated, with the central star surrounded by an expanding ionized shell \citep{Masson1989}, and a thin H$ _{2} $ shell pinpointing the presence of a PDR with signs of recent interaction with collimated outflows \citep{Cox2002}. 
We calculated an average UV photon energy of 10.7 eV in the FUV-MIR region (91--25000 nm), assuming a blackbody spectrum of T$ _{eff} = 2 \times 10^{5}$ K, and thus we utilized the 8 eV charge -- size grid. The location of NGC~7027 in the charge--size grid space (Fig. \ref{fig:multigrid}) is in between the N75 -- C25 and N50 -- C50 tracks, implying a somewhat even distribution of neutral and cationic PAHs in this planetary nebula. 

The average \Nc{} in NGC~7027 is estimated to be $ 149 \pm 34 $ and is moderately larger compared to that of the reflection nebulae in our sample. With the strength of the FUV radiation field, G$_0$, being 6 $10^5$ in NGC~7027 \citep{Justtanont1997}, a larger PAH size is consistent with the results of K19 who reported that the average PAH size depends on the FUV radiation field intensity.

\subsection{M82 and NGC~253} \label{sec:galaxies}

M82 and NGC~253 are the brightest prototypes of nearby starburst galaxies located at similar distances ($ \sim 3.5 $ Mpc; \citealt{Dalcanton2009}; \citealt{Jacobs2009}), with comparable IR luminosities but different chemical compositions \citep{Martin2009b}. For M82, we calculated the average UV photon energy from its modeled spectral energy distribution (SED). The M82 SED decomposition and fitting was performed with the SED-fitting code \textsc{cigale} \citep{Boquien2019} and was obtained from the DustPedia \citep{Davies2017} archive\footnote{\href{http://dustpedia.astro.noa.gr/}{http://dustpedia.astro.noa.gr/}}. \textsc{cigale}, being an energy balance principle based code (i.e. the energy absorbed by dust in the UV-to-near-IR part of the spectrum is re-emitted self-consistently in the mid- and far-IR), is able to deliver both the entire (dust attenuated) SED as well as the unattenuated SED of the young stellar populations. Using  the SED of young stars, the average photon energy in the FUV-MIR region is estimated to be 9.4 eV. Hence, the 10 eV charge -- size grid was used to examine the charge state and the range in PAH size. For NGC~253, no SED modeling was available from the DustPedia archive, due to the absence of photometric coverage in the optical bands for this source. Therefore, we assumed a similar average UV photon energy to M82.

In terms of charge state, both M82 and NGC~253 are located in between the N25C75 and pure cationic tracks, indicating highly ionized environments where PAHs exist predominantly in their cationic state (Fig. \ref{fig:multigrid}). Specifically, NGC~253 resides closer to the N25C75 track while M82 is located closer to the track of pure cationic PAHs, suggesting less intense excitation mechanisms for NGC~253 compared to M82. Indeed, \cite{Martin2009b} presented evidence of NGC~253 being at an earlier starburst evolutionary stage than M82. While at early starburst stages the heating of the ISM is thought to be dominated by shocks affecting the molecular clouds fueling the starburst, the late stages of starburst evolution are vastly dominated by the UV radiation from newly formed massive stars. As such, the UV radiation responsible for pumping and ionizing the PAH molecules is expected to be less dominant in NGC~253 compared to M82, explaining their distinct locations on the charge -- size grid (Fig. \ref{fig:multigrid}). 

With respect to the PAH size distribution, M82 has an average PAH size of $ 86 \pm 11 $ \Nc{} and for NGC~253 the average PAH size is estimated to be $ 123 \pm 25 $ \Nc{}. The high average PAH values in both starburst galaxies are indicative of smaller PAH destruction in harder radiation fields.

\section{Limitations} \label{sec:Limitations}

Among the pure PAHs present in the computational spectroscopic library of PAHdb version 3.0, 2609 species contain between 21 and 50 carbons and 304 species contain between 51 and 386 carbons. PAHdb currently lacks large PAHs containing more than 50 carbons and with a broad array of hydrogen adjacency classes and charge states. For a detailed description, we refer the reader to \cite{Bauschlicher2018}. However, we should note that, based on our study using a sample containing a small number of large PAHs and a large number of small PAHs, large PAHs follow the same trends as those of small PAHs (see e.g. Fig. \ref{fig:calib}). Further improvements are also needed regarding the incorporation of the anharmonicity effects. The PAHdb contains the harmonic frequencies for each species while the applied emission model calculates the maximum temperature a PAH attains after photon absorption using its heat capacity, which is determined by the harmonic vibrational levels. Several studies have recently been reported investigating these anharmonic effects \citep[e.g.][]{Maltseva2015, Mackie2016, Mackie2018}, which will be included in future PAHdb emission models. Finally, we note that we find the same general behaviour for all the applied radiation fields (which differ in the details, see Sect.~\ref{sec:emmodels}). Hence, these results do not seem to be dependent on the applied radiation field.  

\section{Summary and conclusions} \label{sec:Summary}

In this work we examined and determined the most optimal tracers and methods of tracking PAH size, using a variety of molecules with \Nc ranging between 22 and 216, with different ionization fractions, with the only constraint on the symmetry or compactness being the presence of a solo C-H bond. The sample was constructed from the versions 3.00 and 3.20 release of the NASA Ames PAH IR Spectroscopic Database. We generated PAH emission spectra considering radiation fields with average photon energies of 6, 8, 10, and 12 eV, as well as the \cite{Mathis1983} ISRF, and measured the PAH intensities of the 3.3, 6.2, 7.7, 8.6, 11.0, 11.2 \micron, and $\Sigma_{(15-20)}$ bands. The main conclusions of our work can be summarized as follows.

\begin{enumerate}

	\item We illustrated for the first time how the intensity ratio of the solo CH out-of-plane bending mode emission (in the 11.0 -- 11.3 \micron{} region) to the CH stretching mode emission (in the 3.3 \micron{} region) scales with the intrinsic size of cationic and anionic PAH species. Exploring different intensity ratio combinations between the 11.0 \micron{} emission and other emission bands revealed no such dependence with \Nc{} for PAH cations. Similarly, the 11.2/3.3 ratio has the best correlation with \Nc{}, for the case of neutral PAHs, compared to any other intensity ratio.
	
	\item By examining all the individual strong PAH emission bands we show that the intensity of the 3.3 $ \mu $m band has the strongest dependence with \Nc{}, and is the main driver behind correlations of the intensity ratio of the solo CH out-of-plane to CH stretching mode with \Nc{} for neutral, cationic and anionic PAHs.
	
	\item The intensity ratios of 6.2/3.3, 7.7/3.3, 8.6/3.3, and $\Sigma_{(15-20)}$/3.3 also scale with \Nc{} across 2 orders of magnitude for both neutral and cationic PAHs, serving as alternate size tracers. However, the scatter in these correlations is moderately larger compared to the 11.2/3.3 and 11.0/3.3 relations with \Nc.
	
	\item The 6.2/7.7 intensity ratio, previously adopted in the DL01 models to track PAH size, shows no evident scaling with \Nc, for both neutral or cationic PAHs. In addition, no other combination of intensity ratios among the ionic bands, that does not include the 3.3 \micron{} emission, scales with \Nc.
	
	\item Interstellar space contains a mix of PAH charge states, with the observed 11.2 \micron{} band and most of the 3.3 \micron{} band arising in neutral PAHs while the observed 11.0 \micron{} band and a small fraction of the 3.3 \micron{} band originates in cationic PAHs.  For comparison with observations, we defined a new diagnostic space to probe PAH charge and size: the (11.2+11.0)/7.7 -- (11.2+11.0)/3.3 space. We constructed grids in the (11.2+11.0)/7.7 -- (11.2+11.0)/3.3 space for the different radiation fields examined, with tracks corresponding to various ionization fractions and PAH sizes, allowing for the most robust estimation of the charge state and size of astrophysical PAHs to date. We demonstrated the application of the charge -- size grid in a set of astrophysical sources, including planetary nebulae, reflection nebulae, and galaxies.
	
	\item We delivered quantitative sets of scaling relations between PAH intensity ratios and \Nc for a precise PAH size estimation depending on the ionization fraction of the PAHs within a source and for the different radiation fields examined.
\end{enumerate}

With the upcoming launch of the \textit{James Webb Space telescope} (\textit{JWST}) and the various spectroscopic capabilities (integral field unit (IFU), slit and slitless spectroscopy) in both the near Infrared Camera (NIRCam) instrument and Mid-IR Instrument (MIRI), a wealth of spectra and spatially resolved spectroscopic maps in the 0.6 -- 28.5 \micron{} range will become available.  Single-slit spectroscopic observations and/or 3--D data cube maps of the 3.3, 7.7, 11.0, and 11.2 \micron{} emission features will be able to fully utilize the (11.2+11.0)/7.7 -- (11.2+11.0)/3.3 grid diagnostic diagram allowing a detailed characterization of the charge and size distribution of PAHs within sources or for sources as a whole. 

\section*{Acknowledgements}
We would like to thank Kris Sellgren for the constructive comments and suggestions that have improved the clarity of this paper. We sincerely thank Christiaan Boersma for insightful conversations. E.P. acknowledges support from an NSERC Discovery Grant and a Western Science and Engineering Research Board (SERB) Accelerator Award. A.R. gratefully acknowledges support from NASA's APRA program (NNX17AE71G) and from the directed Work Package at NASA Ames titled: ``Laboratory Astrophysics -- The NASA Ames PAH IR Spectroscopic Database''.

\bibliographystyle{mnras}
\bibliography{Bibliography_PAHdb}

\begin{thebibliography}{}
\makeatletter
\relax
\def\mn@urlcharsother{\let\do\@makeother \do\$\do\&\do\#\do\^\do\_\do\%\do\~}
\def\mn@doi{\begingroup\mn@urlcharsother \@ifnextchar [ {\mn@doi@}
  {\mn@doi@[]}}
\def\mn@doi@[#1]#2{\def\@tempa{#1}\ifx\@tempa\@empty \href
  {http://dx.doi.org/#2} {doi:#2}\else \href {http://dx.doi.org/#2} {#1}\fi
  \endgroup}
\def\mn@eprint#1#2{\mn@eprint@#1:#2::\@nil}
\def\mn@eprint@arXiv#1{\href {http://arxiv.org/abs/#1} {{\tt arXiv:#1}}}
\def\mn@eprint@dblp#1{\href {http://dblp.uni-trier.de/rec/bibtex/#1.xml}
  {dblp:#1}}
\def\mn@eprint@#1:#2:#3:#4\@nil{\def\@tempa {#1}\def\@tempb {#2}\def\@tempc
  {#3}\ifx \@tempc \@empty \let \@tempc \@tempb \let \@tempb \@tempa \fi \ifx
  \@tempb \@empty \def\@tempb {arXiv}\fi \@ifundefined
  {mn@eprint@\@tempb}{\@tempb:\@tempc}{\expandafter \expandafter \csname
  mn@eprint@\@tempb\endcsname \expandafter{\@tempc}}}

\bibitem[\protect\citeauthoryear{{Allain}, {Leach}  \& {Sedlmayr}}{{Allain}
  et~al.}{1996}]{Allain1996}
{Allain} T.,  {Leach} S.,   {Sedlmayr} E.,  1996, \aap, \href
  {https://ui.adsabs.harvard.edu/\#abs/1996A&A...305..602A} {305, 602}

\bibitem[\protect\citeauthoryear{{Allamandola}, {Tielens}  \&
  {Barker}}{{Allamandola} et~al.}{1985}]{Allamandola1985}
{Allamandola} L.~J.,  {Tielens} A.~G.~G.~M.,   {Barker} J.~R.,  1985, \mn@doi
  [\apjl] {10.1086/184435}, \href
  {http://adsabs.harvard.edu/abs/1985ApJ...290L..25A} {290, L25}

\bibitem[\protect\citeauthoryear{{Allamandola}, {Tielens}  \&
  {Barker}}{{Allamandola} et~al.}{1989}]{ATB1989}
{Allamandola} L.~J.,  {Tielens} A.~G.~G.~M.,   {Barker} J.~R.,  1989, \mn@doi
  [\apjs] {10.1086/191396}, \href
  {http://adsabs.harvard.edu/abs/1989ApJS...71..733A} {71, 733}

\bibitem[\protect\citeauthoryear{{Allamandola}, {Hudgins}  \&
  {Sandford}}{{Allamandola} et~al.}{1999}]{Allamandola1999}
{Allamandola} L.~J.,  {Hudgins} D.~M.,   {Sandford} S.~A.,  1999, \mn@doi
  [\apj] {10.1086/311843}, \href
  {https://ui.adsabs.harvard.edu/abs/1999ApJ...511L.115A} {511, L115}

\bibitem[\protect\citeauthoryear{{Bakes} \& {Tielens}}{{Bakes} \&
  {Tielens}}{1994}]{BakesTielens1994}
{Bakes} E.~L.~O.,  {Tielens} A.~G.~G.~M.,  1994, \mn@doi [\apj]
  {10.1086/174188}, \href {http://adsabs.harvard.edu/abs/1994ApJ...427..822B}
  {427, 822}

\bibitem[\protect\citeauthoryear{{Bauschlicher}}{{Bauschlicher}}{1998}]{Bauschlicher1998}
{Bauschlicher} Charles~W. J.,  1998, \mn@doi [\apj] {10.1086/311782}, \href
  {https://ui.adsabs.harvard.edu/abs/1998ApJ...509L.125B} {509, L125}

\bibitem[\protect\citeauthoryear{{Bauschlicher}, {Peeters}  \&
  {Allamandola}}{{Bauschlicher} et~al.}{2008}]{Bauschlicher2008}
{Bauschlicher} Charles~W. J.,  {Peeters} E.,   {Allamandola} L.~J.,  2008,
  \mn@doi [\apj] {10.1086/533424}, \href
  {https://ui.adsabs.harvard.edu/abs/2008ApJ...678..316B} {678, 316}

\bibitem[\protect\citeauthoryear{{Bauschlicher}, {Peeters}  \&
  {Allamandola}}{{Bauschlicher} et~al.}{2009}]{Bauschlicher2009}
{Bauschlicher} Charles~W. J.,  {Peeters} E.,   {Allamandola} L.~J.,  2009,
  \mn@doi [\apj] {10.1088/0004-637X/697/1/311}, \href
  {https://ui.adsabs.harvard.edu/abs/2009ApJ...697..311B} {697, 311}

\bibitem[\protect\citeauthoryear{{Bauschlicher} Jr. et~al.,}{{Bauschlicher}
  et~al.}{2010}]{Bauschlicher2010}
{Bauschlicher} Jr. C.~W.,  et~al., 2010, \mn@doi [\apjs]
  {10.1088/0067-0049/189/2/341}, \href
  {http://adsabs.harvard.edu/abs/2010ApJS..189..341B} {189, 341}

\bibitem[\protect\citeauthoryear{{Bauschlicher}, {Ricca}, {Boersma}  \&
  {Allamandola}}{{Bauschlicher} et~al.}{2018}]{Bauschlicher2018}
{Bauschlicher} Charles~W. J.,  {Ricca} A.,  {Boersma} C.,   {Allamandola}
  L.~J.,  2018, \mn@doi [\apjs] {10.3847/1538-4365/aaa019}, \href
  {https://ui.adsabs.harvard.edu/abs/2018ApJS..234...32B} {234, 32}

\bibitem[\protect\citeauthoryear{{Boersma}, {Bauschlicher}, {Allamandola},
  {Ricca}, {Peeters}  \& {Tielens}}{{Boersma} et~al.}{2010}]{Boersma2010}
{Boersma} C.,  {Bauschlicher} C.~W.,  {Allamandola} L.~J.,  {Ricca} A.,
  {Peeters} E.,   {Tielens} A.~G.~G.~M.,  2010, \mn@doi [\aap]
  {10.1051/0004-6361/200912714}, \href
  {http://adsabs.harvard.edu/abs/2010A%26A...511A..32B} {511, A32}

\bibitem[\protect\citeauthoryear{{Boersma}, {Bauschlicher}, {Ricca},
  {Mattioda}, {Peeters}, {Tielens}  \& {Allamandola}}{{Boersma}
  et~al.}{2011}]{Boersma2011}
{Boersma} C.,  {Bauschlicher} Jr. C.~W.,  {Ricca} A.,  {Mattioda} A.~L.,
  {Peeters} E.,  {Tielens} A.~G.~G.~M.,   {Allamandola} L.~J.,  2011, \mn@doi
  [\apj] {10.1088/0004-637X/729/1/64}, \href
  {http://adsabs.harvard.edu/abs/2011ApJ...729...64B} {729, 64}

\bibitem[\protect\citeauthoryear{{Boersma} et~al.,}{{Boersma}
  et~al.}{2014a}]{Boersma2014a}
{Boersma} C.,  et~al., 2014a, \mn@doi [\apjs] {10.1088/0067-0049/211/1/8},
  \href {http://adsabs.harvard.edu/abs/2014ApJS..211....8B} {211, 8}

\bibitem[\protect\citeauthoryear{{Boersma}, {Bregman}  \&
  {Allamandola}}{{Boersma} et~al.}{2014b}]{Boersma2014b}
{Boersma} C.,  {Bregman} J.,   {Allamandola} L.~J.,  2014b, \mn@doi [\apj]
  {10.1088/0004-637X/795/2/110}, \href
  {http://adsabs.harvard.edu/abs/2014ApJ...795..110B} {795, 110}

\bibitem[\protect\citeauthoryear{{Boquien}, {Burgarella}, {Roehlly}, {Buat},
  {Ciesla}, {Corre}, {Inoue}  \& {Salas}}{{Boquien} et~al.}{2019}]{Boquien2019}
{Boquien} M.,  {Burgarella} D.,  {Roehlly} Y.,  {Buat} V.,  {Ciesla} L.,
  {Corre} D.,  {Inoue} A.~K.,   {Salas} H.,  2019, \mn@doi [\aap]
  {10.1051/0004-6361/201834156}, \href
  {https://ui.adsabs.harvard.edu/abs/2019A&A...622A.103B} {622, A103}

\bibitem[\protect\citeauthoryear{{Boulanger}, {Boisssel}, {Cesarsky}  \&
  {Ryter}}{{Boulanger} et~al.}{1998}]{Boulanger1998}
{Boulanger} F.,  {Boisssel} P.,  {Cesarsky} D.,   {Ryter} C.,  1998, \aap,
  \href {https://ui.adsabs.harvard.edu/abs/1998A&A...339..194B} {339, 194}

\bibitem[\protect\citeauthoryear{{Bregman} \& {Temi}}{{Bregman} \&
  {Temi}}{2005}]{BregmanTemi2005}
{Bregman} J.,  {Temi} P.,  2005, \mn@doi [\apj] {10.1086/427738}, \href
  {http://adsabs.harvard.edu/abs/2005ApJ...621..831B} {621, 831}

\bibitem[\protect\citeauthoryear{{Bregman}, {Allamandola}, {Tielens}, {Geballe}
   \& {Witteborn}}{{Bregman} et~al.}{1989}]{Bregman1989}
{Bregman} J.~D.,  {Allamandola} L.~J.,  {Tielens} A.~G.~G.~M.,  {Geballe}
  T.~R.,   {Witteborn} F.~C.,  1989, \mn@doi [\apj] {10.1086/167844}, \href
  {https://ui.adsabs.harvard.edu/abs/1989ApJ...344..791B} {344, 791}

\bibitem[\protect\citeauthoryear{{Calzetti} et~al.,}{{Calzetti}
  et~al.}{2007}]{Calzetti2007}
{Calzetti} D.,  et~al., 2007, \mn@doi [\apj] {10.1086/520082}, \href
  {http://adsabs.harvard.edu/abs/2007ApJ...666..870C} {666, 870}

\bibitem[\protect\citeauthoryear{{Cohen}, {Tielens}, {Bregman}, {Witteborn},
  {Rank}, {Allamandola}, {Wooden}  \& {Jourdain de Muizon}}{{Cohen}
  et~al.}{1989}]{Cohen1989}
{Cohen} M.,  {Tielens} A.~G.~G.~M.,  {Bregman} J.,  {Witteborn} F.~C.,  {Rank}
  D.~M.,  {Allamandola} L.~J.,  {Wooden} D.,   {Jourdain de Muizon} M.,  1989,
  \mn@doi [\apj] {10.1086/167489}, \href
  {https://ui.adsabs.harvard.edu/abs/1989ApJ...341..246C} {341, 246}

\bibitem[\protect\citeauthoryear{{Cortzen} et~al.,}{{Cortzen}
  et~al.}{2019}]{Cortzen2019}
{Cortzen} I.,  et~al., 2019, \mn@doi [\mnras] {10.1093/mnras/sty2777}, \href
  {https://ui.adsabs.harvard.edu/abs/2019MNRAS.482.1618C} {482, 1618}

\bibitem[\protect\citeauthoryear{{Cox}, {Huggins}, {Maillard}, {Habart},
  {Morisset}, {Bachiller}  \& {Forveille}}{{Cox} et~al.}{2002}]{Cox2002}
{Cox} P.,  {Huggins} P.~J.,  {Maillard} J.~P.,  {Habart} E.,  {Morisset} C.,
  {Bachiller} R.,   {Forveille} T.,  2002, \mn@doi [\aap]
  {10.1051/0004-6361:20011780}, \href
  {https://ui.adsabs.harvard.edu/abs/2002A&A...384..603C} {384, 603}

\bibitem[\protect\citeauthoryear{{Croiset}, {Candian}, {Bern{\'e}}  \&
  {Tielens}}{{Croiset} et~al.}{2016}]{Croiset2016}
{Croiset} B.~A.,  {Candian} A.,  {Bern{\'e}} O.,   {Tielens} A.~G.~G.~M.,
  2016, \mn@doi [\aap] {10.1051/0004-6361/201527714}, \href
  {https://ui.adsabs.harvard.edu/#abs/2016A&A...590A..26C} {590, A26}

\bibitem[\protect\citeauthoryear{{Dalcanton} et~al.,}{{Dalcanton}
  et~al.}{2009}]{Dalcanton2009}
{Dalcanton} J.~J.,  et~al., 2009, \mn@doi [The Astrophysical Journal Supplement
  Series] {10.1088/0067-0049/183/1/67}, \href
  {https://ui.adsabs.harvard.edu/abs/2009ApJS..183...67D} {183, 67}

\bibitem[\protect\citeauthoryear{{Davies} et~al.,}{{Davies}
  et~al.}{2017}]{Davies2017}
{Davies} J.~I.,  et~al., 2017, \mn@doi [\pasp]
  {10.1088/1538-3873/129/974/044102}, \href
  {https://ui.adsabs.harvard.edu/abs/2017PASP..129d4102D} {129, 044102}

\bibitem[\protect\citeauthoryear{{Diamond-Stanic} \& {Rieke}}{{Diamond-Stanic}
  \& {Rieke}}{2010}]{Diamond-Stanic2010}
{Diamond-Stanic} A.~M.,  {Rieke} G.~H.,  2010, \mn@doi [\apj]
  {10.1088/0004-637X/724/1/140}, \href
  {http://adsabs.harvard.edu/abs/2010ApJ...724..140D} {724, 140}

\bibitem[\protect\citeauthoryear{{Draine} \& {Li}}{{Draine} \&
  {Li}}{2001}]{Draine2001}
{Draine} B.~T.,  {Li} A.,  2001, \mn@doi [\apj] {10.1086/320227}, \href
  {http://adsabs.harvard.edu/abs/2001ApJ...551..807D} {551, 807}

\bibitem[\protect\citeauthoryear{{Draine} \& {Li}}{{Draine} \&
  {Li}}{2007}]{Draine:07}
{Draine} B.~T.,  {Li} A.,  2007, \apj, 657, 810

\bibitem[\protect\citeauthoryear{{Farrah} et~al.,}{{Farrah}
  et~al.}{2007}]{Farrah2007}
{Farrah} D.,  et~al., 2007, \mn@doi [\apj] {10.1086/520834}, \href
  {https://ui.adsabs.harvard.edu/#abs/2007ApJ...667..149F} {667, 149}

\bibitem[\protect\citeauthoryear{{Fleming}, {France}, {Lupu}  \&
  {McCandliss}}{{Fleming} et~al.}{2010}]{Fleming2010}
{Fleming} B.,  {France} K.,  {Lupu} R.~E.,   {McCandliss} S.~R.,  2010, \mn@doi
  [\apj] {10.1088/0004-637X/725/1/159}, \href
  {https://ui.adsabs.harvard.edu/abs/2010ApJ...725..159F} {725, 159}

\bibitem[\protect\citeauthoryear{{Foley}, {Cazaux}, {Egorov}, {Boschman},
  {Hoekstra}  \& {Schlath{\"o}lter}}{{Foley} et~al.}{2018}]{Foley2018}
{Foley} N.,  {Cazaux} S.,  {Egorov} D.,  {Boschman} L.~M.~P.~V.,  {Hoekstra}
  R.,   {Schlath{\"o}lter} T.,  2018, \mn@doi [\mnras] {10.1093/mnras/sty1528},
  \href {https://ui.adsabs.harvard.edu/abs/2018MNRAS.479..649F} {479, 649}

\bibitem[\protect\citeauthoryear{{Gaia Collaboration} et~al.,}{{Gaia
  Collaboration} et~al.}{2016}]{Gaia2016}
{Gaia Collaboration} et~al., 2016, \mn@doi [\aap]
  {10.1051/0004-6361/201629272}, \href
  {https://ui.adsabs.harvard.edu/abs/2016A&A...595A...1G} {595, A1}

\bibitem[\protect\citeauthoryear{{Gaia Collaboration} et~al.,}{{Gaia
  Collaboration} et~al.}{2018}]{Gaia2018}
{Gaia Collaboration} et~al., 2018, \mn@doi [\aap]
  {10.1051/0004-6361/201833051}, \href
  {https://ui.adsabs.harvard.edu/abs/2018A&A...616A...1G} {616, A1}

\bibitem[\protect\citeauthoryear{{Galliano}, {Madden}, {Tielens}, {Peeters}  \&
  {Jones}}{{Galliano} et~al.}{2008}]{Galliano2008}
{Galliano} F.,  {Madden} S.~C.,  {Tielens} A.~G.~G.~M.,  {Peeters} E.,
  {Jones} A.~P.,  2008, \mn@doi [\apj] {10.1086/587051}, \href
  {http://adsabs.harvard.edu/abs/2008ApJ...679..310G} {679, 310}

\bibitem[\protect\citeauthoryear{{Gordon}, {Engelbracht}, {Rieke}, {Misselt},
  {Smith}  \& {Kennicutt}}{{Gordon} et~al.}{2008}]{Gordon2008}
{Gordon} K.~D.,  {Engelbracht} C.~W.,  {Rieke} G.~H.,  {Misselt} K.~A.,
  {Smith} J.-D.~T.,   {Kennicutt} Jr. R.~C.,  2008, \mn@doi [\apj]
  {10.1086/589567}, \href {http://adsabs.harvard.edu/abs/2008ApJ...682..336G}
  {682, 336}

\bibitem[\protect\citeauthoryear{{Hony}, {Van Kerckhoven}, {Peeters},
  {Tielens}, {Hudgins}  \& {Allamandola}}{{Hony} et~al.}{2001}]{Hony2001}
{Hony} S.,  {Van Kerckhoven} C.,  {Peeters} E.,  {Tielens} A.~G.~G.~M.,
  {Hudgins} D.~M.,   {Allamandola} L.~J.,  2001, \mn@doi [\aap]
  {10.1051/0004-6361:20010242}, \href
  {http://adsabs.harvard.edu/abs/2001A%26A...370.1030H} {370, 1030}

\bibitem[\protect\citeauthoryear{{Hudgins} \& {Allamandola}}{{Hudgins} \&
  {Allamandola}}{1999}]{HudginsAllamandola1999}
{Hudgins} D.~M.,  {Allamandola} L.~J.,  1999, \mn@doi [\apj] {10.1086/311901},
  \href {https://ui.adsabs.harvard.edu/abs/1999ApJ...513L..69H} {513, L69}

\bibitem[\protect\citeauthoryear{{Jacobs}, {Rizzi}, {Tully}, {Shaya}, {Makarov}
   \& {Makarova}}{{Jacobs} et~al.}{2009}]{Jacobs2009}
{Jacobs} B.~A.,  {Rizzi} L.,  {Tully} R.~B.,  {Shaya} E.~J.,  {Makarov} D.~I.,
   {Makarova} L.,  2009, \mn@doi [\aj] {10.1088/0004-6256/138/2/332}, \href
  {https://ui.adsabs.harvard.edu/abs/2009AJ....138..332J} {138, 332}

\bibitem[\protect\citeauthoryear{{Jourdain de Muizon}, {Cox}  \&
  {Lequeux}}{{Jourdain de Muizon} et~al.}{1990}]{Jourdain1990}
{Jourdain de Muizon} M.,  {Cox} P.,   {Lequeux} J.,  1990, \aap, 83, 337

\bibitem[\protect\citeauthoryear{Justtanont, Tielens, Skinner  \&
  Haas}{Justtanont et~al.}{1997}]{Justtanont1997}
Justtanont K.,  Tielens A. G. G.~M.,  Skinner C.~J.,   Haas M.~R.,  1997,
  \mn@doi [The Astrophysical Journal] {10.1086/303623}, 476, 319

\bibitem[\protect\citeauthoryear{{Kounkel} et~al.,}{{Kounkel}
  et~al.}{2018}]{Kounkel2018}
{Kounkel} M.,  et~al., 2018, \mn@doi [\aj] {10.3847/1538-3881/aad1f1}, \href
  {https://ui.adsabs.harvard.edu/abs/2018AJ....156...84K} {156, 84}

\bibitem[\protect\citeauthoryear{{Latter}, {Dayal}, {Bieging}, {Meakin},
  {Hora}, {Kelly}  \& {Tielens}}{{Latter} et~al.}{2000}]{Latter2000}
{Latter} W.~B.,  {Dayal} A.,  {Bieging} J.~H.,  {Meakin} C.,  {Hora} J.~L.,
  {Kelly} D.~M.,   {Tielens} A.~G.~G.~M.,  2000, \mn@doi [\apj]
  {10.1086/309252}, \href
  {https://ui.adsabs.harvard.edu/abs/2000ApJ...539..783L} {539, 783}

\bibitem[\protect\citeauthoryear{{Leger} \& {Puget}}{{Leger} \&
  {Puget}}{1984}]{Leger1984}
{Leger} A.,  {Puget} J.~L.,  1984, \aap, \href
  {http://adsabs.harvard.edu/abs/1984A%26A...137L...5L} {137, L5}

\bibitem[\protect\citeauthoryear{{Lepp} \& {Dalgarno}}{{Lepp} \&
  {Dalgarno}}{1988}]{LeppDalgarno1988}
{Lepp} S.,  {Dalgarno} A.,  1988, \mn@doi [\apj] {10.1086/165915}, \href
  {https://ui.adsabs.harvard.edu/abs/1988ApJ...324..553L} {324, 553}

\bibitem[\protect\citeauthoryear{{Mackie} et~al.,}{{Mackie}
  et~al.}{2016}]{Mackie2016}
{Mackie} C.~J.,  et~al., 2016, \mn@doi [Journal of Chemical Physics]
  {10.1063/1.4961438}, \href
  {https://ui.adsabs.harvard.edu/abs/2016JChPh.145h4313M} {145, 084313}

\bibitem[\protect\citeauthoryear{Mackie, Chen, Candian, Lee  \& Tielens}{Mackie
  et~al.}{2018}]{Mackie2018}
Mackie C.~J.,  Chen T.,  Candian A.,  Lee T.~J.,   Tielens A. G. G.~M.,  2018,
  \mn@doi [The Journal of Chemical Physics] {10.1063/1.5038725}, 149, 134302

\bibitem[\protect\citeauthoryear{Maltseva et~al.,}{Maltseva
  et~al.}{2015}]{Maltseva2015}
Maltseva E.,  et~al., 2015, \mn@doi [The Astrophysical Journal]
  {10.1088/0004-637x/814/1/23}, 814, 23

\bibitem[\protect\citeauthoryear{{Maragkoudakis}, {Ivkovich}, {Peeters},
  {Stock}, {Hemachandra}  \& {Tielens}}{{Maragkoudakis}
  et~al.}{2018}]{Maragkoudakis2018b}
{Maragkoudakis} A.,  {Ivkovich} N.,  {Peeters} E.,  {Stock} D.~J.,
  {Hemachandra} D.,   {Tielens} A.~G.~G.~M.,  2018, \mn@doi [\mnras]
  {10.1093/mnras/sty2658}, \href
  {https://ui.adsabs.harvard.edu/\#abs/2018MNRAS.481.5370M} {481, 5370}

\bibitem[\protect\citeauthoryear{{Mart{\'\i}n}, {Mart{\'\i}n-Pintado}  \&
  {Viti}}{{Mart{\'\i}n} et~al.}{2009}]{Martin2009b}
{Mart{\'\i}n} S.,  {Mart{\'\i}n-Pintado} J.,   {Viti} S.,  2009, \mn@doi [\apj]
  {10.1088/0004-637X/706/2/1323}, \href
  {https://ui.adsabs.harvard.edu/abs/2009ApJ...706.1323M} {706, 1323}

\bibitem[\protect\citeauthoryear{{Masson}}{{Masson}}{1989}]{Masson1989}
{Masson} C.~R.,  1989, \mn@doi [\apj] {10.1086/167011}, \href
  {https://ui.adsabs.harvard.edu/abs/1989ApJ...336..294M} {336, 294}

\bibitem[\protect\citeauthoryear{{Mathis}, {Mezger}  \& {Panagia}}{{Mathis}
  et~al.}{1983}]{Mathis1983}
{Mathis} J.~S.,  {Mezger} P.~G.,   {Panagia} N.,  1983, \aap, \href
  {https://ui.adsabs.harvard.edu/\#abs/1983A&A...128..212M} {500, 259}

\bibitem[\protect\citeauthoryear{{Mattioda}, {Allamandola}  \&
  {Hudgins}}{{Mattioda} et~al.}{2005}]{Mattioda2005}
{Mattioda} A.~L.,  {Allamandola} L.~J.,   {Hudgins} D.~M.,  2005, \mn@doi
  [\apj] {10.1086/431303}, \href
  {https://ui.adsabs.harvard.edu/abs/2005ApJ...629.1183M} {629, 1183}

\bibitem[\protect\citeauthoryear{{Mennella}, {Hornek{\ae}r}, {Thrower}  \&
  {Accolla}}{{Mennella} et~al.}{2012}]{Mannella2012}
{Mennella} V.,  {Hornek{\ae}r} L.,  {Thrower} J.,   {Accolla} M.,  2012,
  \mn@doi [\apj] {10.1088/2041-8205/745/1/L2}, \href
  {https://ui.adsabs.harvard.edu/abs/2012ApJ...745L...2M} {745, L2}

\bibitem[\protect\citeauthoryear{{Molster} et~al.,}{{Molster}
  et~al.}{1996}]{Molster1996}
{Molster} F.~J.,  et~al., 1996, \aap, \href
  {https://ui.adsabs.harvard.edu/abs/1996A&A...315L.373M} {315, L373}

\bibitem[\protect\citeauthoryear{{Mori}, {Sakon}, {Onaka}, {Kaneda}, {Umehata}
  \& {Ohsawa}}{{Mori} et~al.}{2012}]{Mori2012}
{Mori} T.~I.,  {Sakon} I.,  {Onaka} T.,  {Kaneda} H.,  {Umehata} H.,   {Ohsawa}
  R.,  2012, \mn@doi [\apj] {10.1088/0004-637X/744/1/68}, \href
  {https://ui.adsabs.harvard.edu/abs/2012ApJ...744...68M} {744, 68}

\bibitem[\protect\citeauthoryear{{O'Dowd} et~al.,}{{O'Dowd}
  et~al.}{2009}]{ODowd2009}
{O'Dowd} M.~J.,  et~al., 2009, \mn@doi [\apj] {10.1088/0004-637X/705/1/885},
  \href {http://adsabs.harvard.edu/abs/2009ApJ...705..885O} {705, 885}

\bibitem[\protect\citeauthoryear{{Pech}, {Joblin}  \& {Boissel}}{{Pech}
  et~al.}{2002}]{Pech2002}
{Pech} C.,  {Joblin} C.,   {Boissel} P.,  2002, \mn@doi [\aap]
  {10.1051/0004-6361:20020416}, \href
  {https://ui.adsabs.harvard.edu/abs/2002A&A...388..639P} {388, 639}

\bibitem[\protect\citeauthoryear{{Peeters}, {Tielens}, {Roelfsema}  \&
  {Cox}}{{Peeters} et~al.}{1999}]{Peeters1999}
{Peeters} E.,  {Tielens} A.~G.~G.~M.,  {Roelfsema} P.~R.,   {Cox} P.,  1999, in
  {Cox} P.,  {Kessler} M.,  eds,  ESA Special Publication Vol. 427, The
  Universe as Seen by ISO. p.~739

\bibitem[\protect\citeauthoryear{{Peeters}, {Hony}, {Van Kerckhoven},
  {Tielens}, {Allamandola}, {Hudgins}  \& {Bauschlicher}}{{Peeters}
  et~al.}{2002}]{Peeters2002}
{Peeters} E.,  {Hony} S.,  {Van Kerckhoven} C.,  {Tielens} A.~G.~G.~M.,
  {Allamandola} L.~J.,  {Hudgins} D.~M.,   {Bauschlicher} C.~W.,  2002, \mn@doi
  [\aap] {10.1051/0004-6361:20020773}, \href
  {https://ui.adsabs.harvard.edu/abs/2002A&A...390.1089P} {390, 1089}

\bibitem[\protect\citeauthoryear{{Peeters}, {Spoon}  \& {Tielens}}{{Peeters}
  et~al.}{2004}]{Peeters2004}
{Peeters} E.,  {Spoon} H.~W.~W.,   {Tielens} A.~G.~G.~M.,  2004, \mn@doi [\apj]
  {10.1086/423237}, \href {http://adsabs.harvard.edu/abs/2004ApJ...613..986P}
  {613, 986}

\bibitem[\protect\citeauthoryear{{Peeters}, {Tielens}, {Allamandola}  \&
  {Wolfire}}{{Peeters} et~al.}{2012}]{Peeters2012}
{Peeters} E.,  {Tielens} A.~G.~G.~M.,  {Allamandola} L.~J.,   {Wolfire} M.~G.,
  2012, \mn@doi [\apj] {10.1088/0004-637X/747/1/44}, \href
  {http://adsabs.harvard.edu/abs/2012ApJ...747...44P} {747, 44}

\bibitem[\protect\citeauthoryear{{Peeters}, {Bauschlicher}, {Allamandola},
  {Tielens}, {Ricca}  \& {Wolfire}}{{Peeters} et~al.}{2017}]{Peeters2017}
{Peeters} E.,  {Bauschlicher} Jr. C.~W.,  {Allamandola} L.~J.,  {Tielens}
  A.~G.~G.~M.,  {Ricca} A.,   {Wolfire} M.~G.,  2017, \mn@doi [\apj]
  {10.3847/1538-4357/836/2/198}, \href
  {http://adsabs.harvard.edu/abs/2017ApJ...836..198P} {836, 198}

\bibitem[\protect\citeauthoryear{{Regan} et~al.,}{{Regan}
  et~al.}{2006}]{Regan2006}
{Regan} M.~W.,  et~al., 2006, \mn@doi [\apj] {10.1086/505382}, \href
  {https://ui.adsabs.harvard.edu/abs/2006ApJ...652.1112R} {652, 1112}

\bibitem[\protect\citeauthoryear{{Ricca}, {Bauschlicher}, {Boersma}, {Tielens}
  \& {Allamandola}}{{Ricca} et~al.}{2012}]{Ricca2012}
{Ricca} A.,  {Bauschlicher} Charles~W. J.,  {Boersma} C.,  {Tielens} A.
  G.~G.~M.,   {Allamandola} L.~J.,  2012, \mn@doi [\apj]
  {10.1088/0004-637X/754/1/75}, \href
  {https://ui.adsabs.harvard.edu/#abs/2012ApJ...754...75R} {754, 75}

\bibitem[\protect\citeauthoryear{{Ricca}, {Bauschlicher}, {Roser}  \&
  {Peeters}}{{Ricca} et~al.}{2018}]{Ricca2018}
{Ricca} A.,  {Bauschlicher} Charles~W. J.,  {Roser} J.~E.,   {Peeters} E.,
  2018, \mn@doi [\apj] {10.3847/1538-4357/aaa757}, \href
  {https://ui.adsabs.harvard.edu/\#abs/2018ApJ...854..115R} {854, 115}

\bibitem[\protect\citeauthoryear{{Roelfsema} et~al.,}{{Roelfsema}
  et~al.}{1996}]{Roelfsema1996}
{Roelfsema} P.~R.,  et~al., 1996, \aap, \href
  {https://ui.adsabs.harvard.edu/abs/1996A&A...315L.289R} {315, L289}

\bibitem[\protect\citeauthoryear{{Sandstrom} et~al.,}{{Sandstrom}
  et~al.}{2012}]{Sandstrom2012}
{Sandstrom} K.~M.,  et~al., 2012, \mn@doi [\apj] {10.1088/0004-637X/744/1/20},
  \href {http://adsabs.harvard.edu/abs/2012ApJ...744...20S} {744, 20}

\bibitem[\protect\citeauthoryear{{Schutte}, {Tielens}  \&
  {Allamandola}}{{Schutte} et~al.}{1993}]{Schutte1993}
{Schutte} W.~A.,  {Tielens} A.~G.~G.~M.,   {Allamandola} L.~J.,  1993, \mn@doi
  [\apj] {10.1086/173173}, \href
  {http://adsabs.harvard.edu/abs/1993ApJ...415..397S} {415, 397}

\bibitem[\protect\citeauthoryear{{Shannon} \& {Boersma}}{{Shannon} \&
  {Boersma}}{2019}]{Shannon2019}
{Shannon} M.~J.,  {Boersma} C.,  2019, \mn@doi [\apj]
  {10.3847/1538-4357/aaf562}, \href
  {https://ui.adsabs.harvard.edu/\#abs/2019ApJ...871..124S} {871, 124}

\bibitem[\protect\citeauthoryear{{Shipley}, {Papovich}, {Rieke}, {Brown}  \&
  {Moustakas}}{{Shipley} et~al.}{2016}]{Shipley2016}
{Shipley} H.~V.,  {Papovich} C.,  {Rieke} G.~H.,  {Brown} M. J.~I.,
  {Moustakas} J.,  2016, \mn@doi [\apj] {10.3847/0004-637X/818/1/60}, \href
  {https://ui.adsabs.harvard.edu/#abs/2016ApJ...818...60S} {818, 60}

\bibitem[\protect\citeauthoryear{{Smith} et~al.,}{{Smith}
  et~al.}{2007}]{Smith07b}
{Smith} J.~D.~T.,  et~al., 2007, \mn@doi [\apj] {10.1086/510549}, \href
  {http://adsabs.harvard.edu/abs/2007ApJ...656..770S} {656, 770}

\bibitem[\protect\citeauthoryear{{Stierwalt} et~al.,}{{Stierwalt}
  et~al.}{2014}]{Stierwalt2014}
{Stierwalt} S.,  et~al., 2014, \mn@doi [\apj] {10.1088/0004-637X/790/2/124},
  \href {https://ui.adsabs.harvard.edu/abs/2014ApJ...790..124S} {790, 124}

\bibitem[\protect\citeauthoryear{{Stock}, {Choi}, {Moya}, {Otaguro}, {Sorkhou},
  {Allamandola}, {Tielens}  \& {Peeters}}{{Stock} et~al.}{2016}]{Stock2016}
{Stock} D.~J.,  {Choi} W.~D.-Y.,  {Moya} L.~G.~V.,  {Otaguro} J.~N.,  {Sorkhou}
  S.,  {Allamandola} L.~J.,  {Tielens} A.~G.~G.~M.,   {Peeters} E.,  2016,
  \mn@doi [\apj] {10.3847/0004-637X/819/1/65}, \href
  {http://adsabs.harvard.edu/abs/2016ApJ...819...65S} {819, 65}

\bibitem[\protect\citeauthoryear{{Tappe}, {Rho}, {Boersma}  \&
  {Micelotta}}{{Tappe} et~al.}{2012}]{Tappe2012}
{Tappe} A.,  {Rho} J.,  {Boersma} C.,   {Micelotta} E.~R.,  2012, \mn@doi
  [\apj] {10.1088/0004-637X/754/2/132}, \href
  {https://ui.adsabs.harvard.edu/abs/2012ApJ...754..132T} {754, 132}

\bibitem[\protect\citeauthoryear{{Thrower} et~al.,}{{Thrower}
  et~al.}{2012}]{Thrower2012}
{Thrower} J.~D.,  et~al., 2012, \mn@doi [\apj] {10.1088/0004-637X/752/1/3},
  \href {https://ui.adsabs.harvard.edu/abs/2012ApJ...752....3T} {752, 3}

\bibitem[\protect\citeauthoryear{{Tielens}}{{Tielens}}{2008}]{Tielens2008}
{Tielens} A.~G.~G.~M.,  2008, \mn@doi [\araa]
  {10.1146/annurev.astro.46.060407.145211}, \href
  {https://ui.adsabs.harvard.edu/abs/2008ARA&A..46..289T} {46, 289}

\bibitem[\protect\citeauthoryear{{Verstraete} et~al.,}{{Verstraete}
  et~al.}{2001}]{Verstraete2001}
{Verstraete} L.,  et~al., 2001, \mn@doi [\aap] {10.1051/0004-6361:20010515},
  \href {http://adsabs.harvard.edu/abs/2001A%26A...372..981V} {372, 981}

\bibitem[\protect\citeauthoryear{{Young} et~al.,}{{Young}
  et~al.}{2012}]{Young2012}
{Young} E.~T.,  et~al., 2012, \mn@doi [\apj] {10.1088/2041-8205/749/2/L17},
  \href {https://ui.adsabs.harvard.edu/abs/2012ApJ...749L..17Y} {749, L17}

\bibitem[\protect\citeauthoryear{{van Diedenhoven}, {Peeters}, {Van
  Kerckhoven}, {Hony}, {Hudgins}, {Allamandola}  \& {Tielens}}{{van
  Diedenhoven} et~al.}{2004}]{vanDiedenhoven2004}
{van Diedenhoven} B.,  {Peeters} E.,  {Van Kerckhoven} C.,  {Hony} S.,
  {Hudgins} D.~M.,  {Allamandola} L.~J.,   {Tielens} A.~G.~G.~M.,  2004,
  \mn@doi [\apj] {10.1086/422404}, \href
  {https://ui.adsabs.harvard.edu/abs/2004ApJ...611..928V} {611, 928}

\makeatother
\end{thebibliography}

\appendix

\section{Emission model} \label{sec:equations}

When the convolution option is set in the AmesPAHdbIDLSuite\footnote{stable version dated 2015 August 27; \citealt{Bauschlicher2010}; \citealt{Boersma2014a}.} the incident radiation field, which in our case is the ISRF, is considered from $3.1\times 10^{-7} - 13.6$ eV. Upon the absorption of a photon of a given energy ($ hc\nu $) the maximum attained temperature of a PAH molecule is given by the following relation:

\begin{equation} \label{eq:tmax}
\int\displaylimits_{2.73}^{T_{max}} C_V(T)dT = hc\nu
\end{equation} where $ C_{V}(T) $ is the heat capacity of the PAH, described in terms of isolated harmonic oscillators (of frequency $ \nu_{0,i} $ for band i) as:
\begin{equation} \label{eq:heatcap}
C_{V}(T) = k\sum_{i}^{}e^{-\frac{h\nu_{0,i}}{kT}} \Bigg[\frac{\frac{h\nu_{0,i}}{kT}}{1-e^{-\frac{h\nu_{0,i}}{kT}}}\Bigg]^2
\end{equation}

The integrated cross-section for a single band weighted by the total number of incident photons and weighted by the absorption cross sections is given by:

\begin{equation} \label{eq:integcross}
\sigma_{i} = \frac{\sigma_{0,i}}{N} \int\displaylimits_{2.5\times 10^{-3}}^{1.1\times 10^{-5}} N(\nu) \int\displaylimits_{2.73}^{T_{max}(\nu)} B(\nu_{0,i},T) \bigg[\frac{dT}{dt}\bigg]^{-1} dT d\nu
\end{equation} The wavenumber limits of the outer integral ($2.5 \times 10^{-3} - 1.1 \times 10^{5}$ cm$ ^{-1} $) correspond to $3.1\times 10^{-7} - 13.6$ eV; $\sigma_{0,i}$ is the zero-Kelvin absorption cross-section of band i; N($ \nu $) is the number of photons at frequency $ \nu $ and N is the total number of photons absorbed; B($ \nu_{0,i}$, T) [erg s$ ^{-1} $ cm$^{-2}$ Hz$ ^{-1} $ sr$ ^{-1} $] is Planck's function at frequency $ \nu_{0,i} $ [cm$ ^{-1} $] in mode i and temperature T [Kelvin]; dT/dt [Kelvin s$ ^{-1} $] is the cooling-rate defined as:

\begin{equation}
\frac{dT}{dt} = \bigg[\frac{dE}{dT}\bigg]^{-1}_{V} \frac{dE}{dt} = \frac{4\pi}{C_{V}(T)} \sum_{i}^{} \sigma_{0,i}B(\nu_{0,i},T)
\end{equation} The calculation of the number of photons of a given frequency ($ \nu $) absorbed by the PAH is proportional to the absorption cross-section and the incident radiation field at that frequency:

\begin{equation}
N(\nu) = \sigma_{abs}(\nu) \frac{F^{*}(\nu)}{\nu}
\end{equation} where $ \sigma_{abs} $ is the absorption cross sections adopted from \cite{Draine:07} and \cite{Mattioda2005}, and $ F^{*} $ is the intensity of the incident radiation field.

The temperature cascade emission spectrum upon the absorption of a photon of a given energy is calculated considering conservation of energy:
\begin{equation} \label{eq:itegemis}
4\pi\sum_{i}^{}\sigma_{i}\int\displaylimits_{2.73}^{T_{max}} B(\nu_{0,i},T)\bigg[\frac{dT}{dt}\bigg]^{-1} dT = hc\nu
\end{equation} where $\sigma_{i}$ is the integrated absorption cross-section (Eq. \ref{eq:integcross}) of mode i.

\section{Redshifted spectra} \label{sec:zspec}

\begin{figure*}
	\begin{center}
		\hspace*{-0.5cm}\includegraphics[keepaspectratio=true,scale=0.46]{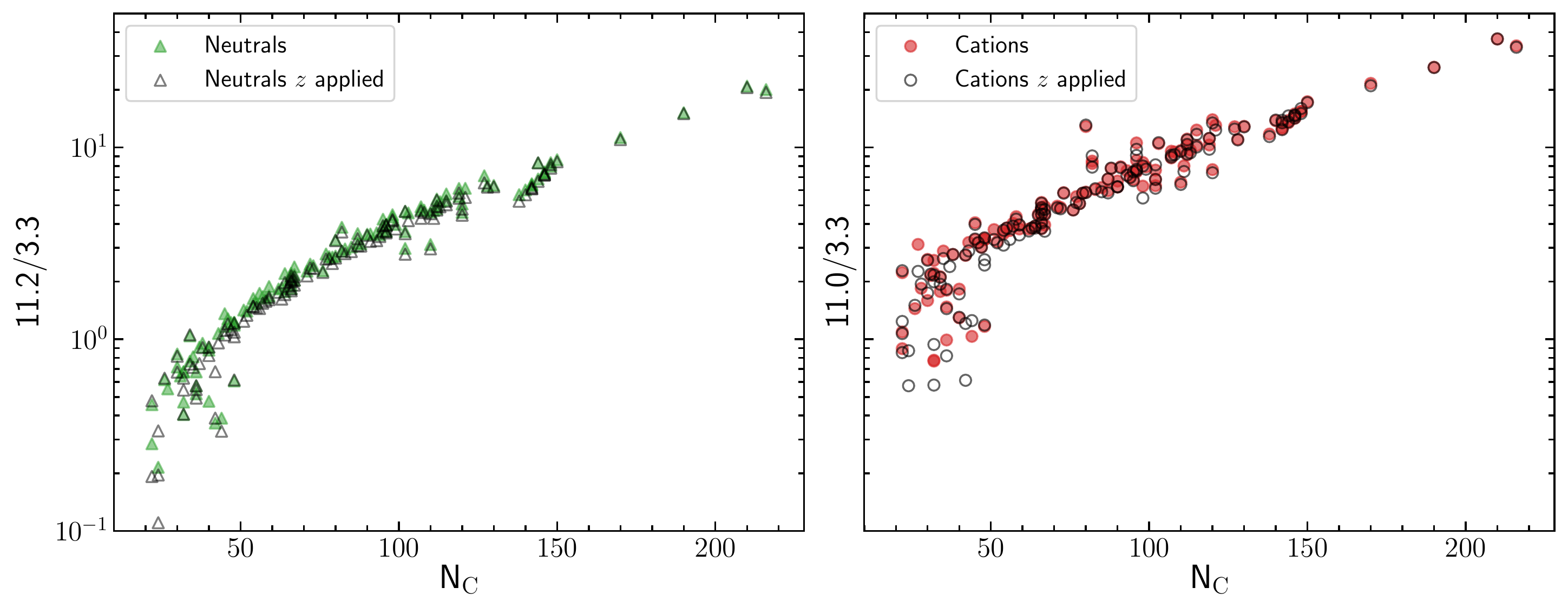}
		\caption{Comparison of PAH intensity ratios 11.2/3.3 for neutral PAHs (left panel) and 11.0/3.3 for cations (right panel) as a function of N$ _{\mathrm{C}} $ calculated from spectra with and without a 15 cm$^{-1}$ redshift application, for the ISRF case.
		}
		\label{fig:Z_comp}
	\end{center}
\end{figure*}

Anharmonicity effects in the spectra of highly vibrationally excited PAH molecules have been typically accounted for by applying a 15 cm$^{-1}$ redshift to the generated emission spectra. However, recent results \citep{Mackie2018} conclude that no redshift correction for anharmonicity should be applied. Here, we demonstrate that our results are independent of whether or not the redshift is applied. Fig. \ref{fig:Z_comp} shows the intensity ratios of 11.2/3.3 and 11.0/3.3 for neutral and cationic PAHs respectively as a function of \Nc, generated from spectra with and without a redshift application. Both families of points (i.e. from spectra with and without a redshift application) have identical distributions and show the same scaling with \Nc{}, with only the smallest molecules slightly shifted to lower intensity ratio values. Consequently, our results and conclusions throughout this work are consistent and unaffected by a redshift application to the generated PAH spectra.

\section{Average ISRF photon energy spectra}

\begin{figure*}
	\begin{center}
		\includegraphics[keepaspectratio=true,scale=0.43]{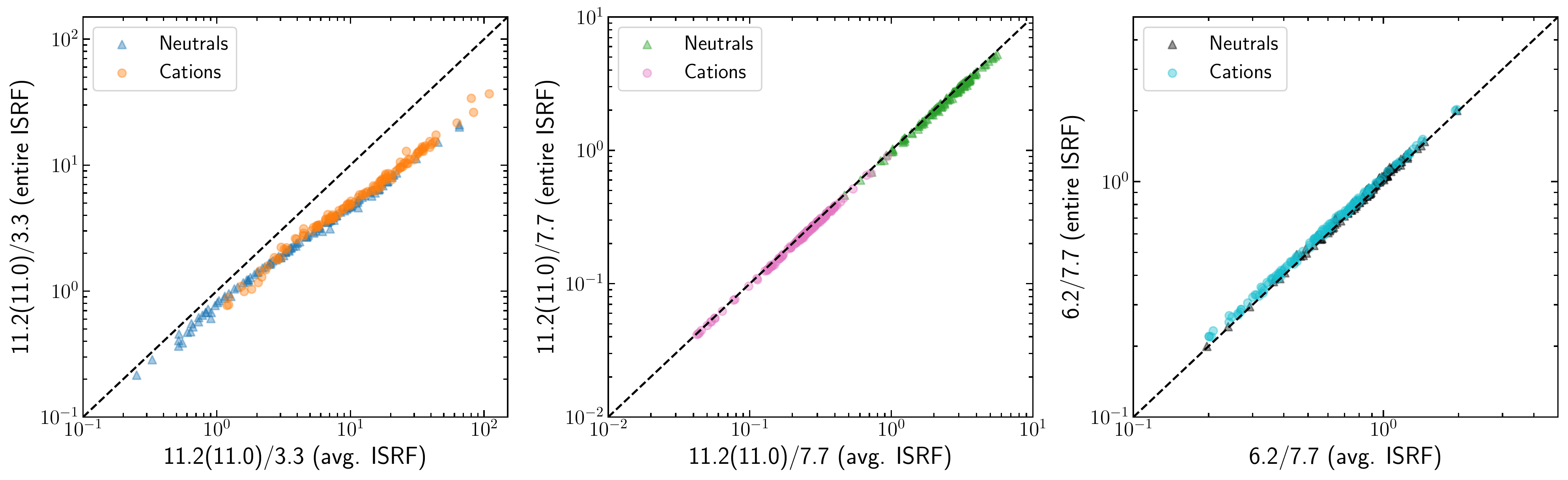}
		\caption{Comparison of PAH intensity ratios 11.2(11.0)/3.3 (left panel), 11.2(11.0)/7.7 (middle panel), and 6.2/7.7 (right panel) calculated from spectra generated with an average ISRF photon energy (filled points) and the entire ISRF (open points).
		}
		\label{fig:ISRFcomp}
	\end{center}
\end{figure*}

\begin{figure*}
	\begin{center}
		\hspace*{-0.5cm}\includegraphics[keepaspectratio=true,scale=0.46]{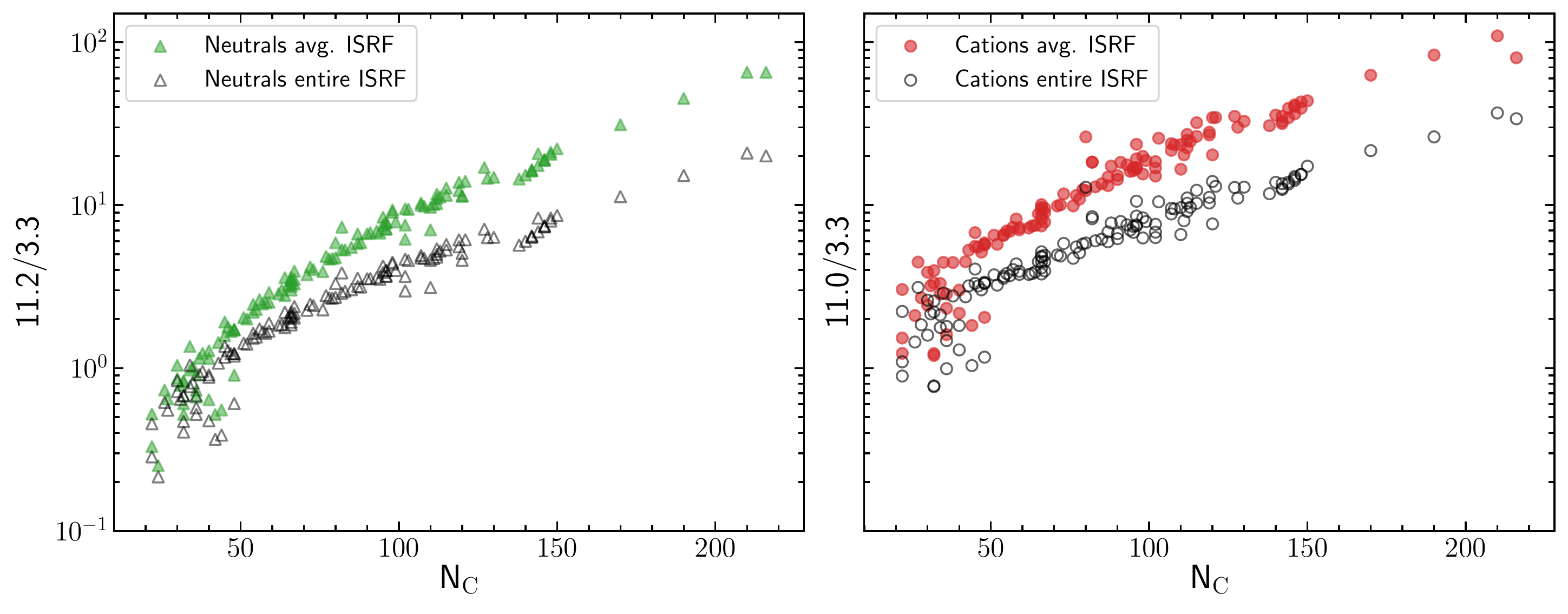}
		\caption{Comparison of PAH intensity ratios as a function of N$ _{\mathrm{C}} $ calculated from spectra generated with an average ISRF photon energy (filled points) and the entire ISRF (open points). Left panel: the intensity ratio of 11.2/3.3 for neutral PAHs; right panel: the intensity ratio of 11.0/3.3 for cations.}
		\label{fig:ISRF_size_comp}
	\end{center}
\end{figure*}

We present a comparison of the 11.2(11.0)/3.3 and 11.2(11.0)/7.7 and 6.2/7.7 intensity ratios from PAH spectra generated considering input for the emission model photons from the entire ISRF spectrum, against those generated assuming the average photon energy of the ISRF. Fig. \ref{fig:ISRFcomp} shows that the 11.2(11.0)/3.3 values of average ISRF photon energy spectra are higher compared to spectra generated considering the entire ISRF spectrum, while the other intensity ratios are unaffected. Fig. \ref{fig:ISRF_size_comp} presents the intensity ratios of 11.2/3.3 and 11.0/3.3 for neutral and cationic PAHs respectively, as a function of \Nc{} using the integrated ISRF and the average photon energy of the ISRF as input, where the difference between these two methods increases with \Nc. Inspecting the individual fluxes reveals that for larger PAHs the 3.3 \micron{} flux is relatively higher in the case of the entire ISRF generated spectra compared to those produced with the average ISRF photon energy, while the 11.2 and 11.0 \micron{} emission are similar in both cases.


\section{The effect of size, charge, and radiation field on PAH spectra} \label{sec:effectonspec}

\begin{figure*}
		\hspace*{-0.2cm}\includegraphics[keepaspectratio=true,scale=0.42]{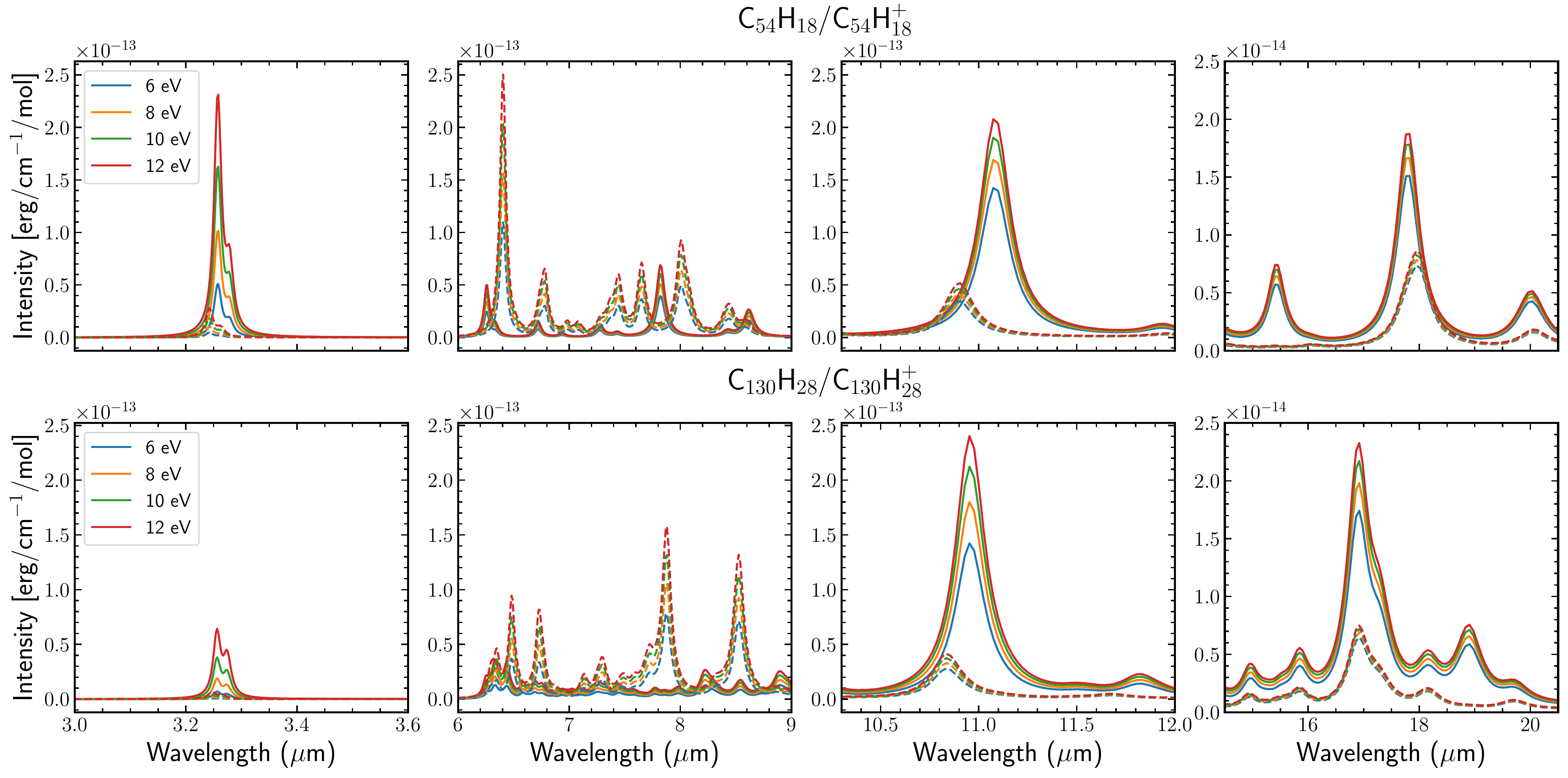}
		\caption{Comparison between the emission features of molecules C$_{54}$H$_{18}$ (top row; solid lines), C$_{54}$H$_{18}^{+}$ (top row; dashed lines), C$_{130}$H$_{28}$ (bottom row; solid lines), and C$_{130}$H$_{28}^{+}$ (bottom row; dashed lines). Different colored lines correspond to spectra generated from radiation fields with average photon energies of 6 (blue line), 8 (orange line), 10 (green line), and 12 eV (red line). The different panels present in order from left to right: the 3.3 \micron{} emission region (1st panel), the 6.2 \micron{}, 7.7 \micron{} complex, and 8.6 \micron{} emission regions (2nd panel), the 11.0 \micron{} and 11.2 \micron{} emission regions (3rd panel), and the $\Sigma_{(15-20)}$ \micron{} emission region (4th panel).}
		\label{fig:speccomp}
\end{figure*}

We demonstrate the variance of the PAH spectral emission features and their dependence on PAH size, charge, and the radiation field to which they are exposed. Figure \ref{fig:speccomp} shows the change of the emission characteristics in the 3.3 \micron, 7.7 \micron{} complex, and 11.2 \micron{} bands, for two different-sized neutral-cation PAH molecule pairs: C$_{54}$H$_{18}$/C$_{54}$H$_{18}^{+}$ and C$_{130}$H$_{28}$/C$_{130}$H$_{28}^{+}$. In all cases, exposure to radiation fields of increasing energy results to a corresponding increase to the intensity of each emission feature. For the smaller molecules (C$_{54}$H$_{18}$ and C$_{54}$H$_{18}^{+}$), the charge state has a significant impact in the shape, peak position, and intensity of almost all the major emission bands, with the 3.3 and 11.2 \micron{} features presenting the most striking contrast in intensity. For the largest molecules (C$_{130}$H$_{28}$ and C$_{130}$H$_{28}^{+}$) the most profound spectral diversities due to charge state are observed mostly at longer wavelengths. Specifically, the 7.7 \micron{} complex emission for cationic PAHs dominates over the corresponding emission of their neutral counterpart, while the opposite is the case for the 11.2 \micron{} emission. PAH size has a great impact on the spectral characteristics at shorter wavelengths and specifically the 3.3 \micron{} band, explaining the observed dependence of the 3.3 \micron{} emission with \Nc{} (Section \ref{sec:33-Nc} and Fig. \ref{fig:fluxcomp}). The 11.2 \micron{} emission is mostly unaffected by PAH size when examining the different-sized neutral and cationic species of C$_{54}$H$_{18}$ and C$_{130}$H$_{28}$ respectively, although a slight increase is observed for the larger neutral molecules, as demonstrated in Fig. \ref{fig:fluxcomp}. 

\section{Examining molecular structure and symmetry}

\begin{figure*}
	\begin{center}
		\includegraphics[keepaspectratio=true,scale=0.4]{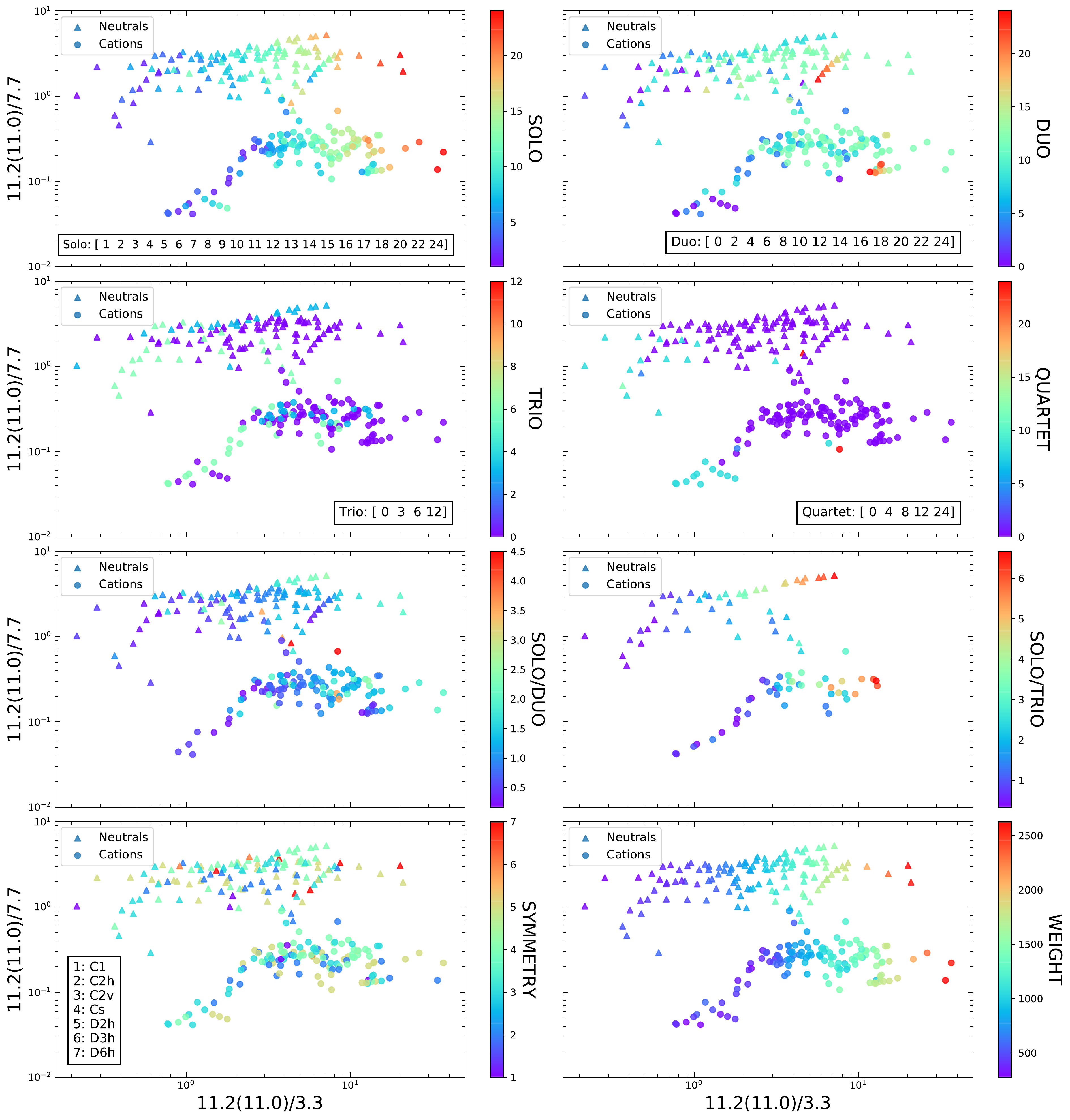}
		\caption{The 11.2(11.0)/7.7 -- 11.2(11.0)/3.3 plane of the PAHdb sample molecules calculated with the ISRF model, color-coded based on the different molecular properties. In order from top left to bottom right the panels and color-coding corresponds to: the number of solos, duos, trios, and quartets C-H groups, the ratio of solo/duo, ratio of solo/trio, symmetry groups, and molecular weight. The discrete values of solos, duos, trios, quartets C-H groups, and molecular symmetry groups are indicated in the legend of the corresponding panels. Neutral PAHs are represented with triangles and cationic PAHs with circles.}
		\label{fig:strucsym}
	\end{center}
\end{figure*}

Figure \ref{fig:strucsym} shows the 11.2(11.0)/7.7 -- 11.2(11.0)/3.3 plane with the PAHs classified according to their edge structure (in terms of number of solo, duo, trio, and quartet CH groups and their ratios), symmetry, and weight. The 11.2(11.0)/3.3 ratio clearly reflects the number of solo CH groups and the molecular weight: the 11.0/11.2 $\mu$m bands are due to the CH$_{oop}$ bending modes of solo CH groups and the molecular weights correlates with \Nc{}. In contrast, the number of duo CH groups is not a good tracer of \Nc{} nor are the number of trio or quartet CH groups. The number of duos first increases but then, for larger sizes, the number of duos remains relatively constant. The number of trios and quartets tends to be small and fairly constant for a broad range of sizes. Likewise, the solo/duo does not show a gradual change with size as more scatter is introduced by adding the duos. In contrast, the solo/trio ratio does follow size. As the number of trios is fairly constant with size, the trends obtained using the solo are comparable to those using the ratio solo/trio. Finally, we do not find any relation between size and  symmetry.

\section{PAH size scaling relations for different radiation fields} \label{sec:scal-rel}

Here we present the parameters of the power-law calibration (eq. \ref{eq:powerlaw}) describing the (11.2+11.0)/3.3 -- \Nc{} relation at different ionization fractions, for radiation fields with average photon energies of 6, 8, 10, and 12 eV. All calibration parameters are given in Tables \ref{tab:calibr6ev}-\ref{tab:calibr12ev}.

\begin{table*}
\parbox{.475\linewidth}{
\centering
		\caption{The 6 eV charge--size grid log parabola and polynomial fit parameters.}
		\label{tab:fitparams6ev}
		\begin{tabular}{@{}lcccc}
			\hline
			\multicolumn{5}{c}{Log parabola parameters} \\
			Charge state & A & x$ _{0} $ & $ \alpha $ & $ \beta $ \\
			\hline
            Neutral & 1.87 & 0.99 & -0.33 & 0.05 \\
            75\% Neutral + 25\% Cation & 0.70 & 0.58 & -0.56 & 0.09 \\
            50\% Neutral + 50\% Cation & 0.64 & 1.74 & -0.40 & 0.10 \\
            25\% Neutral + 75\% Cation & 0.16 & 0.48 & -0.80 & 0.13 \\
            Cation & 0.02 & 0.38 & -1.61 & 0.22 \\
			\hline
			\multicolumn{5}{c}{$3^{\mathrm{rd}}$ degree polynomial parameters} \\
			$\overline{\mathrm{N_{C}}}$ & c0 & c1 & c2 & c3 \\
			\hline
            32 & 34.24 & -141.77 & 181.24 & -66.63 \\
            35 & 55.87 & -98.33 & 54.47 & -9.06 \\
            60 & 89.65 & -67.47 & 16.25 & -1.22 \\
            94 & 146.81 & -47.59 & 5.00 & -0.17 \\
            132 & 255.10 & -35.84 & 1.65 & -0.02 \\
            195 & 493.90 & -30.23 & 0.61 & 0.00 \\
			\hline
		\end{tabular}}
\hfill
\parbox{.475\linewidth}{
\centering
		\caption{The 8 eV charge--size grid log parabola and polynomial fit parameters.}
		\label{tab:fitparams8ev}
		\begin{tabular}{@{}lcccc}
			\hline
			\multicolumn{5}{c}{Log parabola parameters} \\
			Charge state & A & x$ _{0} $ & $ \alpha $ & $ \beta $ \\
			\hline
            Neutral & 2.16 & 1.24 & -0.27 & 0.09 \\
            75\% Neutral + 25\% Cation & 1.18 & 1.52 & -0.30 & 0.11 \\
            50\% Neutral + 50\% Cation & 0.78 & 2.42 & -0.23 & 0.14 \\
            25\% Neutral + 75\% Cation & 0.20 & 0.52 & -0.77 & 0.16 \\
            Cation & 0.02 & 0.36 & -1.59 & 0.26 \\
			\hline
			\multicolumn{5}{c}{$3^{\mathrm{rd}}$ degree polynomial parameters} \\
			$\overline{\mathrm{N_{C}}}$ & c0 & c1 & c2 & c3 \\
			\hline
            32 & 26.54 & -168.91 & 337.48 & -202.03 \\
            35 & 44.89 & -132.15 & 123.83 & -35.68 \\
            59 & 70.82 & -96.79 & 42.60 & -5.89 \\
            94 & 106.52 & -67.84 & 14.05 & -0.93 \\
            133 & 156.42 & -46.60 & 4.55 & -0.14 \\
            196 & 230.38 & -32.21 & 1.49 & -0.02 \\
			\hline
		\end{tabular}}
\vskip\baselineskip
\parbox{.475\linewidth}{
\centering
		\caption{The 10 eV charge--size grid log parabola and polynomial fit parameters.}
		\label{tab:fitparams10ev}
		\begin{tabular}{@{}lcccc}
			\hline
			\multicolumn{5}{c}{Log parabola parameters} \\
			Charge state & A & x$ _{0} $ & $ \alpha $ & $ \beta $ \\
			\hline
            Neutral & 1.93 & 0.78 & -0.31 & 0.10 \\
            75\% Neutral + 25\% Cation & 0.83 & 0.53 & -0.49 & 0.13 \\
            50\% Neutral + 50\% Cation & 0.76 & 2.04 & -0.15 & 0.17 \\
            25\% Neutral + 75\% Cation & 0.22 & 0.49 & -0.73 & 0.18 \\
            Cation & 0.03 & 0.40 & -1.65 & 0.34 \\
			\hline
			\multicolumn{5}{c}{$3^{\mathrm{rd}}$ degree polynomial parameters} \\
			$\overline{\mathrm{N_{C}}}$ & c0 & c1 & c2 & c3 \\
			\hline
            33 & 22.18 & -191.56 & 525.54 & -442.96 \\
            34 & 37.58 & -159.84 & 218.31 & -93.28 \\
            58 & 59.92 & -125.96 & 85.78 & -18.56 \\
            94 & 92.16 & -96.15 & 32.72 & -3.58 \\
            134 & 139.93 & -72.72 & 12.39 & -0.69 \\
            194 & 213.86 & -55.48 & 4.74 & -0.13 \\
			\hline
		\end{tabular}}
\hfill
\parbox{.475\linewidth}{
\centering
		\caption{The 12 eV charge--size grid log parabola and polynomial fit parameters.}
		\label{tab:fitparams12ev}
		\begin{tabular}{@{}lcccc}
			\hline
			\multicolumn{5}{c}{Log parabola parameters} \\
			Charge state & A & x$ _{0} $ & $ \alpha $ & $ \beta $ \\
			\hline
            Neutral & 1.77 & 0.54 & -0.35 & 0.12 \\
            75\% Neutral + 25\% Cation & 0.74 & 0.39 & -0.60 & 0.17 \\
            50\% Neutral + 50\% Cation & 0.73 & 1.97 & -0.04 & 0.19 \\
            25\% Neutral + 75\% Cation & 0.24 & 0.49 & -0.70 & 0.23 \\
            Cation & 0.04 & 0.37 & -1.67 & 0.40 \\
			\hline
			\multicolumn{5}{c}{$3^{\mathrm{rd}}$ degree polynomial parameters} \\
			$\overline{\mathrm{N_{C}}}$ & c0 & c1 & c2 & c3 \\
			\hline
            33 & 19.87 & -216.14 & 752.85 & -819.16 \\
            34 & 34.39 & -193.46 & 351.74 & -202.50 \\
            56 & 55.51 & -162.06 & 154.02 & -46.89 \\
            92 & 85.78 & -130.52 & 64.98 & -10.46 \\
            132 & 130.41 & -103.79 & 27.13 & -2.31 \\
            195 & 199.90 & -83.42 & 11.47 & -0.52 \\
			\hline
		\end{tabular}}
\end{table*}

\begin{table*}
        \begin{center}
                \caption{6 eV charge -- size grid parameters for the average value points.}
                \label{tab:binparams6ev}
                \begin{tabular}{ccccccccc}
                \toprule
                \multicolumn{3}{c}{Neutrals} & \multicolumn{3}{c}{N75C25} & \multicolumn{3}{c}{N50C50} \\
                \cmidrule(lr){1-3} \cmidrule(lr){4-6} \cmidrule(lr){7-9}
                \Nc{} & $\frac{11.2}{3.3}$ & $\frac{11.2}{7.7}$ & \Nc{} & $\frac{(11.2+11.0)}{3.3}$ & $\frac{(11.2+11.0)}{7.7}$  & \Nc{} & $\frac{(11.2+11.0)}{3.3}$ & $\frac{(11.2+11.0)}{7.7}$ \\
                \hline    
                32 & 0.44 & 1.27 & 32 & 0.47 & 0.49 & 30 & 0.51 & 0.24 \\
                36 & 1.07 & 2.08 & 36 & 1.15 & 1.09 & 35 & 1.23 & 0.59 \\
                60 & 2.64 & 2.40 & 60 & 2.80 & 1.40 & 59 & 2.98 & 0.87 \\
                93 & 6.49 & 2.85 & 94 & 6.83 & 1.52 & 92 & 7.23 & 0.88 \\
                134 & 15.97 & 3.02 & 132 & 16.64 & 1.56 & 133 & 17.54 & 0.88 \\
                192 & 39.29 & 3.05 & 196 & 40.57 & 1.46 & 192 & 42.57 & 0.87 \\                \midrule
                \multicolumn{3}{c}{N25C75} & \multicolumn{3}{c}{Cations} & \multicolumn{3}{c}{} \\
                \cmidrule(lr){1-3} \cmidrule(lr){4-6}
                \Nc{} & $\frac{(11.2+11.0)}{3.3}$ & $\frac{(11.2+11.0)}{7.7}$ & \Nc{} & $\frac{11.0}{3.3}$ & $\frac{11.0}{7.7}$ \\
                \hline
                31 & 0.63 & 0.14 & 33 & 1.58 & 0.05 & & & \\
                35 & 1.49 & 0.34 & 35 & 3.27 & 0.19 & & & \\
                60 & 3.53 & 0.53 & 60 & 6.77 & 0.31 & & & \\
                93 & 8.35 & 0.52 & 97 & 14.04 & 0.27 & & & \\
                130 & 19.76 & 0.50 & 131 & 29.10 & 0.23 & & & \\
                196 & 46.75 & 0.47 & 196 & 60.31 & 0.23 & & & \\             
                \bottomrule
                \end{tabular} 
        \end{center}
\end{table*}

\begin{table*}
        \begin{center}
                \caption{8 eV charge -- size grid parameters for the average value points.}
                \label{tab:binparams8ev}
                \begin{tabular}{ccccccccc}
                \toprule
                \multicolumn{3}{c}{Neutrals} & \multicolumn{3}{c}{N75C25} & \multicolumn{3}{c}{N50C50} \\
                \cmidrule(lr){1-3} \cmidrule(lr){4-6} \cmidrule(lr){7-9}
                \Nc{} & $\frac{11.2}{3.3}$ & $\frac{11.2}{7.7}$ & \Nc{} & $\frac{(11.2+11.0)}{3.3}$ & $\frac{(11.2+11.0)}{7.7}$  & \Nc{} & $\frac{(11.2+11.0)}{3.3}$ & $\frac{(11.2+11.0)}{7.7}$ \\
                \hline    
                32 & 0.28 & 1.10 & 32 & 0.31 & 0.44 & 30 & 0.33 & 0.22 \\
                35 & 0.64 & 1.94 & 35 & 0.69 & 0.96 & 35 & 0.74 & 0.53 \\
                59 & 1.44 & 2.12 & 59 & 1.53 & 1.24 & 59 & 1.65 & 0.77 \\
                92 & 3.24 & 2.56 & 94 & 3.42 & 1.35 & 93 & 3.66 & 0.78 \\
                131 & 7.30 & 2.65 & 135 & 7.64 & 1.39 & 132 & 8.13 & 0.80 \\
                196 & 16.45 & 2.37 & 192 & 17.06 & 1.36 & 196 & 18.05 & 0.74 \\
                \midrule
                \multicolumn{3}{c}{N25C75} & \multicolumn{3}{c}{Cations} & \multicolumn{3}{c}{} \\
                \cmidrule(lr){1-3} \cmidrule(lr){4-6}
                \Nc{} & $\frac{(11.2+11.0)}{3.3}$ & $\frac{(11.2+11.0)}{7.7}$ & \Nc{} & $\frac{11.0}{3.3}$ & $\frac{11.0}{7.7}$ \\
                \hline
                32 & 0.42 & 0.12 & 34 & 0.91 & 0.05 & & & \\
                35 & 0.91 & 0.30 & 34 & 1.80 & 0.16 & & & \\
                60 & 1.98 & 0.47 & 59 & 3.56 & 0.27 & & & \\
                95 & 4.29 & 0.46 & 97 & 7.04 & 0.24 & & & \\
                133 & 9.31 & 0.44 & 131 & 13.92 & 0.19 & & & \\
                196 & 20.19 & 0.41 & 196 & 27.52 & 0.19 & & & \\
                \bottomrule
                \end{tabular} 
        \end{center}
\end{table*}

\begin{table*}
        \begin{center}
                \caption{10 eV charge -- size grid parameters for the average value points.}
                \label{tab:binparams10ev}
                \begin{tabular}{ccccccccc}
                \toprule
                \multicolumn{3}{c}{Neutrals} & \multicolumn{3}{c}{N75C25} & \multicolumn{3}{c}{N50C50} \\
                \cmidrule(lr){1-3} \cmidrule(lr){4-6} \cmidrule(lr){7-9}
                \Nc{} & $\frac{11.2}{3.3}$ & $\frac{11.2}{7.7}$ & \Nc{} & $\frac{(11.2+11.0)}{3.3}$ & $\frac{(11.2+11.0)}{7.7}$  & \Nc{} & $\frac{(11.2+11.0)}{3.3}$ & $\frac{(11.2+11.0)}{7.7}$ \\
                \hline    
                34 & 0.21 & 0.90 & 31 & 0.23 & 0.40 & 31 & 0.25 & 0.21 \\
                34 & 0.44 & 1.83 & 35 & 0.47 & 0.86 & 35 & 0.52 & 0.48 \\
                56 & 0.94 & 1.97 & 58 & 0.99 & 1.13 & 58 & 1.08 & 0.71 \\
                91 & 1.98 & 2.29 & 93 & 2.08 & 1.23 & 93 & 2.25 & 0.72 \\
                133 & 4.18 & 2.43 & 134 & 4.36 & 1.29 & 132 & 4.67 & 0.74 \\
                192 & 8.82 & 2.29 & 192 & 9.15 & 1.23 & 196 & 9.70 & 0.67 \\             
                \midrule
                \multicolumn{3}{c}{N25C75} & \multicolumn{3}{c}{Cations} & \multicolumn{3}{c}{} \\
                \cmidrule(lr){1-3} \cmidrule(lr){4-6}
                \Nc{} & $\frac{(11.2+11.0)}{3.3}$ & $\frac{(11.2+11.0)}{7.7}$ & \Nc{} & $\frac{11.0}{3.3}$ & $\frac{11.0}{7.7}$ \\
                \hline
                34 & 0.32 & 0.12 & 34 & 0.62 & 0.04 & & & \\
                34 & 0.64 & 0.26 & 33 & 1.17 & 0.10 & & & \\
                58 & 1.31 & 0.43 & 57 & 2.21 & 0.25 & & & \\
                94 & 2.65 & 0.42 & 98 & 4.18 & 0.22 & & & \\
                136 & 5.37 & 0.40 & 132 & 7.93 & 0.17 & & & \\
                192 & 10.89 & 0.39 & 196 & 15.01 & 0.17 & & & \\
                \bottomrule
                \end{tabular} 
        \end{center}
\end{table*}

\begin{table*}
        \begin{center}
                \caption{12 eV charge -- size grid parameters for the average value points.}
                \label{tab:binparams12ev}
                \begin{tabular}{ccccccccc}
                \toprule
                \multicolumn{3}{c}{Neutrals} & \multicolumn{3}{c}{N75C25} & \multicolumn{3}{c}{N50C50} \\
                \cmidrule(lr){1-3} \cmidrule(lr){4-6} \cmidrule(lr){7-9}
                \Nc{} & $\frac{11.2}{3.3}$ & $\frac{11.2}{7.7}$ & \Nc{} & $\frac{(11.2+11.0)}{3.3}$ & $\frac{(11.2+11.0)}{7.7}$  & \Nc{} & $\frac{(11.2+11.0)}{3.3}$ & $\frac{(11.2+11.0)}{7.7}$ \\
                \hline    
                34 & 0.17 & 0.85 & 32 & 0.18 & 0.34 & 31 & 0.20 & 0.19 \\
                34 & 0.34 & 1.65 & 34 & 0.36 & 0.78 & 34 & 0.40 & 0.43 \\
                55 & 0.68 & 1.92 & 56 & 0.72 & 1.05 & 56 & 0.79 & 0.65 \\
                89 & 1.36 & 2.11 & 90 & 1.44 & 1.13 & 91 & 1.55 & 0.66 \\
                132 & 2.74 & 2.29 & 130 & 2.86 & 1.23 & 131 & 3.07 & 0.70 \\
                192 & 5.50 & 2.09 & 196 & 5.70 & 1.07 & 196 & 6.04 & 0.62 \\
                \midrule
                \multicolumn{3}{c}{N25C75} & \multicolumn{3}{c}{Cations} & \multicolumn{3}{c}{} \\
                \cmidrule(lr){1-3} \cmidrule(lr){4-6}
                \Nc{} & $\frac{(11.2+11.0)}{3.3}$ & $\frac{(11.2+11.0)}{7.7}$ & \Nc{} & $\frac{11.0}{3.3}$ & $\frac{11.0}{7.7}$ \\
                \hline
                34 & 0.26 & 0.11 & 34 & 0.46 & 0.04 & & & \\
                33 & 0.50 & 0.25 & 33 & 0.84 & 0.08 & & & \\
                58 & 0.95 & 0.40 & 56 & 1.53 & 0.23 & & & \\
                94 & 1.83 & 0.39 & 97 & 2.79 & 0.20 & & & \\
                133 & 3.53 & 0.37 & 135 & 5.10 & 0.16 & & & \\
                196 & 6.78 & 0.35 & 192 & 9.30 & 0.16 & & & \\
                \bottomrule
                \end{tabular} 
        \end{center}
\end{table*}

\begin{table*}
\parbox{.475\linewidth}{
\centering
			\caption{6 eV power-law fit parameters for the different PAH charge states, as described by equation \ref{eq:powerlaw}.}
			\label{tab:calibr6ev}
			\begin{tabular}{lcc}
				\hline 
				Charge state & c$_{o}$ & $\alpha$ \\ 
				\hline
			Neutral	& -2.648 & 0.275 \\
			75\% Neutral + 25\% Cation & -2.630 & 0.275 \\
			50\% Neutral + 50\% Cation & -2.615 & 0.276 \\
			25\% Neutral + 75\% Cation & -2.480 & 0.270 \\
			Cation & -2.050 & 0.254 \\
				\hline
			\end{tabular}}
\hfill
\parbox{.475\linewidth}{
\centering
			\caption{8 eV power-law fit parameters for the different PAH charge states, as described by equation \ref{eq:powerlaw}.}
			\label{tab:calibr8ev}
			\begin{tabular}{lcc}
				\hline 
				Charge state & c$_{o}$ & $\alpha$ \\ 
				\hline
			Neutral	& -2.720 & 0.260 \\
			75\% Neutral + 25\% Cation & -2.710 & 0.260 \\
			50\% Neutral + 50\% Cation & -2.620 & 0.256 \\
			25\% Neutral + 75\% Cation & -2.489 & 0.250 \\
			Cation & -2.180 & 0.242 \\
				\hline
			\end{tabular}}
\vskip\baselineskip
\parbox{.475\linewidth}{
\centering
			\caption{10 eV power-law fit parameters for the different PAH charge states, as described by equation \ref{eq:powerlaw}.}
			\label{tab:calibr10ev}
			\begin{tabular}{lcc}
				\hline 
				Charge state & c$_{o}$ & $\alpha$ \\ 
				\hline
			Neutral	& -2.660 & 0.241 \\
			75\% Neutral + 25\% Cation & -2.670 & 0.243 \\
			50\% Neutral + 50\% Cation & -2.620 & 0.241 \\
			25\% Neutral + 75\% Cation & -2.545 & 0.240 \\
			Cation & -2.170 & 0.225 \\
				\hline
			\end{tabular}}
\hfill
\parbox{.475\linewidth}{
\centering
			\caption{12 eV power-law fit parameters for the different PAH charge states, as described by equation \ref{eq:powerlaw}.}
			\label{tab:calibr12ev}
			\begin{tabular}{lcc}
				\hline 
				Charge state & c$_{o}$ & $\alpha$ \\ 
				\hline
			Neutral	& -2.770 & 0.238 \\
			75\% Neutral + 25\% Cation & -2.730 & 0.237 \\
			50\% Neutral + 50\% Cation & -2.670 & 0.234 \\
			25\% Neutral + 75\% Cation & -2.570 & 0.231 \\
			Cation & -2.300 & 0.222 \\
				\hline
			\end{tabular}}
\end{table*}

\section{Alternate PAH charge -- size grids}

\begin{figure*}
	\begin{subfigure}{0.5\textwidth}
		\centering
		\includegraphics[keepaspectratio=true,scale=0.45]{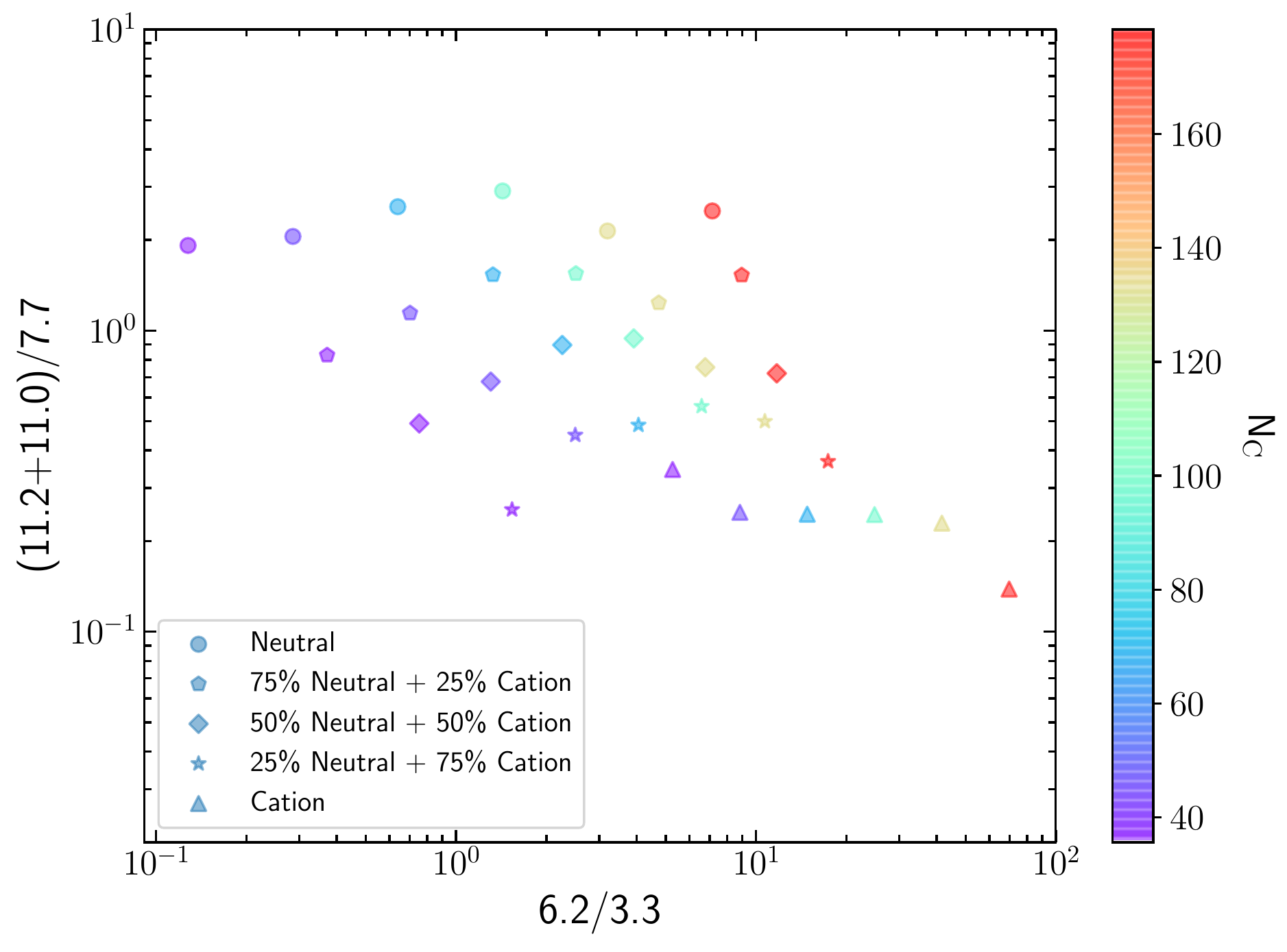}
	\end{subfigure}%
	\hfill
	\begin{subfigure}{0.5\textwidth}
		\centering
		\includegraphics[keepaspectratio=true,scale=0.45]{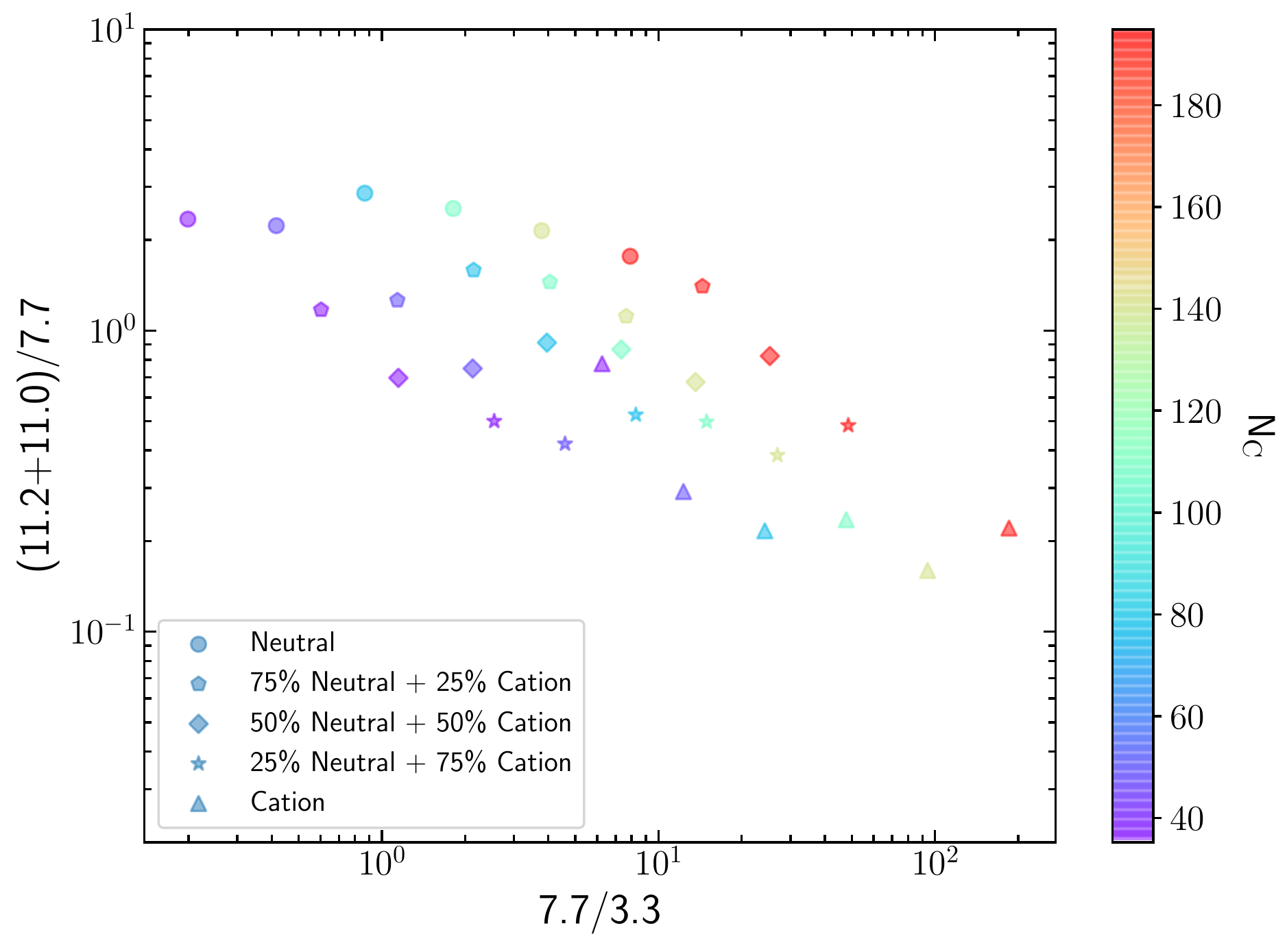}\\
	\end{subfigure}
	\vskip\baselineskip
	\begin{subfigure}{0.5\textwidth}
		\centering
		\includegraphics[keepaspectratio=true,scale=0.45]{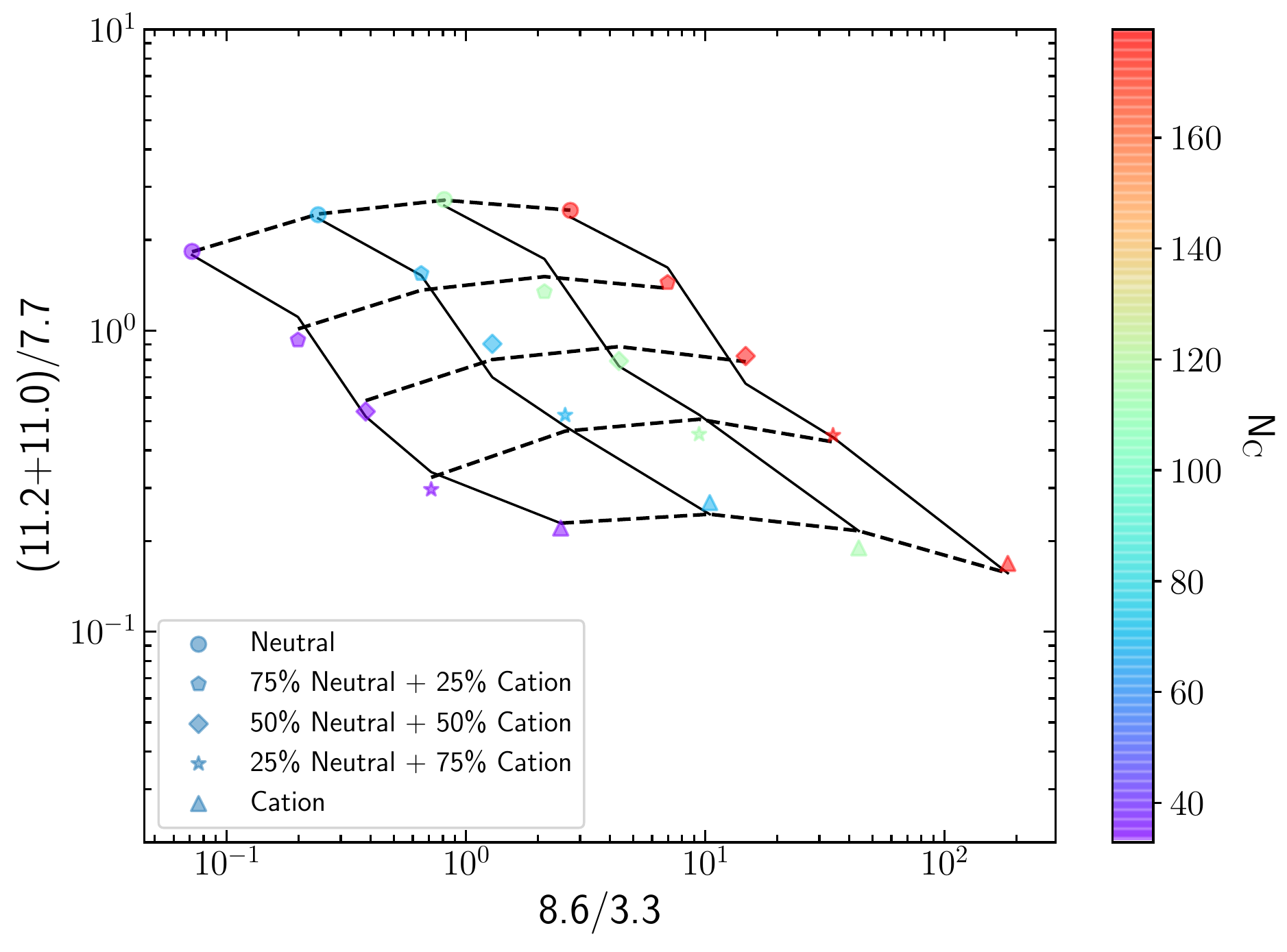}
	\end{subfigure}%
	\hfill
	\begin{subfigure}{0.5\textwidth}
		\centering
		\includegraphics[keepaspectratio=true,scale=0.45]{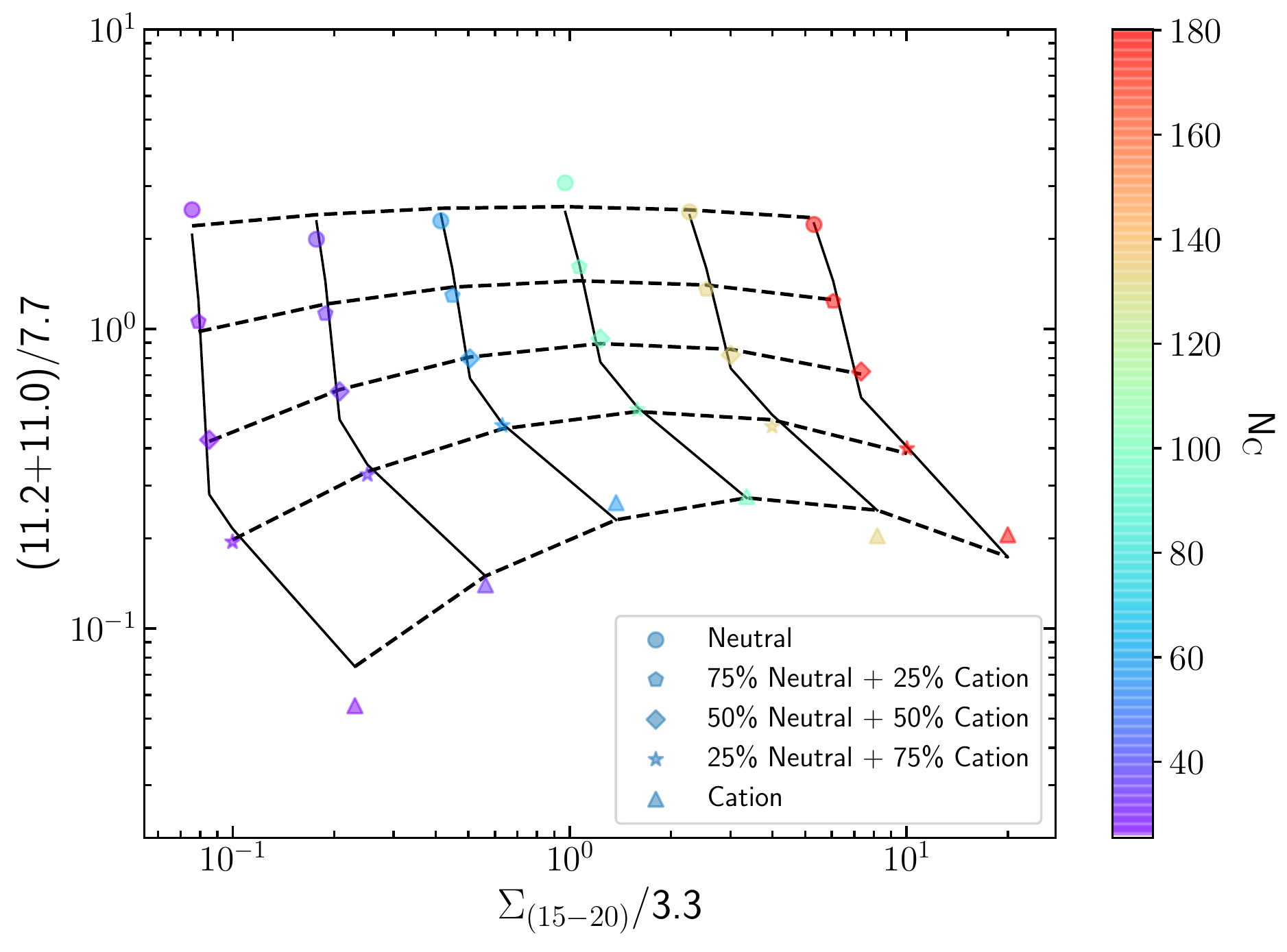}\\
	\end{subfigure}
	\caption{The PAH charge -- size space of spectra generated with the ISRF model, using alternate PAH intensity ratios as size tracers. The (11.2+11.0)/7.7 ratio is used in all panels as PAH charge state tracer, along with the ratios of 6.2/3.3 (top left panel), 7.7/3.3 (top right panel), 8.6/3.3 (bottom left panel), and $\Sigma_{(15-20)}$/3.3 (bottom right panel), for PAH size estimation. The description for the points, color-coding, and grid lines (where applicable) is the same as in Fig. \ref{fig:grid}.}
	\label{fig:altgrids}
\end{figure*}

We examined the applicability of alternate PAH charge -- size grids, keeping the (11.2+11.0)/7.7 ratio as PAH charge tracer, and substituting the (11.2+11.0)/3.3 ratio with the ratios of 6.2/3.3, 7.7/3.3, 8.6/3.3, and $\Sigma_{(15-20)}$/3.3 (Fig. \ref{fig:altgrids}), given their observed scaling with \Nc{} (see Section \ref{sec:33-Nc} and Fig. \ref{fig:newratiosAll}). We followed the methodology described in Section \ref{sec:grid} using spectra generated with the ISRF model. A grid construction was feasible for the 8.6/3.3 and $\Sigma_{(15-20)}$/3.3 cases where the wavelength separation, and thus the carrier size, between the corresponding features increased. The shape of the (11.2+11.0)/7.7 -- $\Sigma_{(15-20)}$/3.3 grid is similar to the (11.2+11.0)/7.7 -- (11.2+11.0)/3.3 grid, as a result of the tight scaling of $\Sigma_{(15-20)}$/3.3 with \Nc. The (11.2+11.0)/7.7 -- $\Sigma_{(15-20)}$/3.3 grid fit parameters, \Nc{} values, and intensity ratios for the binned points, are given in Tables \ref{tab:altgridfitparams} and \ref{tab:altgridbinparams}.

\begin{table}
		\caption{The ISRF (11.2+11.0)/7.7 -- $\Sigma_{(15-20)}$/3.3 charge--size grid log parabola and polynomial fit parameters (presented in the bottom right panel of Fig. \ref{fig:altgrids}) as described by equations \ref{eq:logparab} and \ref{eq:polynom}.}
		\label{tab:altgridfitparams}
		\hspace{-1.0cm}
		\begin{tabular}{@{}lcccc}
			\hline
			\multicolumn{5}{c}{ISRF Log parabola parameters (eq. \ref{eq:logparab})} \\
			Charge state & A & x$ _{0} $ & $ \alpha $ & $ \beta $ \\
			\hline
			Neutral & 2.43 & 0.20 & -0.07 & 0.03 \\
            75\% Neutral + 25\% Cation & 1.16 & 0.15 & -0.22 & 0.05 \\
            50\% Neutral + 50\% Cation & 0.88 & 0.90 & -0.09 & 0.09 \\
            25\% Neutral + 75\% Cation & 0.34 & 0.25 & -0.46 & 0.12 \\
            Cation & 0.01 & 0.04 & -1.55 & 0.17 \\
			\hline
			\multicolumn{5}{c}{ISRF $3^{\mathrm{rd}}$ degree polynomial parameters (eq. \ref{eq:polynom})} \\
			$\overline{\mathrm{N_{C}}}$ & c0 & c1 & c2 & c3 \\
			\hline
			25 & 97.67 & -2535.82 & 20557.80 & -49430.48  \\
            33 & 58.20 & -605.80 & 1980.31 & -1931.98 \\
            57 & 38.03 & -157.96 & 207.96 & -82.26 \\
            96 & 27.05 & -44.84 & 23.79 & -3.82 \\
            133 & 20.84 & -13.89 & 2.98 & -0.19 \\
            180 & 17.00 & -4.61 & 0.40 & -0.01 \\
			\hline
		\end{tabular} 
\end{table}

\begin{table*}
    \begin{center}
    \begin{minipage}{125mm}
                \caption{ISRF (11.2+11.0)/7.7 -- $\Sigma_{(15-20)}$/3.3 charge -- size grid average points parameters.}
                \label{tab:altgridbinparams}
                \begin{tabular}{ccccccccc}
                \toprule
                \multicolumn{3}{c}{Neutrals} & \multicolumn{3}{c}{N75C25} & \multicolumn{3}{c}{N50C50} \\
                \cmidrule(lr){1-3} \cmidrule(lr){4-6} \cmidrule(lr){7-9}
                \Nc{} & $\frac{11.2}{3.3}$ & $\frac{11.2}{7.7}$ & \Nc{} & $\frac{(11.2+11.0)}{3.3}$ & $\frac{(11.2+11.0)}{7.7}$  & \Nc{} & $\frac{(11.2+11.0)}{3.3}$ & $\frac{(11.2+11.0)}{7.7}$ \\
                \cmidrule(lr){1-3} \cmidrule(lr){4-6} \cmidrule(lr){7-9}    
                26 & 0.08 & 2.50 & 26 & 0.08 & 1.06 & 26 & 0.09 & 0.43 \\
                36 & 0.18 & 1.99 & 34 & 0.19 & 1.12 & 34 & 0.21 & 0.62 \\
                60 & 0.41 & 2.30 & 59 & 0.45 & 1.29 & 58 & 0.51 & 0.80 \\
                99 & 0.97 & 3.08 & 98 & 1.07 & 1.61 & 96 & 1.23 & 0.93 \\
                132 & 2.27 & 2.46 & 132 & 2.54 & 1.36 & 132 & 3.01 & 0.82 \\
                184 & 5.31 & 2.23 & 184 & 6.05 & 1.24 & 184 & 7.33 & 0.72 \\
                \cmidrule(lr){1-3} \cmidrule(lr){4-6} \cmidrule(lr){7-9}
                \multicolumn{3}{c}{N25C75} & \multicolumn{3}{c}{Cations} & \multicolumn{3}{c}{} \\
                \cmidrule(lr){1-3} \cmidrule(lr){4-6}
                \Nc{} & $\frac{(11.2+11.0)}{3.3}$ & $\frac{(11.2+11.0)}{7.7}$ & \Nc{} & $\frac{11.0}{3.3}$ & $\frac{11.0}{7.7}$ \\
                \cmidrule(lr){1-3} \cmidrule(lr){4-6}
                26 & 0.10 & 0.19 & 24 & 0.23 & 0.06 & & & \\
                30 & 0.25 & 0.33 & 31 & 0.56 & 0.14 & & & \\
                55 & 0.63 & 0.48 & 52 & 1.37 & 0.26 & & & \\
                94 & 1.59 & 0.54 & 94 & 3.35 & 0.28 & & & \\
                134 & 3.99 & 0.47 & 133 & 8.18 & 0.20 & & & \\
                175 & 10.03 & 0.40 & 173 & 19.97 & 0.21 & & & \\
                \bottomrule
                \end{tabular}
    \end{minipage}
    \end{center}
\end{table*}

\section{PAHdb sample UIDs}

Tables \ref{tab:UIDsv300} - \ref{tab:UIDsv310An} present the PAHdb UIDs of our sample molecules.

\begin{table*}
		\caption{The PAHdb v3.00 UIDs for neutral (N) and cation (C) molecules.}
		\label{tab:UIDsv300}
		\begin{tabular}{@{}cccccccccccc}
  \hline
  \multicolumn{1}{|c|}{PAH} & \multicolumn{2}{c}{UIDs} &
  \multicolumn{1}{|c|}{PAH} & \multicolumn{2}{c}{UIDs} &
  \multicolumn{1}{|c|}{PAH} & \multicolumn{2}{c}{UIDs} &
  \multicolumn{1}{|c|}{PAH} & \multicolumn{2}{c}{UIDs} \\
    & N & C & & N & C & & N & C & & N & C \\
  \hline
  C$_{32}$H$_{14}$ & 4 & 5 & C$_{130}$H$_{28}$ & 168 & 169 & C$_{66}$H$_{20}$ & 600 & 601 & C$_{45}$H$_{15}$ & 718 & 722\\
  C$_{48}$H$_{18}$ & 35 & 36 & C$_{102}$H$_{26}$ & 177 & 178 & C$_{66}$H$_{20}$ & 603 & 604 & C$_{80}$H$_{20}$ & 719 & 730\\
  C$_{54}$H$_{18}$ & 37 & 38 & C$_{102}$H$_{26}$ & 180 & 181 & C$_{112}$H$_{26}$ & 606 & 607 & C$_{63}$H$_{21}$ & 727 & 731\\
  C$_{27}$H$_{13}$ & 66 & 65 & C$_{110}$H$_{30}$ & 183 & 184 & C$_{150}$H$_{30}$ & 612 & 613 & C$_{112}$H$_{28}$ & 728 & 738\\
  C$_{47}$H$_{17}$ & 76 & 77 & C$_{120}$H$_{36}$ & 186 & 187 & C$_{216}$H$_{36}$ & 615 & 621 & C$_{210}$H$_{36}$ & 742 & 740\\
  C$_{59}$H$_{19}$ & 89 & 88 & C$_{24}$H$_{14}$ & 204 & 205 & C$_{170}$H$_{32}$ & 619 & 623 & C$_{148}$H$_{30}$ & 743 & 744\\
  C$_{48}$H$_{20}$ & 100 & 101 & C$_{24}$H$_{14}$ & 206 & 207 & C$_{40}$H$_{16}$ & 625 & 626 & C$_{146}$H$_{30}$ & 746 & 747\\
  C$_{96}$H$_{24}$ & 108 & 111 & C$_{24}$H$_{14}$ & 208 & 209 & C$_{128}$H$_{28}$ & 631 & 632 & C$_{146}$H$_{30}$ & 749 & 750\\
  C$_{66}$H$_{20}$ & 115 & 117 & C$_{22}$H$_{14}$ & 301 & 302 & C$_{190}$H$_{34}$ & 634 & 636 & C$_{146}$H$_{30}$ & 752 & 753\\
  C$_{78}$H$_{22}$ & 120 & 122 & C$_{22}$H$_{14}$ & 305 & 306 & C$_{48}$H$_{18}$ & 635 & 639 & C$_{144}$H$_{30}$ & 756 & 757\\
  C$_{36}$H$_{16}$ & 128 & 129 & C$_{22}$H$_{14}$ & 307 & 308 & C$_{90}$H$_{24}$ & 638 & 642 & C$_{142}$H$_{30}$ & 760 & 761\\
  C$_{40}$H$_{18}$ & 131 & 132 & C$_{40}$H$_{22}$ & 533 & 540 & C$_{144}$H$_{30}$ & 641 & 644 & C$_{142}$H$_{30}$ & 763 & 764\\
  C$_{42}$H$_{22}$ & 137 & 138 & C$_{36}$H$_{20}$ & 534 & 542 & C$_{57}$H$_{19}$ & 646 & 647 & C$_{142}$H$_{30}$ & 766 & 767\\
  C$_{44}$H$_{20}$ & 140 & 141 & C$_{32}$H$_{18}$ & 535 & 544 & C$_{87}$H$_{23}$ & 649 & 651 & C$_{140}$H$_{30}$ & 769 & 770\\
  C$_{48}$H$_{22}$ & 143 & 144 & C$_{32}$H$_{18}$ & 536 & 546 & C$_{119}$H$_{27}$ & 653 & 654 & C$_{138}$H$_{30}$ & 772 & 773\\
  C$_{48}$H$_{20}$ & 146 & 147 & C$_{82}$H$_{24}$ & 561 & 562 & C$_{67}$H$_{21}$ & 656 & 657 & C$_{26}$H$_{16}$ & 814 & 815\\
  C$_{30}$H$_{14}$ & 152 & 153 & C$_{98}$H$_{28}$ & 565 & 566 & C$_{71}$H$_{21}$ & 659 & 694 & C$_{30}$H$_{18}$ & 817 & 818\\
  C$_{36}$H$_{16}$ & 154 & 155 & C$_{98}$H$_{28}$ & 567 & 568 & C$_{96}$H$_{23}$ & 693 & 697 & C$_{34}$H$_{20}$ & 820 & 821\\
  C$_{42}$H$_{22}$ & 156 & 158 & C$_{32}$H$_{14}$ & 591 & 592 & C$_{96}$H$_{23}$ & 696 & 700 &  &  & \\
  C$_{110}$H$_{26}$ & 162 & 163 & C$_{66}$H$_{20}$ & 594 & 595 & C$_{96}$H$_{22}$ & 699 & 715 &  &  & \\
  C$_{112}$H$_{26}$ & 165 & 166 & C$_{66}$H$_{20}$ & 597 & 598 & C$_{66}$H$_{18}$ & 714 & 721 &  &  & \\
\hline\end{tabular}
\end{table*}

\begin{table*}
		\caption{The PAHdb v3.20 UIDs for neutral (N) and cation (C) molecules.}
		\label{tab:UIDsv310}
		\begin{tabular}{@{}cccccccccccc}
  \hline
  \multicolumn{1}{|c|}{PAH} & \multicolumn{2}{c}{UIDs} &
  \multicolumn{1}{|c|}{PAH} & \multicolumn{2}{c}{UIDs} &
  \multicolumn{1}{|c|}{PAH} & \multicolumn{2}{c}{UIDs} &
  \multicolumn{1}{|c|}{PAH} & \multicolumn{2}{c}{UIDs} \\
    & N & C & & N & C & & N & C & & N & C \\
  \hline
  C$_{34}$H$_{16}$ & 3161 & 3162 & C$_{82}$H$_{24}$ & 3203 & 3204 & C$_{55}$H$_{19}$ & 3244 & 3245 & C$_{95}$H$_{25}$ & 3277 & 3278\\
  C$_{38}$H$_{16}$ & 3164 & 3165 & C$_{88}$H$_{24}$ & 3206 & 3207 & C$_{59}$H$_{21}$ & 3247 & 3248 & C$_{95}$H$_{27}$ & 3280 & 3281\\
  C$_{46}$H$_{18}$ & 3169 & 3170 & C$_{94}$H$_{24}$ & 3211 & 3212 & C$_{65}$H$_{21}$ & 3249 & 3250 & C$_{99}$H$_{25}$ & 3282 & 3283\\
  C$_{52}$H$_{18}$ & 3173 & 3174 & C$_{102}$H$_{26}$ & 3214 & 3215 & C$_{67}$H$_{21}$ & 3252 & 3253 & C$_{103}$H$_{27}$ & 3285 & 3286\\
  C$_{54}$H$_{20}$ & 3176 & 3177 & C$_{108}$H$_{26}$ & 3217 & 3218 & C$_{67}$H$_{23}$ & 3255 & 3256 & C$_{107}$H$_{27}$ & 3287 & 3288\\
  C$_{56}$H$_{20}$ & 3179 & 3180 & C$_{120}$H$_{28}$ & 3220 & 3221 & C$_{73}$H$_{21}$ & 3257 & 3258 & C$_{107}$H$_{27}$ & 3289 & 3290\\
  C$_{58}$H$_{20}$ & 3182 & 3183 & C$_{148}$H$_{30}$ & 3223 & 3224 & C$_{77}$H$_{23}$ & 3260 & 3261 & C$_{111}$H$_{27}$ & 3291 & 3292\\
  C$_{62}$H$_{20}$ & 3185 & 3186 & C$_{31}$H$_{15}$ & 3226 & 3227 & C$_{79}$H$_{23}$ & 3263 & 3264 & C$_{113}$H$_{27}$ & 3293 & 3294\\
  C$_{64}$H$_{20}$ & 3188 & 3189 & C$_{35}$H$_{15}$ & 3229 & 3230 & C$_{83}$H$_{23}$ & 3265 & 3266 & C$_{115}$H$_{29}$ & 3296 & 3297\\
  C$_{64}$H$_{22}$ & 3191 & 3192 & C$_{37}$H$_{15}$ & 3232 & 3234 & C$_{85}$H$_{23}$ & 3268 & 3269 & C$_{115}$H$_{27}$ & 3298 & 3299\\
  C$_{72}$H$_{22}$ & 3194 & 3195 & C$_{43}$H$_{17}$ & 3235 & 3236 & C$_{87}$H$_{25}$ & 3271 & 3272 & C$_{119}$H$_{29}$ & 3300 & 3301\\
  C$_{76}$H$_{22}$ & 3197 & 3198 & C$_{45}$H$_{17}$ & 3238 & 3239 & C$_{91}$H$_{25}$ & 3273 & 3274 & C$_{121}$H$_{29}$ & 3302 & 3303\\
  C$_{80}$H$_{22}$ & 3200 & 3201 & C$_{51}$H$_{19}$ & 3241 & 3242 & C$_{93}$H$_{25}$ & 3275 & 3276 & C$_{127}$H$_{31}$ & 3304 & 3305\\
\hline\end{tabular}
\end{table*}

\begin{table*}
		\caption{PAHdb v3.20 anion molecule UIDs}
		\label{tab:UIDsv310An}
		\begin{tabular}{@{}cccccc}
  \hline
    PAH & UID & PAH & UID & PAH & UID \\
  \hline
  C$_{22}$H$_{12}^{-}$ & 3157 & C$_{64}$H$_{22}^{-}$ & 3193 & C$_{37}$H$_{15}^{-}$ & 3233\\
  C$_{28}$H$_{14}^{-}$ & 3159 & C$_{72}$H$_{22}^{-}$ & 3196 & C$_{43}$H$_{17}^{-}$ & 3237\\
  C$_{30}$H$_{14}^{-}$ & 3160 & C$_{76}$H$_{22}^{-}$ & 3199 & C$_{45}$H$_{17}^{-}$ & 3240\\
  C$_{34}$H$_{16}^{-}$ & 3163 & C$_{80}$H$_{22}^{-}$ & 3202 & C$_{51}$H$_{19}^{-}$ & 3243\\
  C$_{38}$H$_{16}^{-}$ & 3166 & C$_{82}$H$_{24}^{-}$ & 3205 & C$_{55}$H$_{19}^{-}$ & 3246\\
  C$_{42}$H$_{16}^{-}$ & 3168 & C$_{88}$H$_{24}^{-}$ & 3208 & C$_{65}$H$_{21}^{-}$ & 3251\\
  C$_{46}$H$_{18}^{-}$ & 3171 & C$_{90}$H$_{24}^{-}$ & 3210 & C$_{67}$H$_{21}^{-}$ & 3254\\
  C$_{48}$H$_{18}^{-}$ & 3172 & C$_{94}$H$_{24}^{-}$ & 3213 & C$_{73}$H$_{21}^{-}$ & 3259\\
  C$_{52}$H$_{18}^{-}$ & 3175 & C$_{102}$H$_{26}^{-}$ & 3216 & C$_{77}$H$_{23}^{-}$ & 3262\\
  C$_{54}$H$_{20}^{-}$ & 3178 & C$_{108}$H$_{26}^{-}$ & 3219 & C$_{83}$H$_{23}^{-}$ & 3267\\
  C$_{56}$H$_{20}^{-}$ & 3181 & C$_{120}$H$_{28}^{-}$ & 3222 & C$_{85}$H$_{23}^{-}$ & 3270\\
  C$_{58}$H$_{20}^{-}$ & 3184 & C$_{148}$H$_{30}^{-}$ & 3225 & C$_{95}$H$_{25}^{-}$ & 3279\\
  C$_{62}$H$_{20}^{-}$ & 3187 & C$_{31}$H$_{15}^{-}$ & 3228 & C$_{99}$H$_{25}^{-}$ & 3284\\
  C$_{64}$H$_{20}^{-}$ & 3190 & C$_{35}$H$_{15}^{-}$ & 3231 & C$_{113}$H$_{27}^{-}$ & 3295\\
\hline\end{tabular}
\end{table*}

\end{document}